\documentclass[12pt,preprint]{aastex}
\usepackage{epsfig}

\def\gtorder{\mathrel{\raise.3ex\hbox{$>$}\mkern-14mu
                \lower0.6ex\hbox{$\sim$}}}
\def\ltorder{\mathrel{\raise.3ex\hbox{$<$}\mkern-14mu
                \lower0.6ex\hbox{$\sim$}}}
\newcommand{\hii}{H~{\small II} }

\shorttitle{High Res. MIR Imaging of UC \hii Regions}
\shortauthors{Giveon et al.}

\begin{document}

\title{High Resolution Mid-Infrared Imaging of Radio Ultra-Compact \hii Regions}

\author{Giveon, U. and Richter, M.J.}
\affil{Department of Physics, University of California, Davis, CA 95616;\\ giveon@physics.ucdavis.edu, richter@physics.ucdavis.edu}
\author{Becker, R.H.}
\affil{University of California, Davis \& \\
Lawrence Livermore National Laboratory, Livermore, CA 94566;\\
bob@igpp.ucllnl.org}
\and
\author{White, R.L.}
\affil{Space Telescope Science Institute, 3700 San Martin Drive, Baltimore, MD 21218;\\ rlw@stsci.edu}

\begin{abstract}

We present data from mid-infrared Keck Telescope imaging of 18 radio-selected ultra-compact \hii region candidates at diffraction-limited resolution\footnote{The data presented herein were obtained at the W.M. Keck Observatory, which is operated as a scientific partnership among the California Institute of Technology, the University of California and the National Aeronautics and Space Administration. The Observatory was made possible by the generous financial support of the W.M. Keck Foundation.}. The goal of these observations is to determine the sizes, luminosities, and morphologies of the mid-infrared emitting dust surrounding the stellar sources. All 18 sources were imaged at 11.7 $\mu$m and at 17.65 $\mu$m, and 10 of them were imaged also at 24.5 $\mu$m. All the sources were  resolved. We have generated dust temperature and optical depth maps and combine them with radial velocity measurements and radio data (1.4 and 5 GHz) to constrain the properties of these star-forming regions. Half of our objects are excited by B-stars, and all our objects have derived types that are later than an O6 star. We find a significant correlation between infrared and radio flux densities, and a weaker one between infrared diameters and the central source ionizing photon rates. This latter correlation suggests that the more compact sources result from later spectral types rather than young age. Our new data may suggest a revision to infrared color selection criteria of ultra-compact \hii regions at resolutions $\ltorder 1''$. These 18 sources are part of a sample of 687 sources dominated by ultra-compact \hii regions selected by matching radio and infrared maps of the first Galactic quadrant by Giveon and coworkers. The new mid-infrared images constitute a significant improvement in resolving sub-structure at these wavelengths. If applied to all of this sample our analysis will improve our understanding of embedded star-formation in the Galaxy.
\end{abstract}

\keywords{Galaxy: general --- \hii regions --- infrared: ISM --- ISM: dust}

\section{Introduction}
\label{intro}

Embedded star-forming regions constitute an important contribution to the overall Galactic star-formation (Wood \& Churchwell 1989a; Churchwell 2002). The energetic radiation fields of massive stars ionize the surrounding interstellar medium, producing  radio free-free radiation, and heat their natal dust cocoons, generating mid-infrared (MIR) radiation. Measuring the radio and IR luminosities of \hii regions can therefore constrain the properties of the stellar objects embedded in them. The physical basis for the radio and IR emissions determines the spectral range most suited for these observations.
The shape of the thermal free-free spectrum in the radio is such that the optical depth decreases at higher frequencies, enabling more sources to be detected at these frequencies. In the IR, the longer the wavelength, the cooler the objects that can be observed, making the observations more sensitive to more embedded sources in the mid and far-IR (FIR). Understanding the processes and the various evolutionary phases of embedded high-mass stars (e.g., protostars, young stellar objects, and ultra-compact [UC] \hii regions) requires the study of large, complete samples. Statistically complete samples can establish the distribution of properties such as IR colors (e.g., Wood \& Churchwell 1989b), radio spectral indices, spatial distributions (e.g., Zoonematkermani et al. 1990; Becker et al. 1994; Giveon et al. 2005a; White, Becker, \& Helfand 2005), and ultimately, the spectral types and the initial mass-function and formation-rate of massive stars (e.g., Wood \& Churchwell 1989b).

Numerous studies have addressed the issue of determining the stellar content, the evolutionary phase, and the dust properties of individual embedded sources or small to medium ($<100$ sources) samples of them, by combining IR and radio observations: embedded sources and molecular clouds (Brand \& Wouterloot 1991; Mookerjea et al. 1999; Saito et al. 2006), protostars and star-forming regions (Homeier 2003; Kraemer et al. 2003; Klein et al. 2005; van der Tak, Tuthill, \& Danchi 2005; Williams, Fuller, \& Sridharan 2005), UC and compact \hii regions (Chini, Kruegel, \& Wargau 1987; Wood \& Churchwell 1989a,b; Ball et al. 1996; Mart\'{i}n-Hern\'{a}ndez, van der Hulst, \& Tielens 2003; Crowther \& Conti 2003; Alvarez et al. 2004), and masers (Testi et al. 1994; De Buizer, Pi\~{n}a, \& Telesco 2000; De Buizer 2003a, 2005a; Ellington 2006; Wu et al. 2006).
Relatively few studies (e.g., Zoonematkermani et al. 1990; Becker et al. 1994; Giveon et al. 2005a; White, Becker, \& Helfand 2005; Beltr\'{a}n et al. 2006) combined IR and radio observations to determine the characteristics of massive star-formation on a Galactic scale by analyzing large samples ($>100$ sources). One of the major drawbacks of these larger samples was the poor angular resolution the IR surveys used -- IRAS with $\sim 1'$, and the Midcourse Space Experiment (MSX) with $18.3''$, compared to the resolution of the radio surveys (typically $1''$ to a few arcseconds). Not only does the resolution mismatch severely limit comparison between the IR and radio data, it also increases the risk of false positives in IR-radio samples.

This work presents high-resolution diffraction-limited MIR observations of 18 UC \hii region candidates. These objects are part of a sample of 687 candidates selected by matching IR and radio maps of the first Galactic quadrant (Giveon et al. 2005a). The 18 sources sample almost triples the number of long wavelength ($\ge 18 \mu$m) MIR observations of UC \hii regions observed at high-resolution (De Buizer et al. 2002a,b, 2003b, 2005b, and Kraemer et al. 2003). The IR maps for finding these 687 matches were taken from the MSX archive (Egan, Price, \& Kraemer 2003), presenting a relative improvement in angular resolution compared to IRAS. The radio maps were generated from VLA observations (White, Becker, \& Helfand 2005) at an unprecedented sensitivity ($>90\%$ complete for sources with $F_{5 \rm GHz}\ge 3$ mJy) and angular resolution ($6''$).
This sample is the most complete of its kind so far, due to the quality and spatial extent of the radio maps. Giveon et al. (2005a) showed that the radio survey is $>90\%$ complete in detecting all embedded O stars across the Galaxy in the survey's area, assuming ionization-bounded nebulae.
The still relatively poor resolution of MSX can lead to confusion regarding IR-radio associations and prohibits the study of source morphologies. With higher IR resolution, it becomes feasible to study the relationships between the central stellar sources, the ionized gas, and the surrounding dust. The goal of the present observations is to verify our selection of UC \hii regions by determining their sizes, luminosities, morphologies, and stellar types of the ionizing stars.

The SPITZER program, Galactic Legacy Infrared Mid-Plane Survey Extraordinaire (GLIMPSE and GLIMPSE {\small II}), will provide maps of large parts of the Galactic plane ($|l|\le 65^{\circ}$, $|b|\le 1^{\circ}$; Benjamin et al. 2003) at wavelengths $<8\ \mu$m. The major advantages of this survey compared to the MSX survey is its higher angular resolution ($\sim 2''$), and its much higher sensitivity (0.4 mJy compared to 0.1 Jy at 8$\mu$m). While GLIMPSE will be ideal in confirming the radio and IR coincidences, and will help improve our understanding of the morphologies of our full sample of candidates, observations at longer wavelengths are needed for evaluating the dust properties.

The new Keck observations present a factor of 50 improvement in angular resolution compared to MSX. While we have begun our study in a limited sub-sample, especially for our longest wavelength (24.5 $\mu$m), in the future, we hope to include more sources in the analysis presented in this paper.
In \S \ref{obs} we describe the observations and present the sample. In \S \ref{results} we present flux density maps, dust temperature maps, and dust optical depth maps of our sample, and we derive MIR luminosities and spectral types. In \S \ref{discuss} we discuss correlations between the measured and calculated properties. In \S \ref{conc} we present our conclusions.

\section{Observations}
\label{obs}

Observations were performed in two half-nights in August 2004 on the Keck I 10 meter telescope on Mauna Kea, using the Long-Wavelength Spectrometer (LWS). We obtained images at 11.7 $\mu$m ($\Delta\lambda=1.1$ $\mu$m), 17.65 $\mu$m ($\Delta\lambda=0.85$ $\mu$m), and 24.5 $\mu$m ($\Delta\lambda=0.74$ $\mu$m). The LWS detector is a $128\times 128$ pixel Si:As array manufactured by Boeing. LWS has a field of view of $10.84''\times 10.84''$, with a scale of $0.0847''$ pixel$^{-1}$. Sky and optics background offsets were subtracted using secondary chopping throws of $10''$--$15''$ at 2.5 Hz and by nodding the telescope every 10 seconds. Frame times of 10--30 ms were used for all observations.
The standard stars $\beta$ Oph, $\alpha$ Aql, HD 2486, and HD 25477 were observed in the 11.7 $\mu$m and the 17.65 $\mu$m bands, and in a range of airmasses, during the first night. The standard stars $\beta$ Oph and $\gamma$ Aql were observed in all 3 bands, and in a range of airmasses, during the second night.
The FWHM of the standard stars was $0.3''$ and $0.4''$ at 11.7 and 17.65 $\mu$m, respectively, in the first night, and $0.4''$, $0.5''$, and $0.6''$ at 11.7, 17.65, and 24.5 $\mu$m, respectively, in the second night. These values are very close to the  diffraction limits -- $0.3''$, $0.4''$, and $0.6''$ at 11.7, 17.65, and 24.5 $\mu$m, respectively.

We designed our list of objects in a manner that allowed an efficient source acquisition, combining the sample properties and the observing methodology with the LWS instrument.
The original sample from Giveon et al. (2005a) spans longitudes $340^{\circ}$--$42^{\circ}$ and latitudes $|b|\le 0.4^{\circ}$.
We chose from the original sample all point sources ($<5''$ at 5 GHz) with IR-radio matching reliability $>95\%$ (see definition in Giveon et al. 2005a) to minimize the probability for chance coincidences. We chose sources with rising IR spectrum (8--21 $\mu$m), to include only sources with dominant IR continuum associated with blackbody temperatures $\ltorder 200$ K.
We limited ourselves to longitudes $\ge 10^{\circ}$ to overlap with the GLIMPSE {\small I} survey.
We selected only sources that were isolated (within $30''$) in the MSX maps to avoid confusion of the chop-nod procedure, used to subtract the background from the images. Since time allowed, two additional sources were added to the observations from a matching of the MSX6C catalog and a 1.4 GHz radio survey (Giveon et al. 2005b), which included later targets.

These criteria resulted in a list of 54 targets at 5 GHz. From these, 22 fields were observed, out of which only 5 yielded no MIR source within $\sim 5''$ of the radio positions. The two 1.4 GHz sources yielded one LWS detection, summing up to 24 observed fields with 18 detections. All of these 24 sources have MSX-radio counterparts within $\le 12''$ and 16 of them have their counterparts within $\le 4''$.
We suspect that the non-detections result from the large difference in angular resolution between LWS and MSX ($\times 50$), combined with the small ($10''$) field of view of LWS.
This either caused missing the source, or a false match between the IR source and a radio source, or if the match is real, the IR source may be a low surface brightness source, below the detection limit of LWS.
Because the targets were invisible in the optical, we offset the telescope to the target fields from visible stars with known coordinates, resulting in pointing accuracy typically better than $1''$. This makes the non-detections less likely to be a result of the telescope.
The GLIMPSE survey will provide higher-resolution images of our entire survey of 687 sources, and will allow a better understanding of the relation between the IR and the radio emission from these regions.
Table \ref{pos1} lists the 24 RA-sorted radio positions (the centers of the LWS fields) in columns (2)--(3); the MIR total flux densities at the corresponding MSX bands (8.3, 12.1, 14.7, and 21.3 $\mu$m) in columns (4)--(7); the integrated radio flux densities at 1.4 and 5 GHz (White, Becker, \& Helfand 2005; Helfand, Becker, \& White 2006) in columns (8)--(9); and an LWS detection flag in column (10). The sources from the 1.4 GHz survey have only 1.4 GHz data listed as their radio flux densities.
\begin{deluxetable}{cccccccccc}
\tabletypesize{\tiny}
\rotate
\tablecaption{Observed sources and their total MSX and radio flux densities \label{pos1}}
\tablewidth{0pt}
\tablehead{
 & & & \colhead{$F_{\nu}(8.3 \mu{\rm m})$} & \colhead{$F_{\nu}(12.1 \mu{\rm m})$} & \colhead{$F_{\nu}(14.7 \mu{\rm m})$} & \colhead{$F_{\nu}(21.3 \mu{\rm m})$} &  \colhead{$S_{1.4\ {\rm GHz}}$} & \colhead{$S_{5\ {\rm GHz}}$} & \\
 \colhead{Source} & \multicolumn{2}{c}{R.A. (2000) DEC.} & \colhead{(Jy)} & \colhead{(Jy)} & \colhead{(Jy)} & \colhead{(Jy)} & \colhead{(mJy)} & \colhead{(mJy)} & \colhead{LWS?} \\
\colhead{(1)} & \colhead{(2)} & \colhead{(3)} & \colhead{(4)} & \colhead{(5)} & \colhead{(6)} & \colhead{(7)} & \colhead{(8)} & \colhead{(9)} & \colhead{(10)}}
\startdata
 11.11198$-$0.39795 & 18 11 31.781 & $-$19 30 38.56 &  5.05 & 11.20 &  16.47 &  68.05 &   253.2\tablenotemark{a} &   111.9 & + \\
 11.94545$-$0.03634 & 18 11 53.107 & $-$18 36 21.85 &  7.47 & 34.51 &  47.85 & 110.71 &   250.9 &   453.4 & + \\
 16.94565$-$0.07319 & 18 21 55.944 & $-$14 13 26.80 &  2.35 &  5.96 &  11.01 &  30.11 &   258.5\tablenotemark{a} &   420.5 & $-$ \\
 18.71179+0.00085   & 18 25 04.046 & $-$12 37 45.26 &  0.46 &  1.64 &   3.31 &   9.91 &    41.0\tablenotemark{a} &   102.8 & + \\
 19.75611$-$0.12775 & 18 27 31.548 & $-$11 45 55.30 &  2.49 &  9.14 &  16.57 &  51.92 &    10.6\tablenotemark{a} &    28.8 & + \\
 21.38654$-$0.25346 & 18 31 03.794 & $-$10 22 45.88 &  1.40 &  4.51 &   8.06 &  28.98 &    51.1\tablenotemark{a} &   106.6 & + \\
 24.38654+0.28741   & 18 34 43.702 & $-$07 28 06.64 &  0.18 &  1.04 &   1.94 &   4.88 &    11.4\tablenotemark{a} &     9.4 & $-$ \\
 23.86964$-$0.12038 & 18 35 13.740 & $-$08 06 54.43 &  3.65 & 10.86 &  13.35 &  24.40 &    60.2 &    36.9 & $-$ \\
 25.39918$-$0.14081 & 18 38 08.270 & $-$06 45 57.82 & 12.51 & 38.86 &  62.48 & 235.88 &  1353.4 &   819.9 & + \\
 25.80211$-$0.15640 & 18 38 56.270 & $-$06 24 54.65 &  2.26 &  8.55 &  18.27 &  68.66 &    19.2\tablenotemark{a} &    35.2 & + \\
 27.18725$-$0.08095 & 18 41 13.166 & $-$05 08 57.73 &  1.34 &  3.06 &   5.23 &  16.77 &    28.3\tablenotemark{a} &    14.3 & + \\
 28.28875$-$0.36359 & 18 44 14.986 & $-$04 17 56.36 & 33.67 & 71.74 & 117.07 & 600.89 &   410.9\tablenotemark{a} &   527.7 & + \\
 28.30644$-$0.38385 & 18 44 21.269 & $-$04 17 33.04 &  4.52 & 16.31 &  24.12 &  72.72 & \nodata &    11.4 & $-$ \\
 30.04343$-$0.14200 & 18 46 40.207 & $-$02 38 12.19 & 12.55 & 14.38 &  14.59 &  13.81 &     4.9\tablenotemark{a} &     2.1 & + \\
 30.86744+0.11493   & 18 47 15.605 & $-$01 47 10.50 &  3.89 & 17.58 &  31.69 &  53.25 &   137.2\tablenotemark{a} &   255.3 & + \\
 30.58991$-$0.04231 & 18 47 18.797 & $-$02 06 17.86 &  1.29 &  1.45 &   3.63 &   7.96 &     7.9\tablenotemark{a} &    54.7 & + \\
 30.66808$-$0.33134 & 18 48 29.134 & $-$02 10 02.06 &  1.29 &  3.57 &   7.97 &  20.02 &    49.0\tablenotemark{a} &   153.2 & + \\
 31.24557$-$0.11285 & 18 48 45.689 & $-$01 33 13.21 &  2.01 &  4.70 &   7.91 &  21.29 &    42.9\tablenotemark{a} &   452.1 & $-$ \\
 33.91585+0.11111   & 18 52 50.117 &   +00 55 29.78 & 10.02 & 32.78 &  48.23 & 160.77 &   465.0 &   354.7 & + \\
 33.81104$-$0.18582 & 18 53 42.072 &   +00 41 46.54 &  8.43 & 18.79 &  28.55 &  55.50 & \nodata &    73.5 & + \\
 35.46832+0.13984   & 18 55 33.994 &   +02 19 10.63 & 12.40 & 20.05 &  30.20 & 155.53 &   235.1 &   321.5 & + \\
 35.13988$-$0.76237 & 18 58 10.766 &   +01 37 03.58 &  5.93 & 11.71 &  14.18 &  74.60 &    17.4 & \nodata & $-$ \\
 37.87411$-$0.39866 & 19 01 53.398 &   +04 12 48.82 & 10.06 & 45.41 &  84.51 & 149.15 &  1279.7 &   595.8 & + \\
111.28293$-$0.66355 & 23 16 04.164 &   +60 02 00.46 &  7.20 & 12.08 &  14.05 &  94.06 &   112.0 & \nodata & + \\
\enddata
\tablecomments{UC HII regions candidates observed, sorted by RA. Listed are the radio positions, total MIR flux densities of the IR counterparts measured by the MSX point-source catalog (8.3, 12.1, 14.7, and 21.3 $\mu$m), integrated radio flux densities (1.4 and 5 GHz), and LWS detection flag.}
\tablenotetext{a}{More reliable flux densities from a new multi-configuration VLA observations (Helfand, Becker, \& White 2006). See http://third.ucllnl.org/gps/index.html for details.}
\end{deluxetable}

\section{Results and Data Reduction}
\label{results}

\subsection{Flux Densities}
\label{flux}

Standard stars were observed throughout the nights in all three bands and at varying airmass values. The best-fit photometric solution was obtained for all stars of a given night simultaneously. The color terms are altogether insignificant, and only small airmass corrections, which increase with wavelength, are required. The absolute photometric accuracy was derived by summing in quadrature the standard deviation of the standard stars flux densities from the least-squares airmass fit and the Poisson error of the measured counts. The $1\sigma$ accuracies are typically $4\%$ at 11.7 $\mu$m, $5\%$ at 17.65 $\mu$m, and $20\%$ at 24.5 $\mu$m. In some cases, the subtraction of the chop-nod images resulted in artificial offsets, making parts of the image have negative counts. These residual offsets were removed manually by setting regions with no apparent emission to zero in each image.
A cut-off of $3\sigma$ above the sky level was applied to all images, and only local peaks above this cut-off are considered. Total flux densities were calculated by summing-up all pixels above the $3\sigma$ cut-off and applying the absolute calibration.

\begin{figure}
\plotone{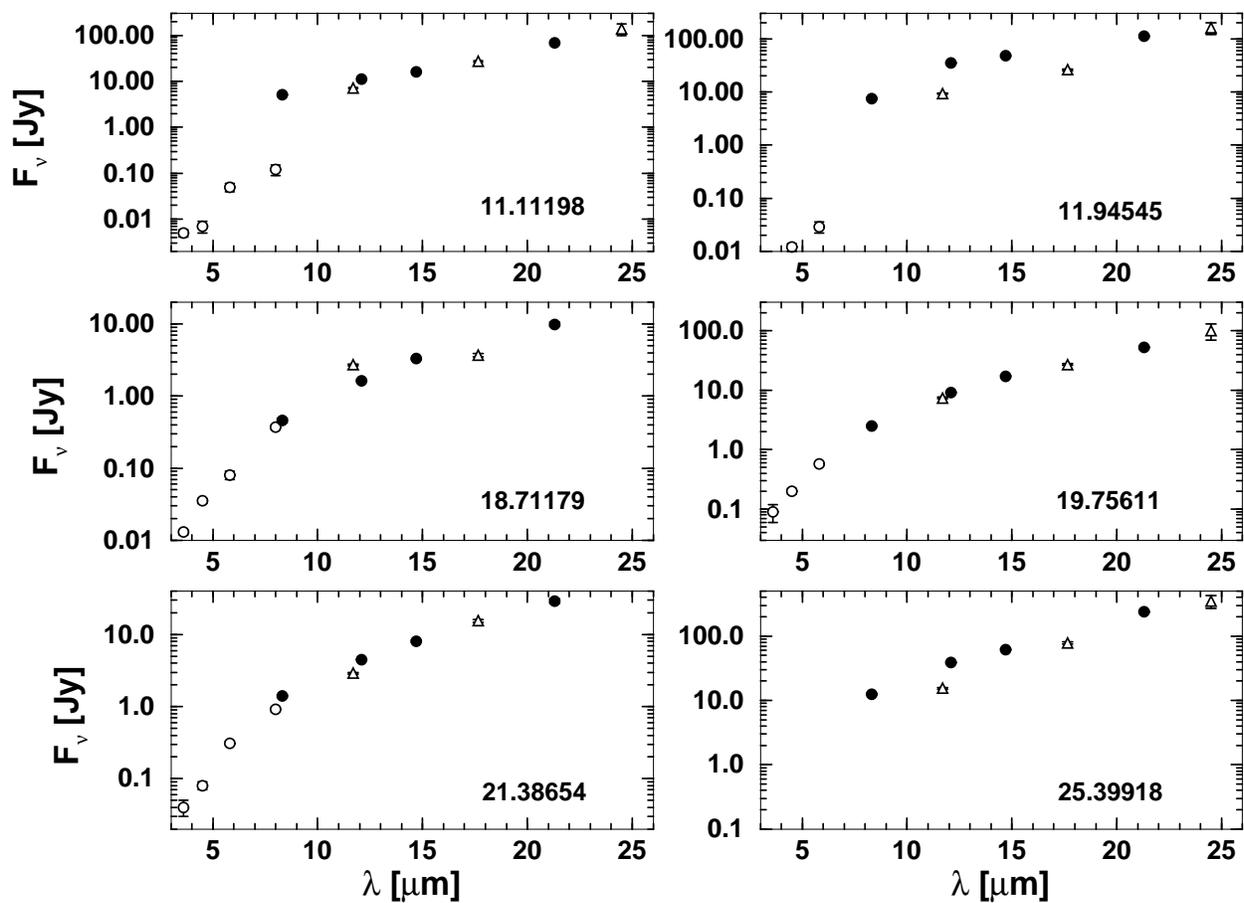}
\caption{Combined GLIMPSE--MSX--LWS SEDs of our sources with their 1$\sigma$ uncertainties: empty circles indicate GLIMPSE data points, filled circles indicate MSX data points, and triangles indicate LWS data points. Source identification is given in each panel.}
\label{sed1}
\end{figure}

\begin{figure}
\plotone{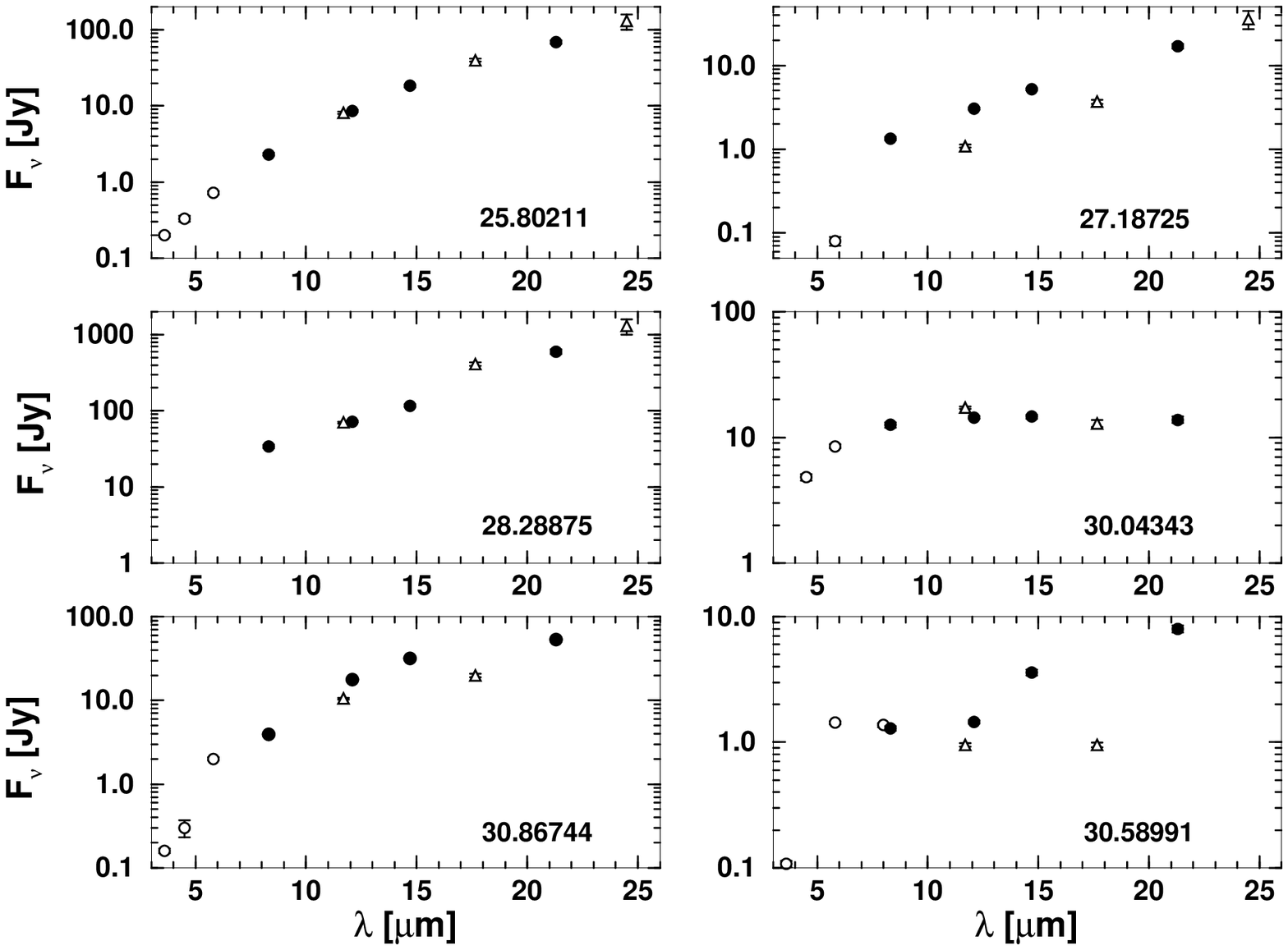}
\caption{Combined GLIMPSE--MSX--LWS SEDs of our sources with their 1$\sigma$ uncertainties: empty circles indicate GLIMPSE data points, filled circles indicate MSX data points, and triangles indicate LWS data points. Source identification is given in each panel.}
\label{sed2}
\end{figure}

\begin{figure}
\plotone{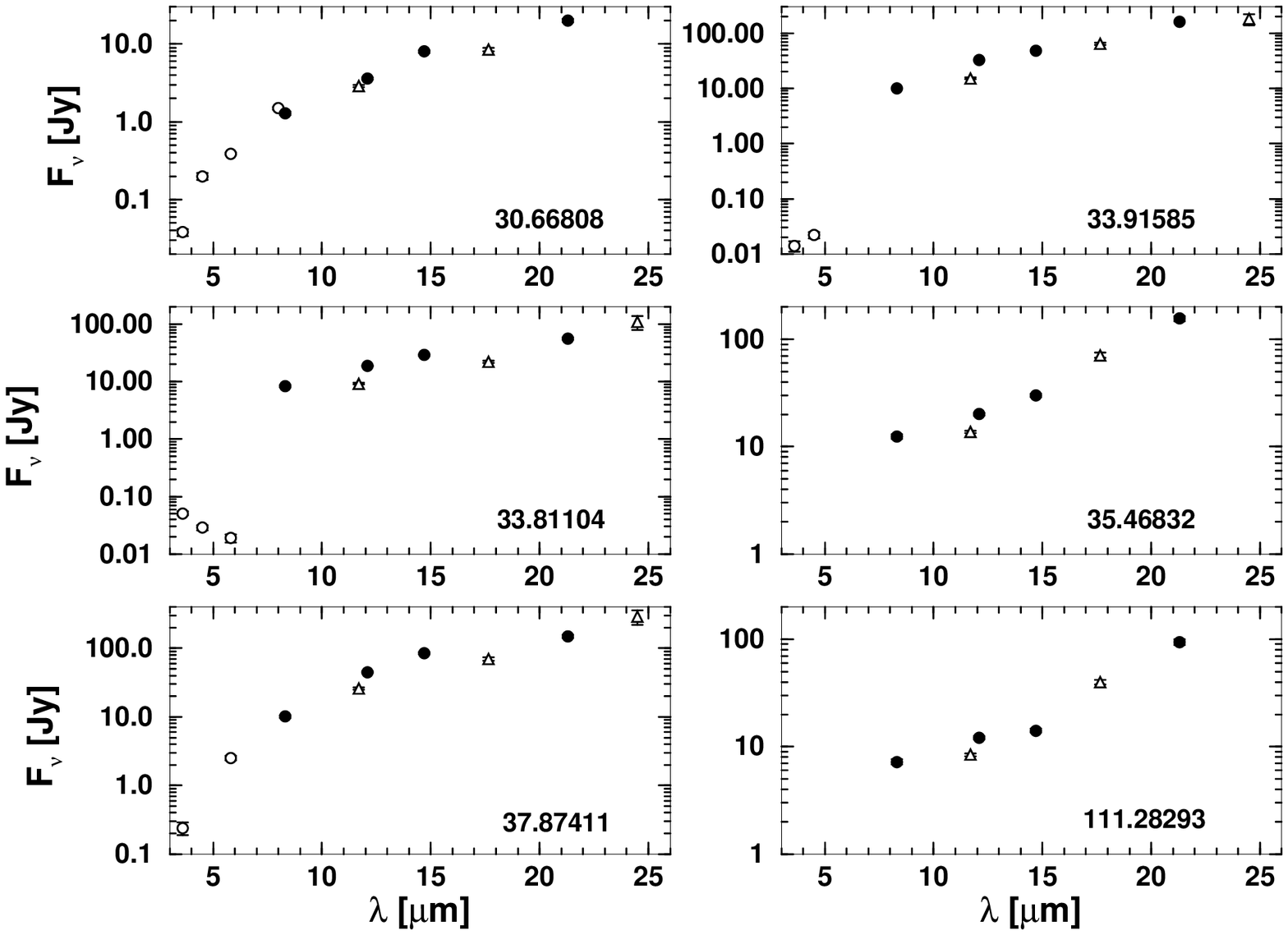}
\caption{Combined GLIMPSE--MSX--LWS SEDs of our sources with their 1$\sigma$ uncertainties: empty circles indicate GLIMPSE data points, filled circles indicate MSX data points, and triangles indicate LWS data points. Source identification is given in each panel.}
\label{sed3}
\end{figure}

Figures \ref{sed1}, \ref{sed2}, and \ref{sed3} show the LWS spectral energy distributions (SEDs) of our sources combined with GLIMPSE (flux densities at 3.6, 4.5, 5.8, and 8.0 $\mu$m; see Table \ref{glimpse}) and MSX data (Table \ref{pos1}): empty circles indicate GLIMPSE data points, filled circles indicate MSX data points, and triangles indicate LWS data points. These SEDs are not extinction corrected.

Most SEDs demonstrate the consistency of our calibrated flux densities and those from GLIMPSE and MSX, mainly for sources that are MSX point sources and are confined to the field of view in the LWS images (see LWS images, Figures \ref{map1_12} -- \ref{map18_12}). MSX had a beam of $18.3''$, while LWS has a field of view of only $10''$ across, which leads to some inconsistencies for sources with emission extending the LWS field of view (e.g., 11.94545$-$0.03634 in Figure \ref{sed1}). These plots show that our sources have rising MIR spectra typical of warm dust peaking at longer wavelengths ($60-100$ $\mu$m), but we note the irregularly flat SED of 30.04343$-$0.14200 (Figure \ref{sed2}) between 10 and 21 $\mu$m, consistently seen in both the LWS and the MSX data. An extinction correction will raise the flux density at the shorter wavelengths, making 30.04343$-$0.14200 an inverted MIR spectrum source, indicative of a hotter source (see Table \ref{tmap_data} in \S \ref{temp} for temperature information). This source is the weakest radio emitter in our sample, and it has a non-inverted radio spectrum (i.e., its 1.4 GHz flux density is greater than its 5 GHz flux density; see Table \ref{pos1}). This is possibly not an UC \hii region but another type of source, such as an AGB star, a Wolf-Rayet star, or a supernova remnant. Additional observations are required to determine the nature of this source.

\begin{deluxetable}{ccccccc}
\tabletypesize{\scriptsize}
\tablecaption{GLIMPSE flux densities of our detected sources \label{glimpse}}
\tablewidth{0pt}
\tablehead{
 & \colhead{GLIMPSE} &
\colhead{$F_{\nu}(3.6 \mu{\rm m})$} &
\colhead{$F_{\nu}(4.5 \mu{\rm m})$} &
\colhead{$F_{\nu}(5.8 \mu{\rm m})$} &
\colhead{$F_{\nu}(8.0 \mu{\rm m})$} &
\colhead{$D$} \\
\colhead{Source} &
\colhead{Name} &
\colhead{(Jy)} &
\colhead{(Jy)} &
\colhead{(Jy)} &
\colhead{(Jy)} &
\colhead{(arcsec)} \\
\colhead{(1)} & \colhead{(2)} & \colhead{(3)} & \colhead{(4)} & \colhead{(5)} & \colhead{(6)} & \colhead{(7)} }
\startdata
 11.11198$-$0.39795 & G011.1093-00.3961 & 0.005   & 0.007   & 0.05    & 0.12    &  8.8    \\
 11.94545$-$0.03634 & G011.9429-00.0393 & \nodata & 0.01    & 0.03    & \nodata & 10.4    \\
 18.71179+0.00085   & G018.7106+00.0002 & 0.01    & 0.03    & 0.08    & 0.37    &  1.6    \\
 19.75611$-$0.12775 & G019.7548-00.1283 & 0.09    & 0.20    & 0.57    & \nodata &  0.9    \\
 21.38654$-$0.25346 & G021.3857-00.2543 & 0.04    & 0.08    & 0.31    & 0.91    &  2.9    \\
 25.39918$-$0.14081 &                   & \nodata & \nodata & \nodata & \nodata & \nodata \\
 25.80211$-$0.15640 & G025.8013-00.1569 & 0.20    & 0.33    & 0.72    & \nodata &  2.6    \\
 27.18725$-$0.08095 & G027.1851-00.0817 & \nodata & \nodata & 0.08    & \nodata &  3.0    \\
 28.28875$-$0.36359 &                   & \nodata & \nodata & \nodata & \nodata & \nodata \\
 30.04343$-$0.14200 & G030.0422-00.1428 & \nodata & 4.8     & 8.5     & \nodata &  1.9    \\
 30.86744+0.11493   & G030.8663+00.1143 & 0.16    & 0.30    & 2.0     & \nodata &  1.8    \\
 30.58991$-$0.04231 & G030.5889-00.0428 & 0.11    & \nodata & 1.4     & 1.4     &  1.9    \\
 30.66808$-$0.33134 & G030.6670-00.3319 & 0.04    & 0.20    & 0.39    & 1.5     &  1.7    \\
 33.91585+0.11111   & G033.9122+00.1119 & 0.01    & 0.02    & \nodata & \nodata &  8.8    \\
 33.81104$-$0.18582 & G033.8087-00.1832 & 0.05    & 0.03    & 0.02    & \nodata & 10.7    \\
 35.46832+0.13984   &                   & \nodata & \nodata & \nodata & \nodata & \nodata \\
 37.87411$-$0.39866 & G037.8732-00.3995 & 0.24    & \nodata & 2.5     & \nodata &  3.1    \\
111.28293$-$0.66355 &                   & \nodata & \nodata & \nodata & \nodata & \nodata \\
\enddata
\tablecomments{GLIMPSE counterparts of our observed sources. We list the GLIMPSE name designation, the 3.6, 4.5, 5.8, and 8.0 $\mu$m flux densities in Janskys and the distance $D$ in arcseconds from the LWS source.}
\end{deluxetable}
Table \ref{glimpse} lists GLIMPSE counterparts of our 18 LWS sources. The GLIMPSE name designations are given in column (2); the flux densities (not corrected for extinction) at 3.6, 4.5, 5.8, and 8.0 $\mu$m in Janskys are listed in columns (3)--(6); and the distances in arcseconds from the LWS source are given in column (7).
Table 3 lists the detected sources and their LWS flux densities: multiple peaks are numbered in column (2); peak flux densities in the three bands observed are given in columns (3), (6), and (9); peak locations (RA and Dec offsets from the coordinates of the field centers given in Table \ref{pos1}) are listed in columns (4), (7), and (10); and their total flux densities in the bands observed are listed in columns (5), (8), and (11). For fields with multiple sources, we label them `1', `2', `3', etc, in decreasing order of peak flux density at 11.7 $\mu$m. The same numerical labels at different frequencies correspond to the same peaks.

\clearpage
\pagestyle{empty}
\clearpage
\pagestyle{plaintop}

Flux density maps in the observed bands are shown in Figures \ref{map1_12}--\ref{map18_12} with sub-structures indicated by a cross as listed in Table 3. The images shown are smoothed to make their resolution compatible with each other for the subsequent analysis (\S \ref{temp}). The smoothed resolution is given in the individual images. All images were rotated, so that north is up and east is to the left. The directions of increasing Galactic longitude $l$ and latitude $b$ are indicated in each map. Eight of our sources have radio maps with similar resolution in the literature:
The UC \hii regions 11.11198$-$0.39795, 28.28875$-$0.36359, 37.87411$-$0.39866, and 111.28293$-$0.66355 (Kurtz, Churchwell, \& Wood 1994) selected using IRAS colors; the radio-selected sources 25.39918$-$0.14081, 33.91585+0.11111, and 35.46832+0.13984 (Fey, Spangler, \& Cordes 1991); and 27.18725$-$0.08095 (Sridharan et al. 2002) selected as a high-mass protostellar object using FIR, radio continuum, and molecular line data. The radio maps show a good overall match in shape when compared to our LWS images.

\subsection{Dust Temperatures}
\label{temp}

Dust color temperatures maps for each source were generated from the flux density maps. When a source is observed in three bands, both $T_{17.65/11.7}$ and $T_{24.5/17.65}$ are calculated. These temperatures are usually not the same, because the sources are not necessarily perfect blackbodies. Detailed modeling of the dust properties, the geometry, and the possible spectral features in the bands observed for each source are beyond the scope of this paper. Even though our calculation is less accurate, the color temperatures and the optical depths (\S \ref{taus}) are used to understand the spatial trends in our sources, and their face values are secondary. See Li \& Chen (1996), De Buizer et al. (2002a), and De Buizer (2005a) for similar analysis.

A few preparatory steps are required in order to construct the temperature maps properly. First, all images were aligned relative to the 11.7 $\mu$m images by employing a centroid-fitting algorithm to the peak or peaks. The images were then smoothed to the resolution of the image with the poorest resolution in each pair of flux density maps (17.65 $\mu$m or 24.5 $\mu$m). The smoothing was done by convolving the images with the point-spread function (PSF) of one of the reference stars in the corresponding band. For each of the two temperatures, we have also convolved the poorer resolution image with the PSF of the better resolution band. For example, the 11.7 $\mu$m image was convolved with the 17.65 $\mu$m PSF, and the 17.65 $\mu$m image was convolved with the 11.7 $\mu$m PSF. This ensured that both images had the same effective resolution (De Buizer et al. 2002a).
A $3 \sigma$ cut-off above the noise level was applied to all images, to ensure using only pixels with minimally good signal-to-noise. Temperatures were then calculated only for diffraction limited spatial resolution elements which were above this cut-off at both bands.

We have tested the robustness of our temperature maps by shifting one of the flux density maps of each temperature by $\pm 3$ pixels ($0.25''$) in RA and Dec -- an unlikely large uncertainty in the image alignment. The absolute values of the temperature maxima changed by as much as $30\%$, but the main components remained unchanged in their shapes and relative positions. The average temperatures are much less sensitive to misalignments, and for the above shifts in the flux density maps, they typically vary by $<1\%$. This test also shows that aligning the images using the peak intensity does not bias the temperature and optical depth peaks.

\begin{figure}
\epsscale{0.8}
\plotone{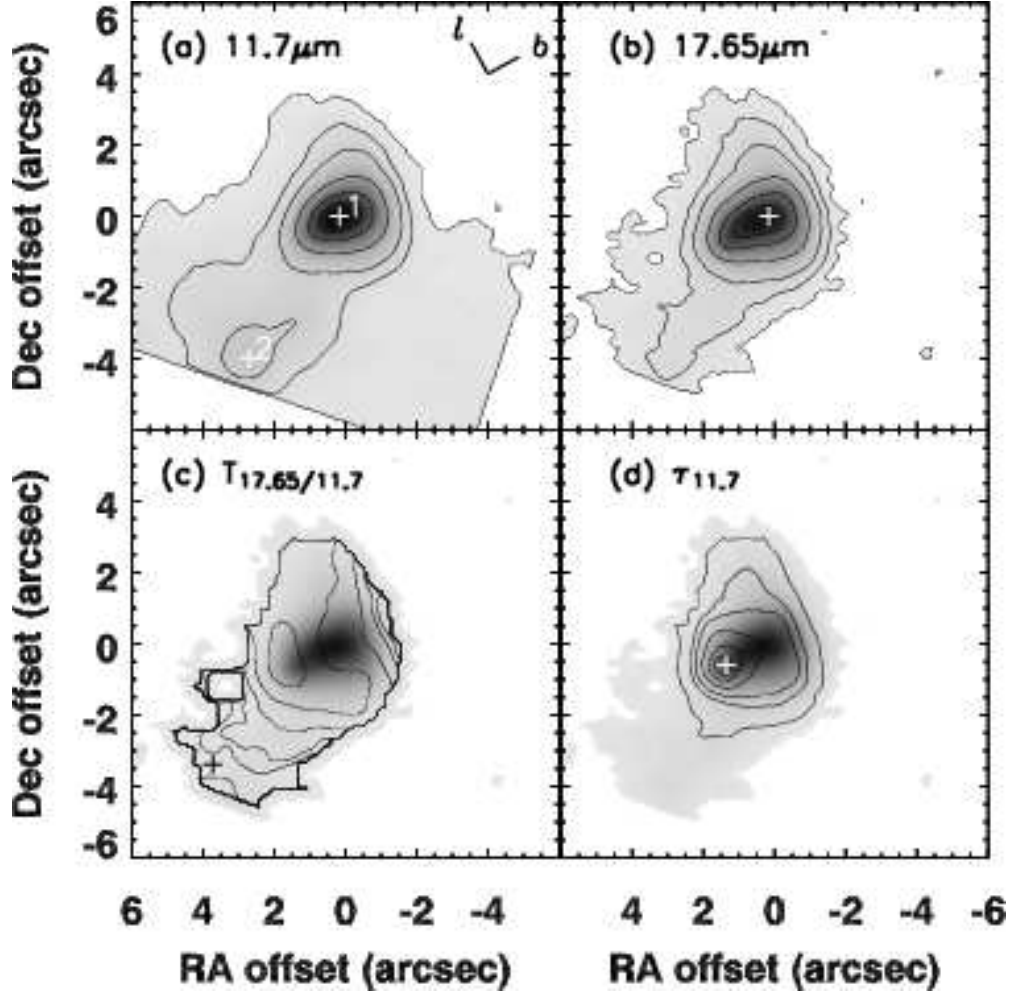}
\caption{Maps of the source 11.11198$-$0.39795: (a) The 11.7 $\mu$m flux density image smoothed to a resolution of $0.4''$ with overplotted contours. The directions of increasing Galactic longitude $l$ and latitude $b$ are indicated. (b) The 17.65 $\mu$m flux density image with overplotted contours. (c) $T_{17.65/11.7}$ plotted as contours overlaid on the 17.65 $\mu$m flux density map. The region near the flux density peak is the temperature minimum. (d) $\tau_{11.7}$ plotted as contours overlaid on the 17.65 $\mu$m flux density map. Sharp edges result from array rotation to orient north up and east to the left. Flux density contour levels are 90, 75, 50, 25, 10, 5, and 1 percent of the map peak -- 1.12, 0.93, 0.62, 0.31, 0.12, 0.06, 0.01 Jansky/arcsec$^2$ at 11.7 $\mu$m with a peak of 1.24 Jansky/arcsec$^2$ and 5.1, 4.3, 2.8, 1.4, 0.6, 0.3, 0.06 Jansky/arcsec$^2$ at 17.65 $\mu$m with a peak of 5.7 Jansky/arcsec$^2$. Temperature contour levels are 95, 90, 80, 70, 60, and 50 percent of the map peak (184 K) -- 175, 166, 147, 129, 110, 92 K. Optical depth contour levels are 75, 50, 25, 10, 5, and 1 percent of the map peak (0.07) -- 0.053, 0.035, 0.018, 0.007, 0.004, 0.0007. Peaks are indicated by a cross. Multiple peaks are sorted numerically or alphabetically (see Tables \ref{tmap_data} and \ref{taumap_data}).}
\label{map1_12}
\end{figure}

\begin{figure}
\epsscale{0.8}
\plotone{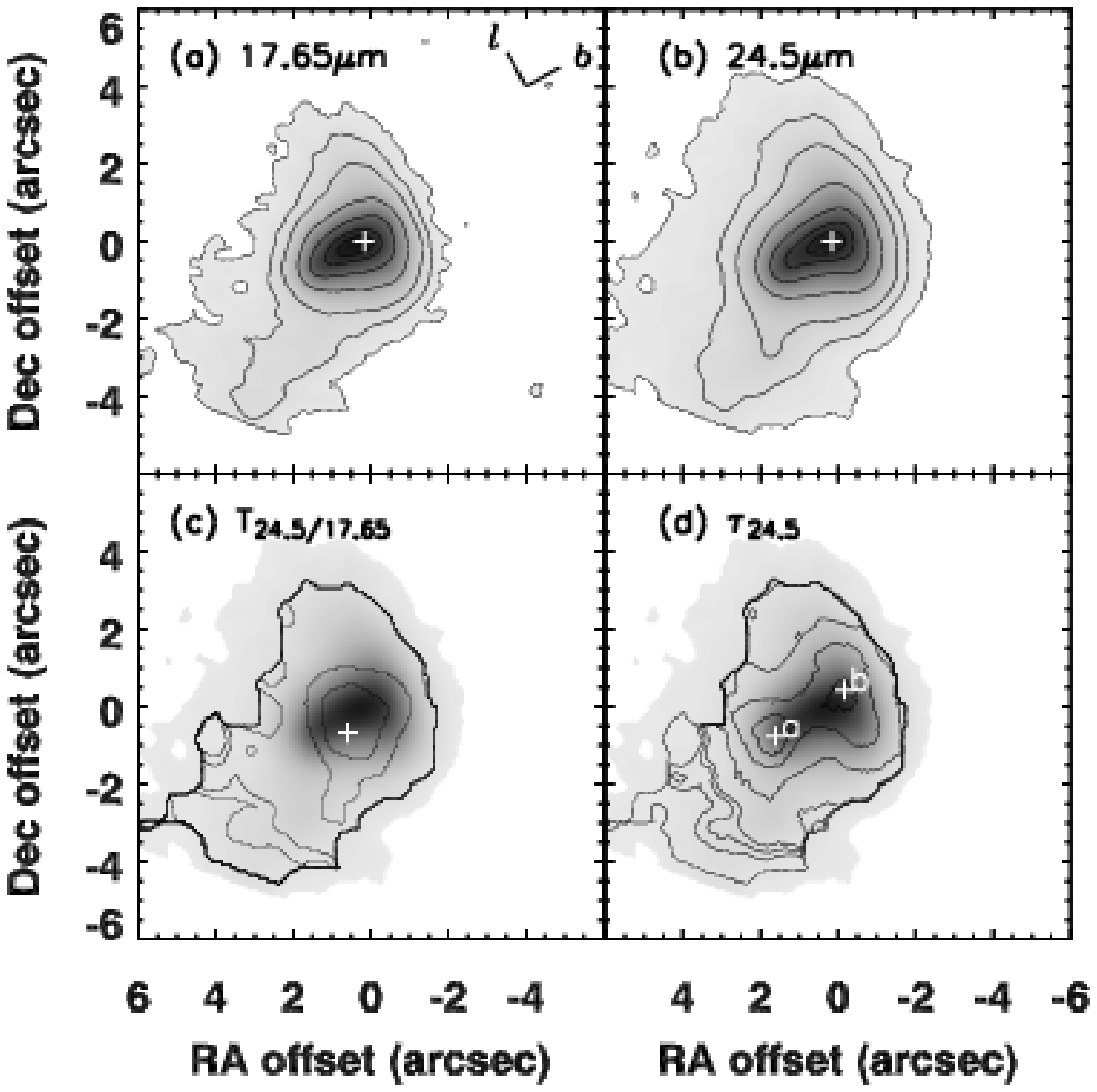}
\caption{Maps of the source 11.11198$-$0.39795: (a) The 17.65 $\mu$m flux density image smoothed to a resolution of $0.6''$ with overplotted contours. The directions of increasing Galactic longitude $l$ and latitude $b$ are indicated. (b) The 24.5 $\mu$m flux density image with overplotted contours. (c) $T_{24.5/17.65}$ plotted as contours overlaid on the 24.5 $\mu$m flux density map. (d) $\tau_{24.5}$ plotted as contours overlaid on the 24.5 $\mu$m flux density map. Sharp edges result from array rotation to orient north up and east to the left. The 17.65$\mu$m flux density contour levels are 5.1, 4.3, 2.8, 1.4, 0.6, 0.3, 0.06 Jansky/arcsec$^2$ with a peak of 5.7 Jansky/arcsec$^2$. The 24.5$\mu$m flux density contour levels are 19.8, 16.5, 11.0, 5.5, 2.2, 1.1, 0.2 Jansky/arcsec$^2$ with a peak of 22 Jansky/arcsec$^2$. Temperature contour levels are 77, 73, 65, 57, 49, 41 K, with a temperature peak of 81 K. Optical depth contour levels are 0.53, 0.35, 0.18, 0.07, 0.04, 0.007 with a peak of 0.7. Peaks are indicated by a cross. Multiple peaks are sorted numerically or alphabetically (see Tables \ref{tmap_data} and \ref{taumap_data}).}
\label{map1_25}
\end{figure}

\begin{figure}
\epsscale{0.8}
\plotone{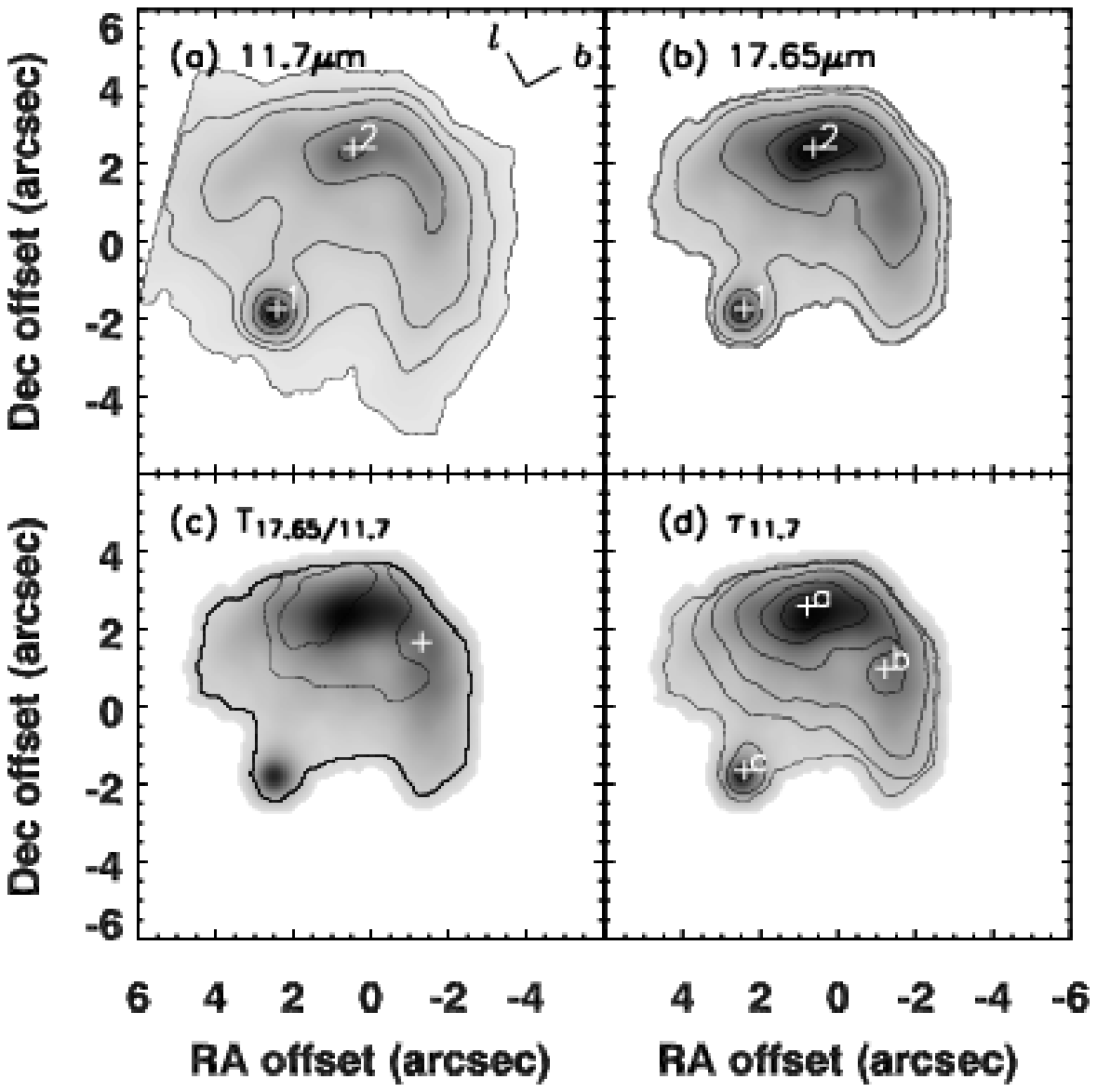}
\caption{Maps of the source 11.94545$-$0.03634: (a), (b), (c), and (d) are the same as in Figure \ref{map1_12}. The 11.7$\mu$m flux density contour levels are 1.11, 0.92, 0.62, 0.31, 0.12, 0.06, 0.01 Jansky/arcsec$^2$ with a peak of 1.23 Jansky/arcsec$^2$. The 17.65$\mu$m flux density contour levels are 2.36, 1.96, 1.31, 0.65, 0.26, 0.13, 0.03 Jansky/arcsec$^2$ with a peak of 2.62 Jansky/arcsec$^2$. Temperature contour levels are 151, 143, 127, 111, 95, 80 K, with a temperature peak of 159 K. Optical depth contour levels are 0.0049, 0.0033, 0.0016, 0.0007, 0.00033, $6.5\cdot 10^{-5}$ with a peak of 0.0065. Sharp edges result from array rotation to orient north up and east to the left. In the temperature map, the region near the flux density peak is the temperature minimum.}
\label{map2}
\end{figure}

\begin{figure}
\epsscale{0.8}
\plotone{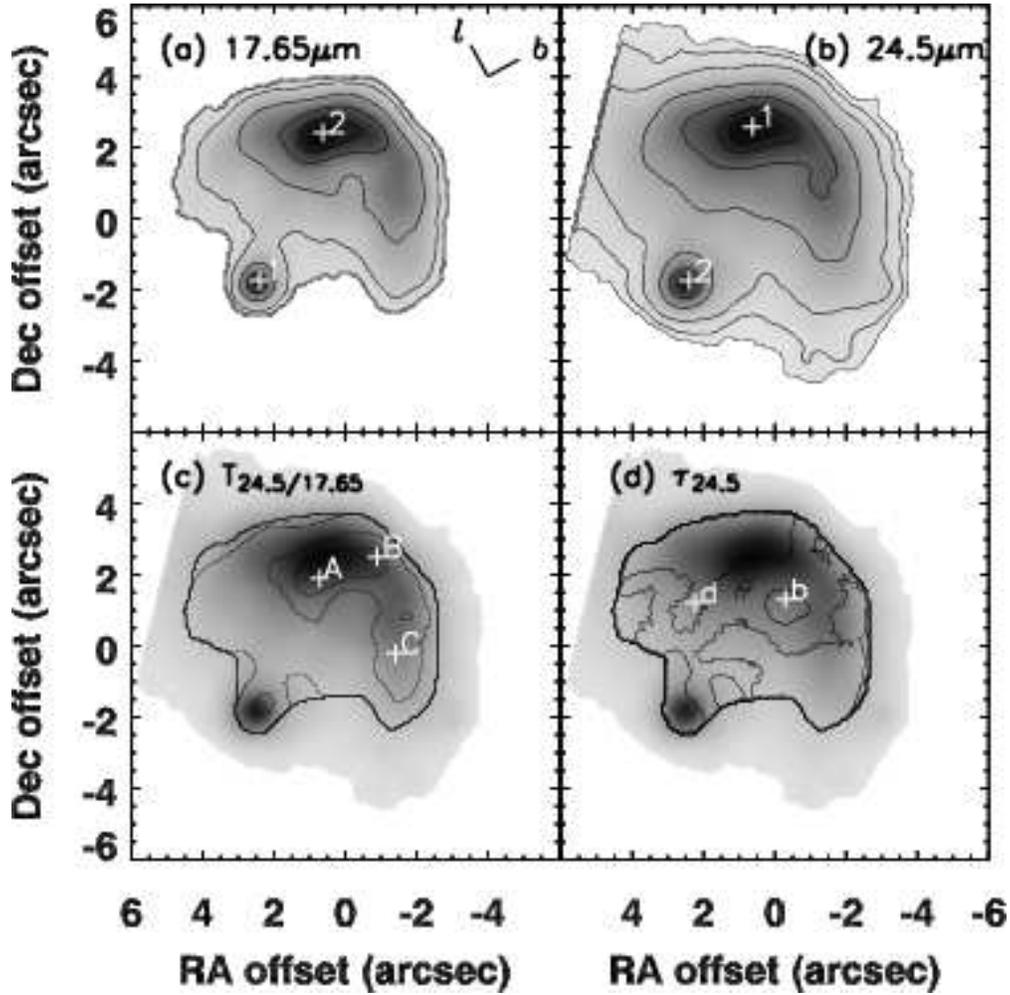}
\caption{Maps of the source 11.94545$-$0.03634: (a), (b), (c), and (d) are the same as in Figure \ref{map1_25}. The 17.65$\mu$m flux density contour levels are 2.36, 1.96, 1.31, 0.65, 0.26, 0.13, 0.03 Jansky/arcsec$^2$ with a peak of 2.62 Jansky/arcsec$^2$. The 24.5$\mu$m flux density contour levels are 8.93, 7.44, 4.96, 2.48, 0.99, 0.50, 0.10 Jansky/arcsec$^2$ with a peak of 9.92 Jansky/arcsec$^2$. Temperature contour levels are 77, 73, 65, 57, 49, 41 K, with a temperature peak of 81 K. Optical depth contour levels are 0.25, 0.17, 0.08, 0.03, 0.02, 0.003 with a peak of 0.33. Sharp edges result from array rotation to orient north up and east to the left.}
\label{map2_25}
\end{figure}

\begin{figure}
\epsscale{0.8}
\plotone{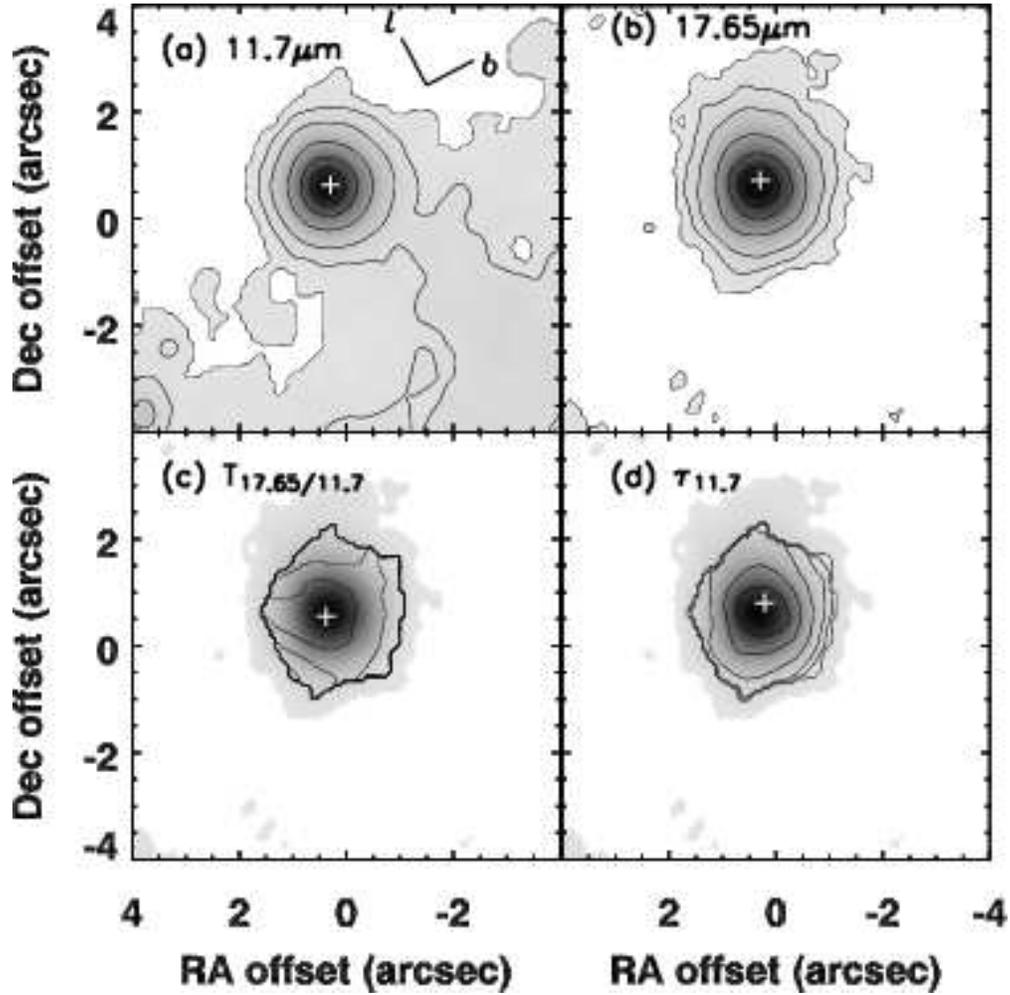}
\caption{Maps of the source 18.71179+0.00085: (a), (b), (c), and (d) are the same as in Figure \ref{map1_12}. The 11.7$\mu$m flux density contour levels are 0.53, 0.44, 0.29, 0.15, 0.06, 0.03, 0.006 Jansky/arcsec$^2$ with a peak of 0.59 Jansky/arcsec$^2$. The 17.65$\mu$m flux density contour levels are 1.7, 1.4, 0.9, 0.5, 0.2, 0.1, 0.02 Jansky/arcsec$^2$ with a peak of 1.89 Jansky/arcsec$^2$. Temperature contour levels are 146, 137, 123, 108, 92, 77 K, with a temperature peak of 154 K. Optical depth contour levels are 0.0023, 0.0015, 0.0008, 0.0003, 0.0002, $3.0\cdot 10^{-5}$ with a peak of 0.003.}
\label{map3}
\end{figure}

\begin{figure}
\epsscale{0.8}
\plotone{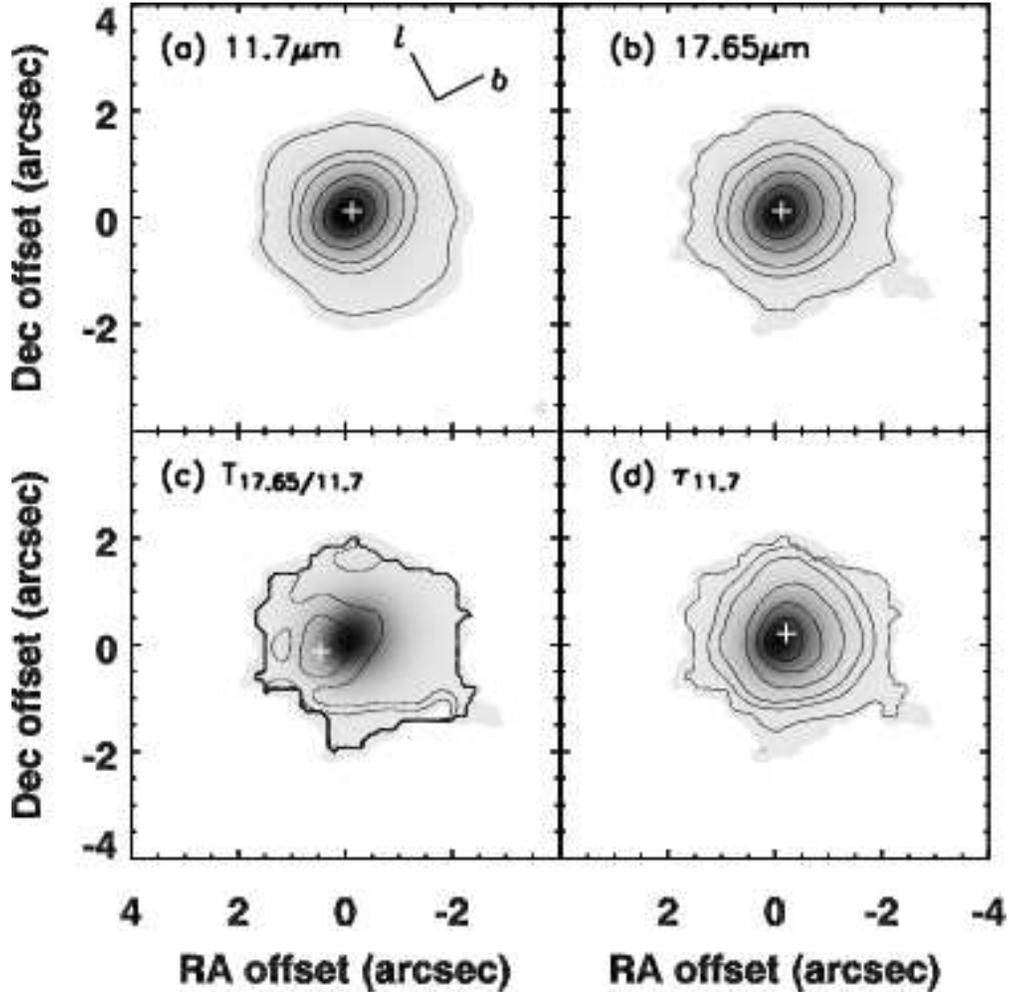}
\caption{Maps of the source 19.75611$-$0.12775: (a), (b), (c), and (d) are the same as in Figure \ref{map1_12}. The 11.7$\mu$m flux density contour levels are 5.0, 4.2, 2.8, 1.4, 0.6, 0.3, 0.06 Jansky/arcsec$^2$ with a peak of 5.6 Jansky/arcsec$^2$. The 17.65$\mu$m flux density contour levels are 16.6, 13.8, 9.2, 4.6, 1.8, 0.9, 0.2 Jansky/arcsec$^2$ with a peak of 18.4 Jansky/arcsec$^2$. Temperature contour levels are 151, 143, 127, 111, 95, 80 K, with a temperature peak of 159 K. Optical depth contour levels are 0.022, 0.015, 0.007, 0.003, 0.001, 0.0003 with a peak of 0.029.}
\label{map4_12}
\end{figure}

\begin{figure}
\epsscale{0.8}
\plotone{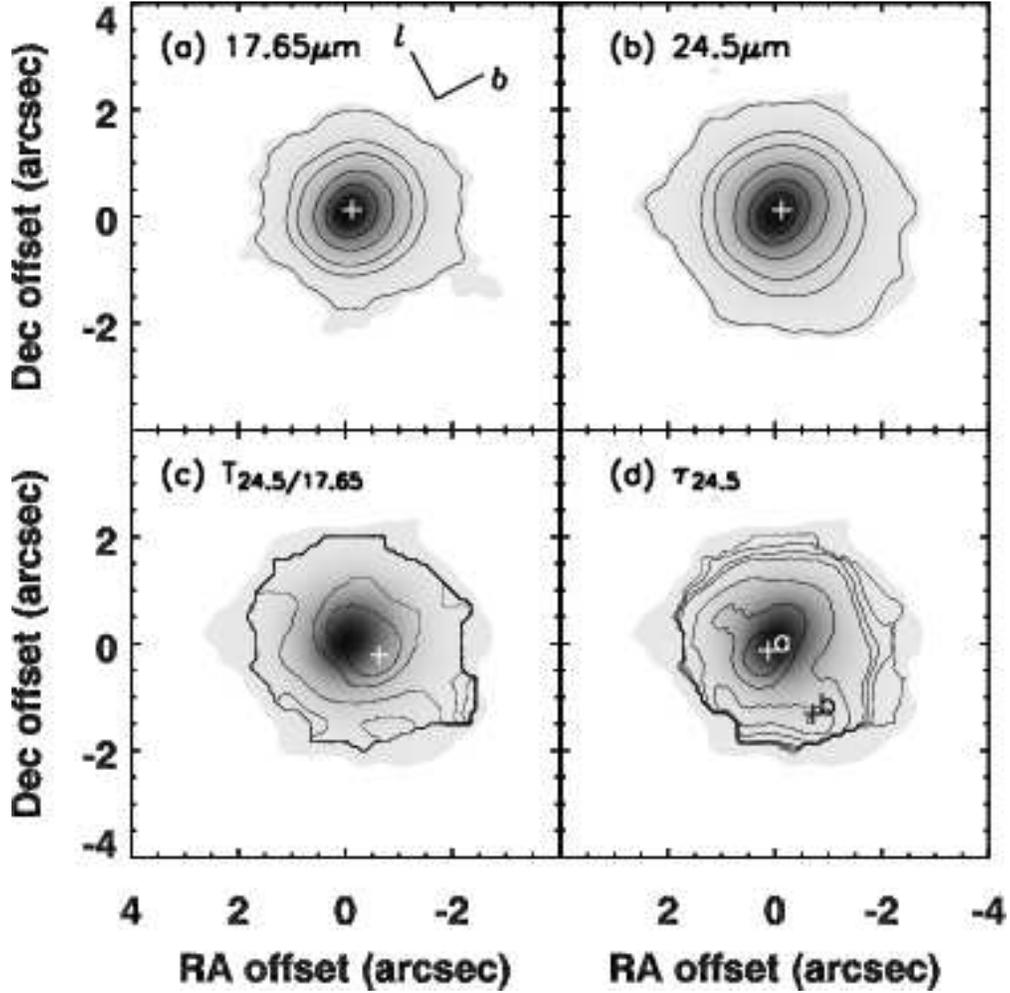}
\caption{Maps of the source 19.75611$-$0.12775: (a), (b), (c), and (d) are the same as in Figure \ref{map1_25}. The 17.65$\mu$m flux density contour levels are 16.6, 13.8, 9.2, 4.6, 1.8, 0.9, 0.2 Jansky/arcsec$^2$ with a peak of 18.4 Jansky/arcsec$^2$. The 24.5$\mu$m flux density contour levels are 42.3, 35.3, 23.5, 11.8, 4.7, 2.4, 0.5 Jansky/arcsec$^2$ with a peak of 47 Jansky/arcsec$^2$. Temperature contour levels are 86, 82, 73, 64, 55, 46 K, with a temperature peak of 91 K. Optical depth contour levels are 0.45, 0.30, 0.15, 0.06, 0.03, 0.006 with a peak of 0.60.}
\label{map4_25}
\end{figure}

\begin{figure}
\epsscale{0.8}
\plotone{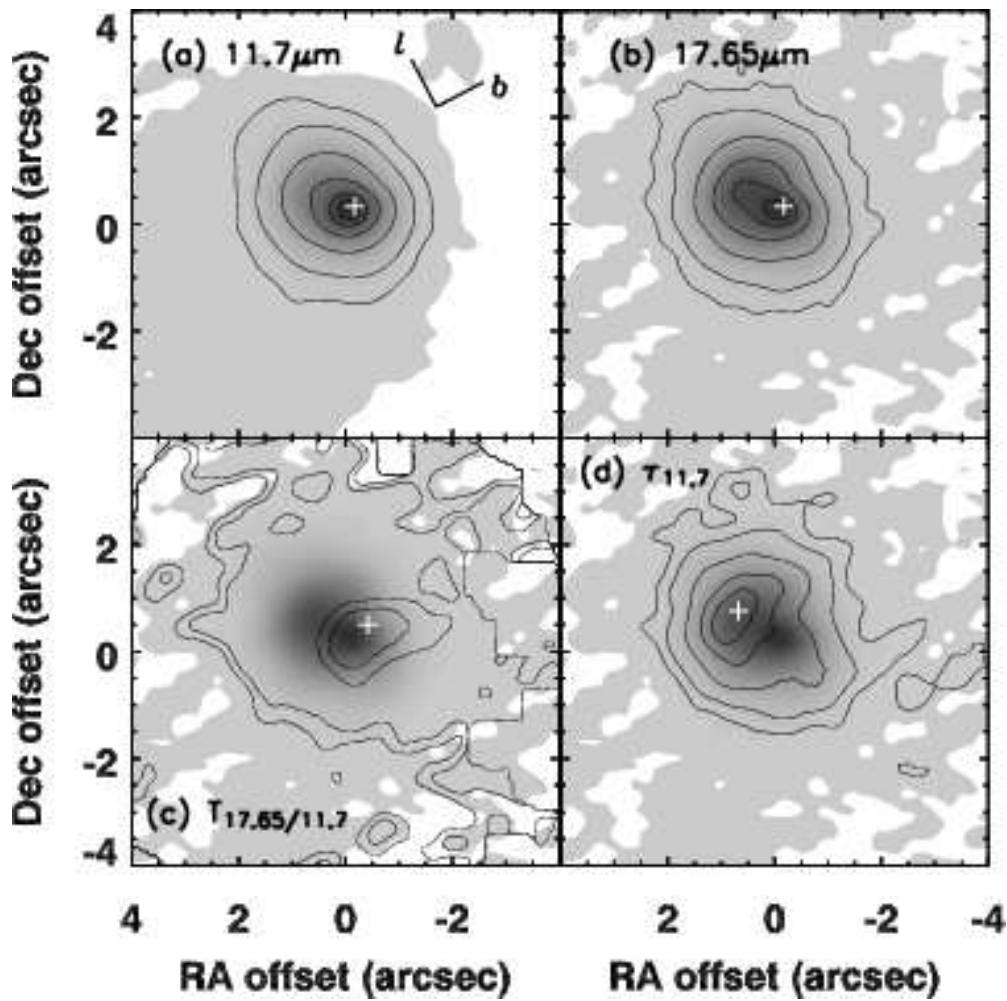}
\caption{Maps of the source 21.38654$-$0.25346: (a), (b), (c), and (d) are the same as in Figure \ref{map1_12}. The 11.7$\mu$m flux density contour levels are 1.22, 1.01, 0.68, 0.34, 0.14, 0.07 Jansky/arcsec$^2$ with a peak of 1.35 Jansky/arcsec$^2$. The 17.65$\mu$m flux density contour levels are 3.6, 3.0, 2.0, 1.0, 0.4, 0.2 Jansky/arcsec$^2$ with a peak of 4 Jansky/arcsec$^2$. Temperature contour levels are 151, 143, 127 K, with a temperature peak of 159 K. Optical depth contour levels are 0.0082, 0.0055, 0.0027, 0.0011, 0.0005 with a peak of 0.011.}
\label{map5}
\end{figure}

\begin{figure}
\epsscale{0.8}
\plotone{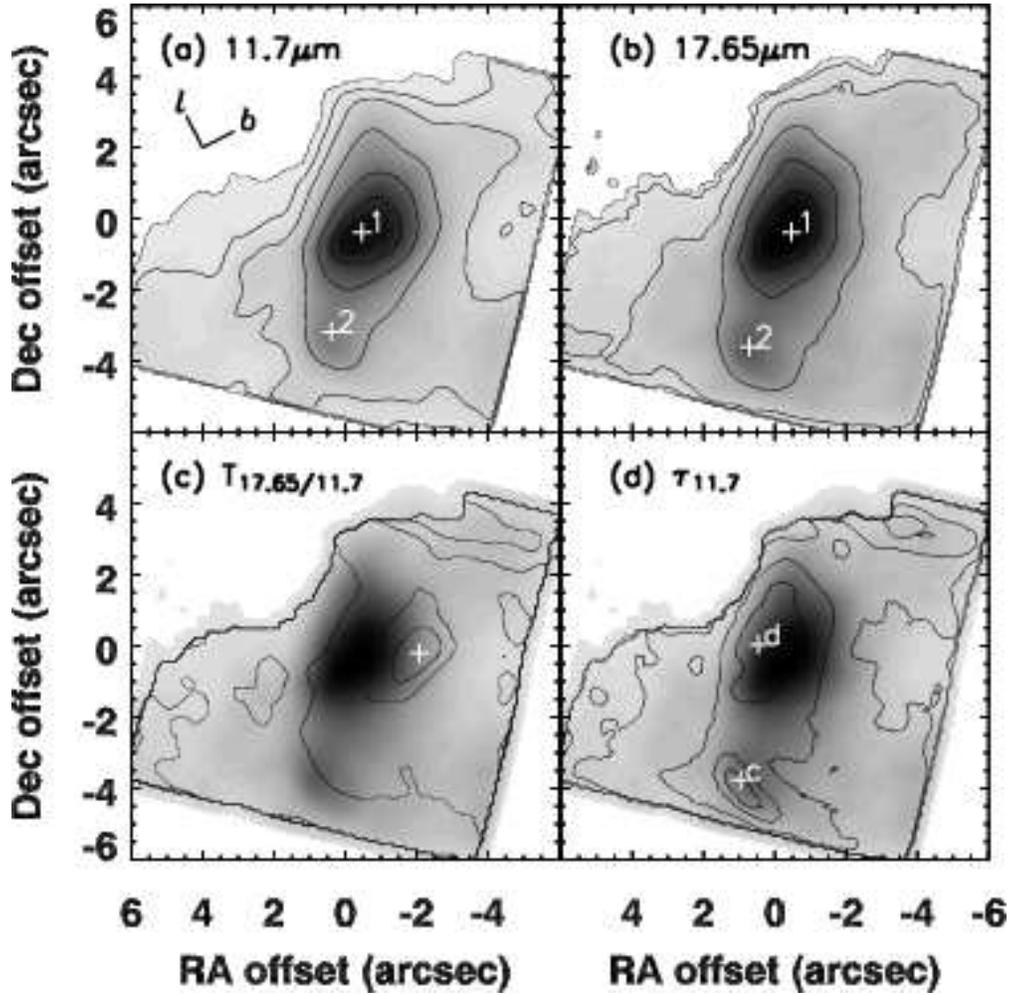}
\caption{Maps of the source 25.39918$-$0.14081: (a), (b), (c), and (d) are the same as in Figure \ref{map1_12}. The 11.7$\mu$m flux density contour levels are 0.9, 0.75, 0.5, 0.25, 0.1, 0.05, 0.01 Jansky/arcsec$^2$ with a peak of 1.0 Jansky/arcsec$^2$. The 17.65$\mu$m flux density contour levels are 3.8, 3.1, 2.1, 1.1, 0.4, 0.2, 0.04 Jansky/arcsec$^2$ with a peak of 4.2 Jansky/arcsec$^2$. Temperature contour levels are 150, 142, 126, 111, 95, 79 K, with a temperature peak of 158 K. Optical depth contour levels are 0.014, 0.009, 0.005, 0.002, 0.001, 0.0002 with a peak of 0.019. Sharp edges result from array rotation to orient north up and east to the left.}
\label{map6_12}
\end{figure}

\begin{figure}
\epsscale{0.8}
\plotone{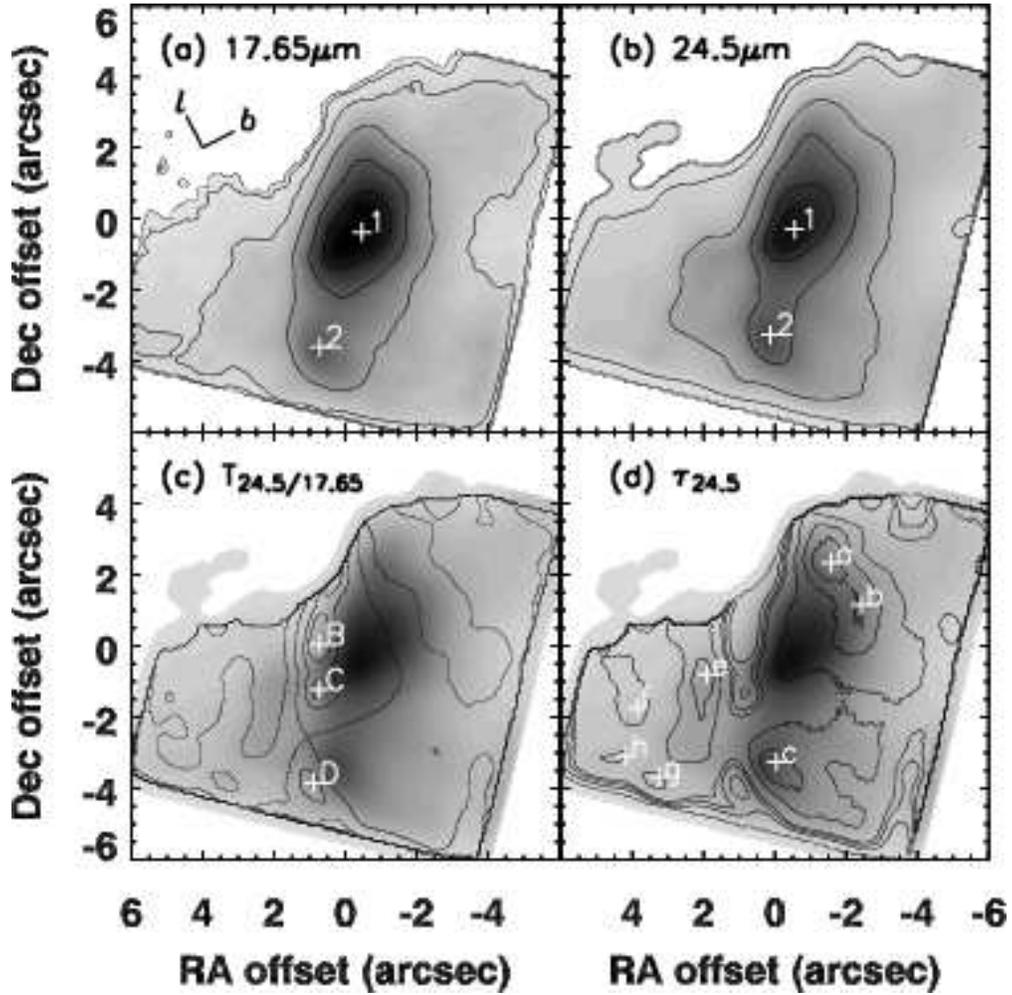}
\caption{Maps of the source 25.39918$-$0.14081: (a), (b), (c), and (d) are the same as in Figure \ref{map1_25}. The 17.65$\mu$m flux density contour levels are 3.8, 3.1, 2.1, 1.1, 0.4, 0.2, 0.04 Jansky/arcsec$^2$ with a peak of 4.2 Jansky/arcsec$^2$. The 24.5$\mu$m flux density contour levels are 12.6, 10.5, 7.0, 3.5, 1.4, 0.7, 0.1 Jansky/arcsec$^2$ with a peak of 14 Jansky/arcsec$^2$. Temperature contour levels are 96, 91, 81, 71, 61, 51 K, with a temperature peak of 101 K. Optical depth contour levels are 0.60, 0.40, 0.20, 0.08, 0.04, 0.008 with a peak of 0.80. Sharp edges result from array rotation to orient north up and east to the left.}
\label{map6_25}
\end{figure}

\begin{figure}
\epsscale{0.8}
\plotone{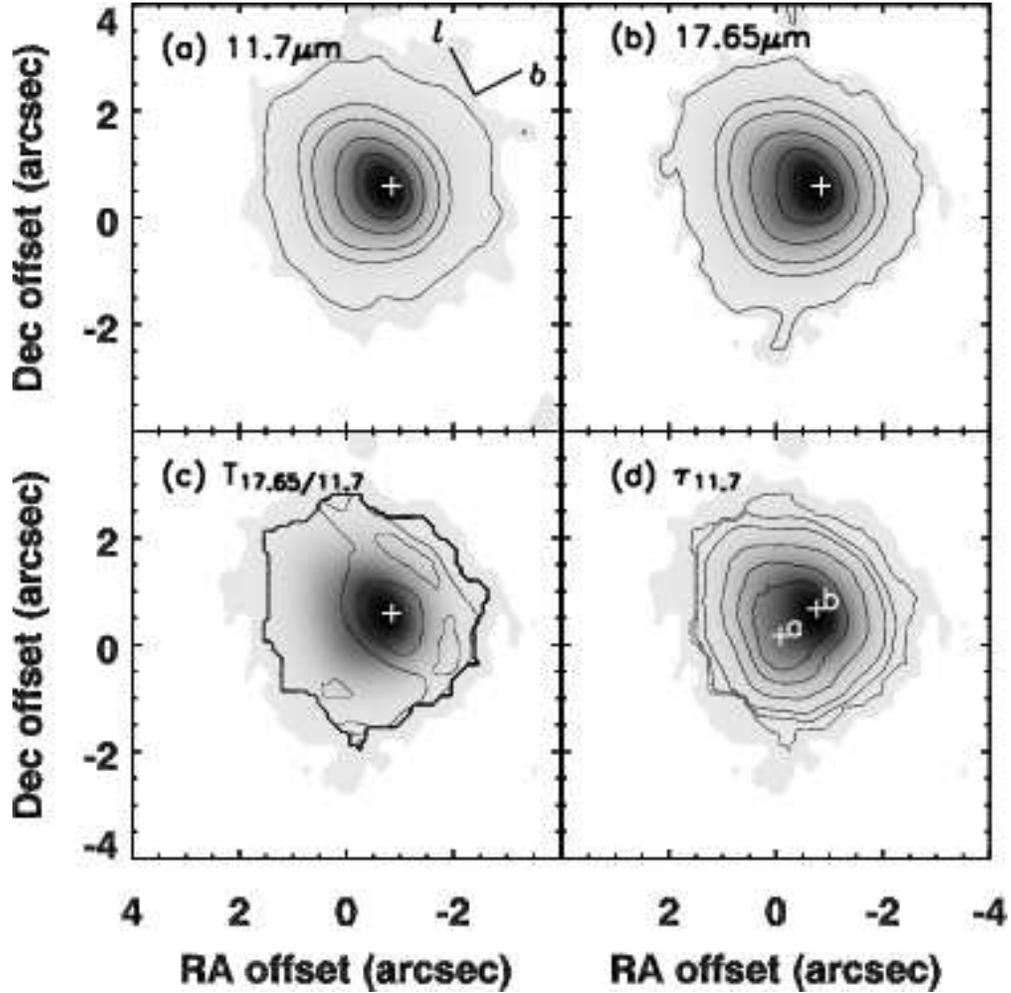}
\caption{Maps of the source 25.80211$-$0.15640: (a), (b), (c), and (d) are the same as in Figure \ref{map1_12}. The 11.7$\mu$m flux density contour levels are 3.7, 3.1, 2.1, 1.0, 0.4, 0.2, 0.04 Jansky/arcsec$^2$ with a peak of 4.1 Jansky/arcsec$^2$. The 17.65$\mu$m flux density contour levels are 14.3, 11.9, 8.0, 4.0, 1.6, 0.8, 0.2 Jansky/arcsec$^2$ with a peak of 15.9 Jansky/arcsec$^2$. Temperature contour levels are 136, 129, 114, 100, 86, 72 K, with a temperature peak of 143 K. Optical depth contour levels are 0.029, 0.020, 0.010, 0.004, 0.002, 0.0004 with a peak of 0.039.}
\label{map7_12}
\end{figure}

\begin{figure}
\epsscale{0.8}
\plotone{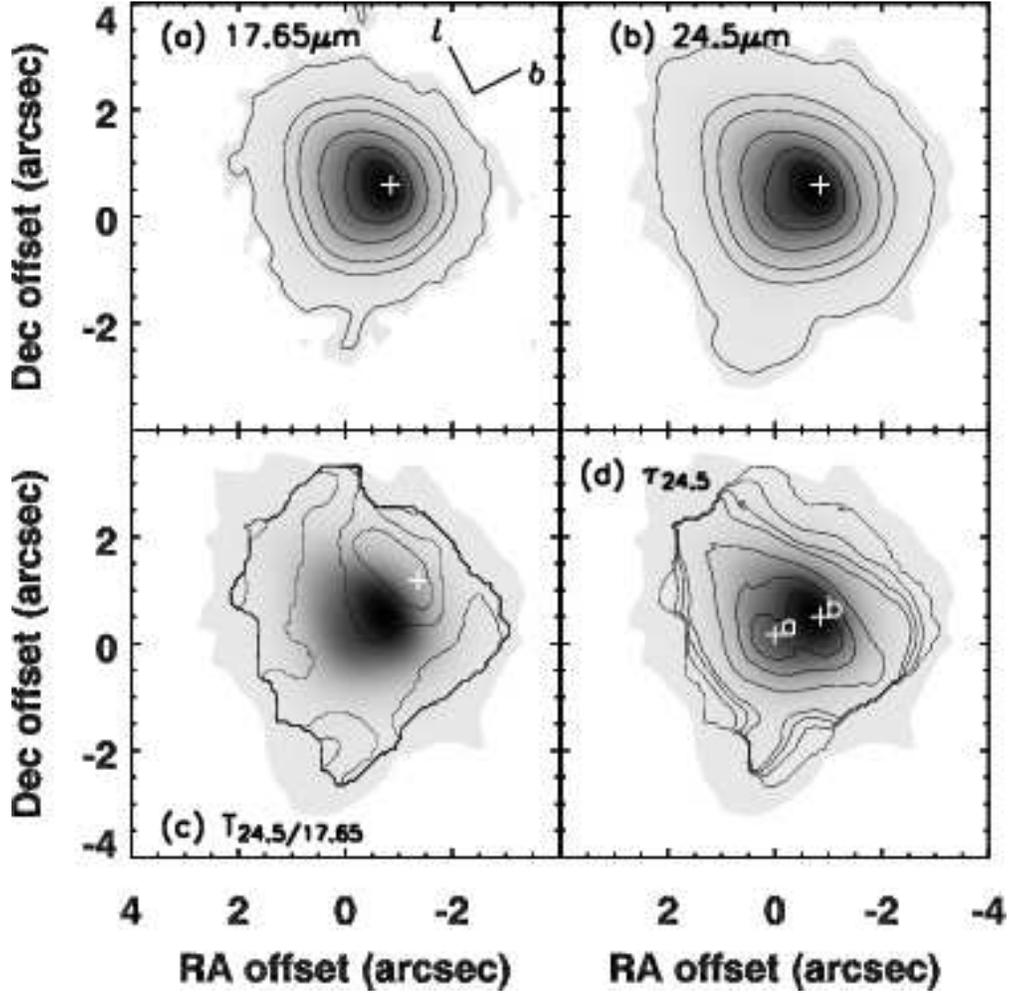}
\caption{Maps of the source 25.80211$-$0.15640: (a), (b), (c), and (d) are the same as in Figure \ref{map1_25}. The 17.65$\mu$m flux density contour levels are 14.3, 11.9, 8.0, 4.0, 1.6, 0.8, 0.2 Jansky/arcsec$^2$ with a peak of 15.9 Jansky/arcsec$^2$. The 24.5$\mu$m flux density contour levels are 36.0, 30.0, 20.0, 10.0, 4.0, 2.0, 0.4 Jansky/arcsec$^2$ with a peak of 40 Jansky/arcsec$^2$. Temperature contour levels are 96, 91, 81, 71, 61, 51 K, with a temperature peak of 101 K. Optical depth contour levels are 0.26, 0.18, 0.09, 0.04, 0.02, 0.003 with a peak of 0.35.}
\label{map7_25}
\end{figure}

\begin{figure}
\epsscale{0.8}
\plotone{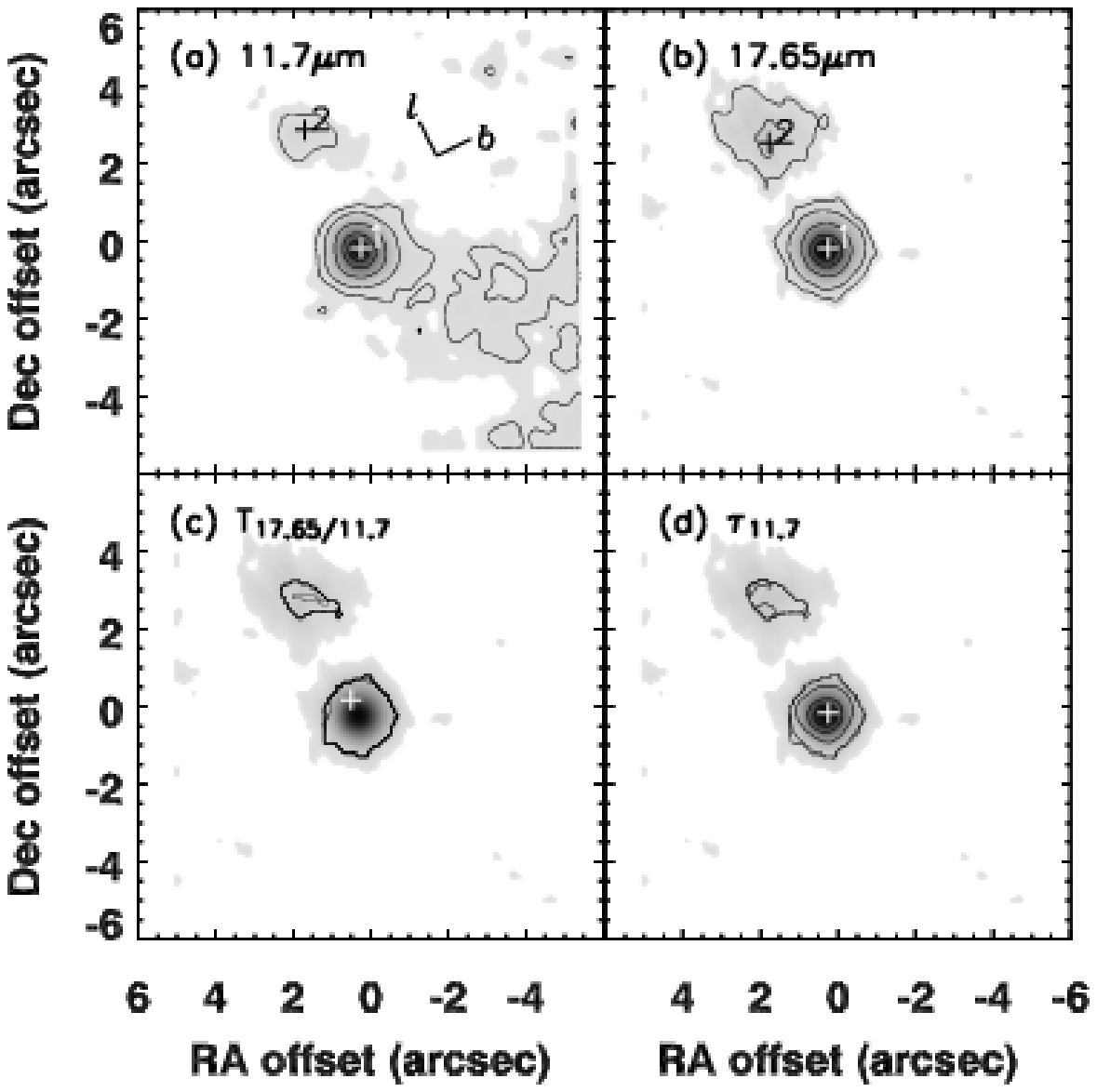}
\caption{Maps of the source 27.18725$-$0.08095: (a), (b), (c), and (d) are the same as in Figure \ref{map1_12}. The 11.7$\mu$m flux density contour levels are 0.39, 0.32, 0.22, 0.11, 0.04, 0.02, 0.004 Jansky/arcsec$^2$ with a peak of 0.43 Jansky/arcsec$^2$. The 17.65$\mu$m flux density contour levels are 1.75, 1.46, 0.97, 0.49, 0.19, 0.10, 0.02 Jansky/arcsec$^2$ with a peak of 1.94 Jansky/arcsec$^2$. Temperature contour levels are 133, 126, 112, 98, 84, 70 K, with a temperature peak of 140 K. Optical depth contour levels are 0.0038, 0.0025, 0.0013, 0.0005, 0.0003, $5\cdot 10^{-5}$ with a peak of 0.005.}
\label{map8_12}
\end{figure}

\begin{figure}
\epsscale{0.8}
\plotone{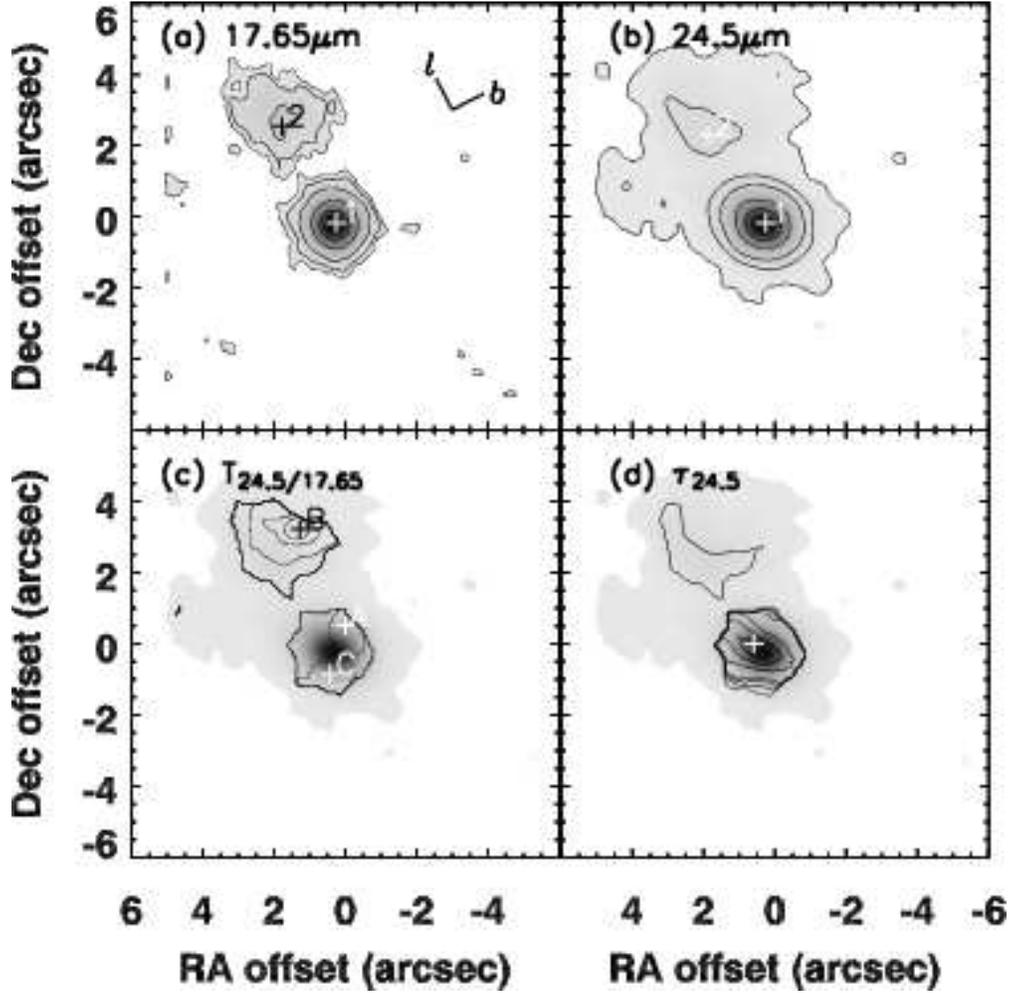}
\caption{Maps of the source 27.18725$-$0.08095: (a), (b), (c), and (d) are the same as in Figure \ref{map1_25}. The 17.65$\mu$m flux density contour levels are 1.75, 1.46, 0.97, 0.49, 0.19, 0.10, 0.02 Jansky/arcsec$^2$ with a peak of 1.94 Jansky/arcsec$^2$. The 24.5$\mu$m flux density contour levels are 16.2, 13.5, 9.0, 4.5, 1.8, 0.9, 0.2 Jansky/arcsec$^2$ with a peak of 18 Jansky/arcsec$^2$. Temperature contour levels are 80, 76, 67, 59, 50, 42 K, with a temperature peak of 84 K. Optical depth contour levels are 5.3, 3.5, 1.8, 0.7, 0.35, 0.01 with a peak of 7.}
\label{map8_25}
\end{figure}

\begin{figure}
\epsscale{0.8}
\plotone{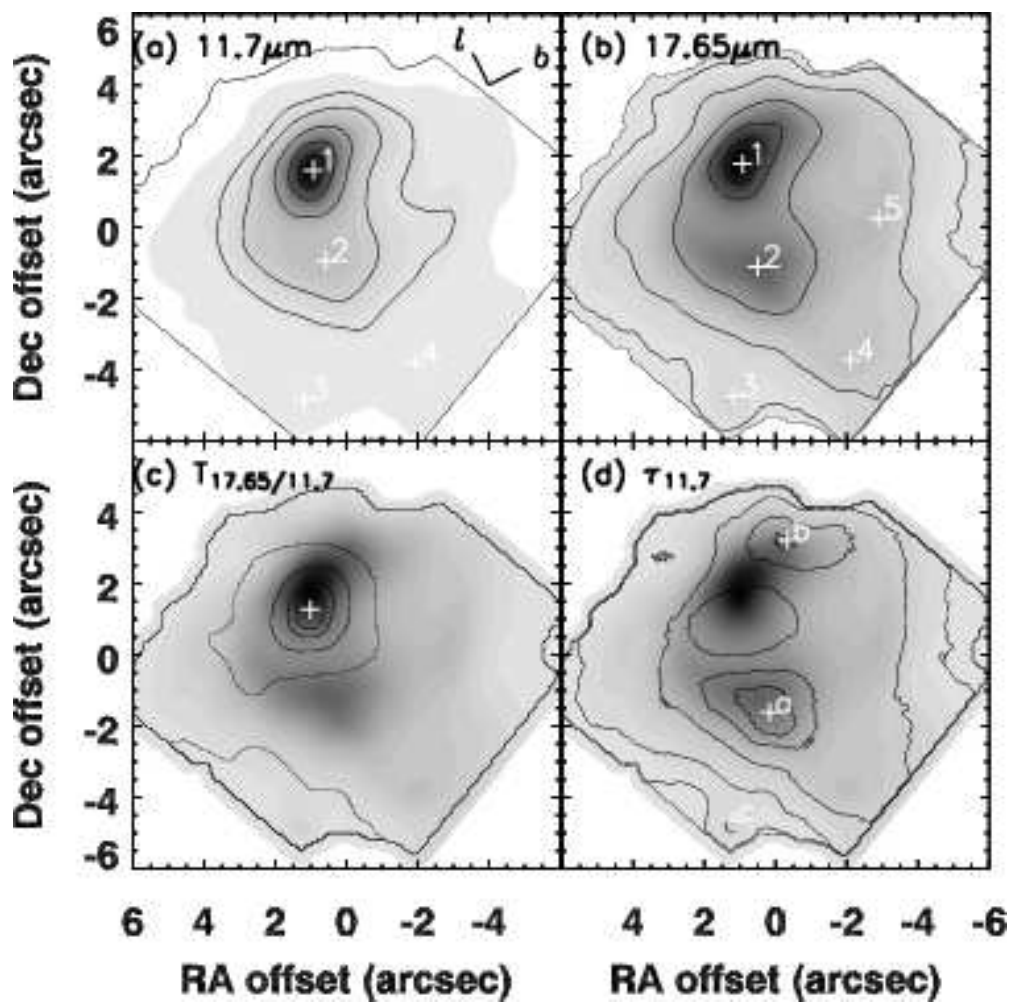}
\caption{Maps of the source 28.28875$-$0.36359: (a), (b), (c), and (d) are the same as in Figure \ref{map1_12}. The 11.7$\mu$m flux density contour levels are 9.2, 7.6, 5.1, 2.6, 1.0, 0.5, 0.1 Jansky/arcsec$^2$ with a peak of 10.2 Jansky/arcsec$^2$. The 17.65$\mu$m flux density contour levels are 23.4, 19.5, 13.0, 6.5, 2.6, 1.3, 0.3 Jansky/arcsec$^2$ with a peak of 26 Jansky/arcsec$^2$. Temperature contour levels are 133, 126, 112, 98, 84, 70 K, with a temperature peak of 174 K. Optical depth contour levels are 0.09, 0.06, 0.03, 0.012, 0.006, 0.001 with a peak of 0.12. Sharp edges result from array rotation to orient north up and east to the left. The central bulb between the two peaks in the $\tau_{11.7}$ map is a local minimum.}
\label{map9_12}
\end{figure}

\begin{figure}
\epsscale{0.8}
\plotone{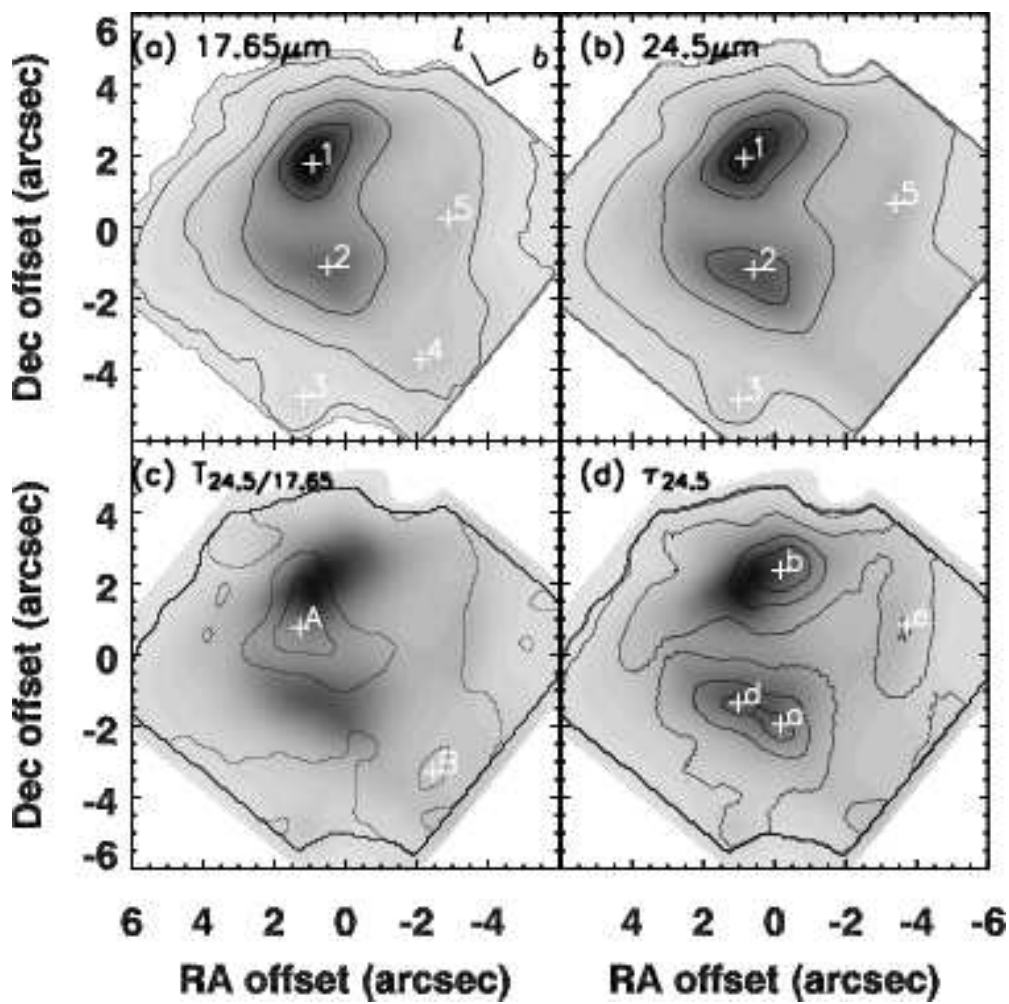}
\caption{Maps of the source 28.28875$-$0.36359: (a), (b), (c), and (d) are the same as in Figure \ref{map1_25}. The 17.65$\mu$m flux density contour levels are 23.4, 19.5, 13.0, 6.5, 2.6, 1.3, 0.3 Jansky/arcsec$^2$ with a peak of 26 Jansky/arcsec$^2$. The 24.5$\mu$m flux density contour levels are 54.0, 45.0, 30.0, 15.0, 6.0, 3.0, 0.6 Jansky/arcsec$^2$ with a peak of 60 Jansky/arcsec$^2$. Temperature contour levels are 99, 94, 83, 73, 62, 52 K, with a temperature peak of 104 K. Optical depth contour levels are 0.43, 0.29, 0.14, 0.06,  0.03, 0.006 with a peak of 0.57. Sharp edges result from array rotation to orient north up and east to the left.}
\label{map9_25}
\end{figure}

\begin{figure}
\epsscale{0.8}
\plotone{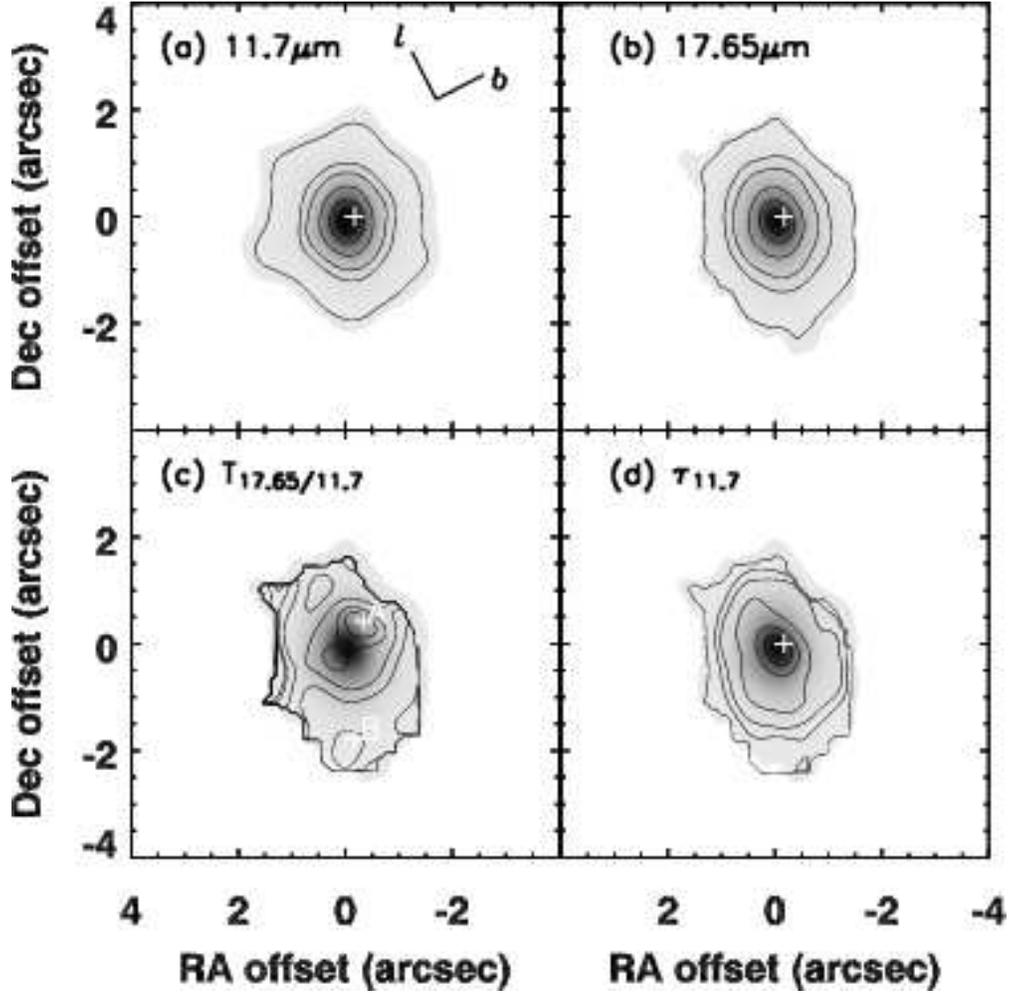}
\caption{Maps of the source 30.04343$-$0.14200: (a), (b), (c), and (d) are the same as in Figure \ref{map1_12}. The 11.7 $\mu$m flux density image is smoothed to a resolution of $0.5''$. The 11.7$\mu$m flux density contour levels are 14.8, 12.3, 8.2, 4.1, 1.6, 0.8, 0.2 Jansky/arcsec$^2$ with a peak of 16.4 Jansky/arcsec$^2$. The 17.65$\mu$m flux density contour levels are 13.1, 11.0, 7.3, 3.7, 1.5, 0.7, 0.1 Jansky/arcsec$^2$ with a peak of 14.6 Jansky/arcsec$^2$. Temperature contour levels are 311, 294, 262, 229, 196, 164 K, with a temperature peak of 372 K. Optical depth contour levels are 0.00113, 0.00075, 0.00038, 0.00015, $7.5\cdot 10^{-5}$, $1.5\cdot 10^{-5}$ with a peak of 0.0015.}
\label{map10}
\end{figure}

\begin{figure}
\epsscale{0.8}
\plotone{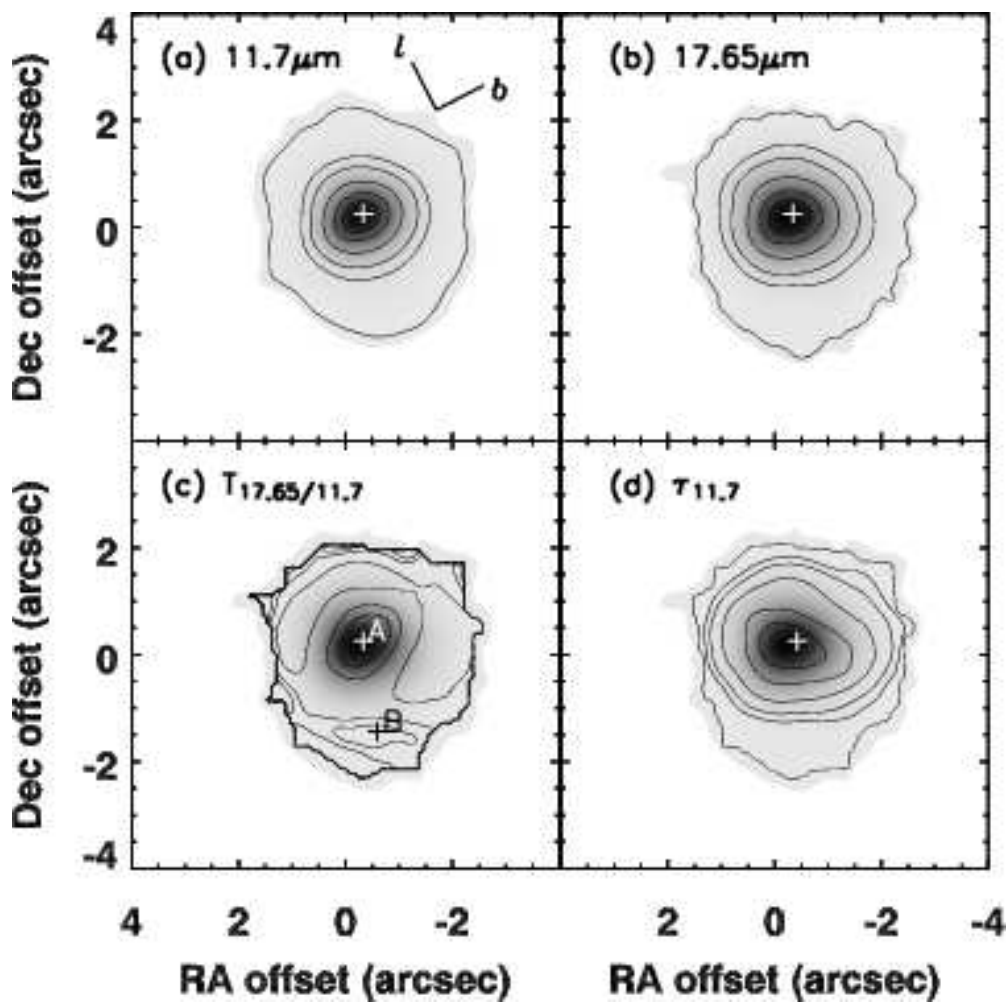}
\caption{Maps of the source 30.86744+0.11493: (a), (b), (c), and (d) are the same as in Figure \ref{map1_12}. The 11.7 $\mu$m flux density image is smoothed to a resolution of $0.5''$. The 11.7$\mu$m flux density contour levels are 6.3, 5.3, 3.5, 1.8, 0.7, 0.4, 0.07 Jansky/arcsec$^2$ with a peak of 7.0 Jansky/arcsec$^2$. The 17.65$\mu$m flux density contour levels are 9.5, 7.9, 5.3, 2.6, 1.1, 0.5, 0.1 Jansky/arcsec$^2$ with a peak of 10.5 Jansky/arcsec$^2$. Temperature contour levels are 204, 193, 172, 150, 129, 108 K, with a temperature peak of 215 K. Optical depth contour levels are 0.0026, 0.0017, 0.0009, 0.0003, 0.0002, $3.4\cdot 10^{-5}$ with a peak of 0.0034.}
\label{map11}
\end{figure}

\begin{figure}
\epsscale{0.8}
\plotone{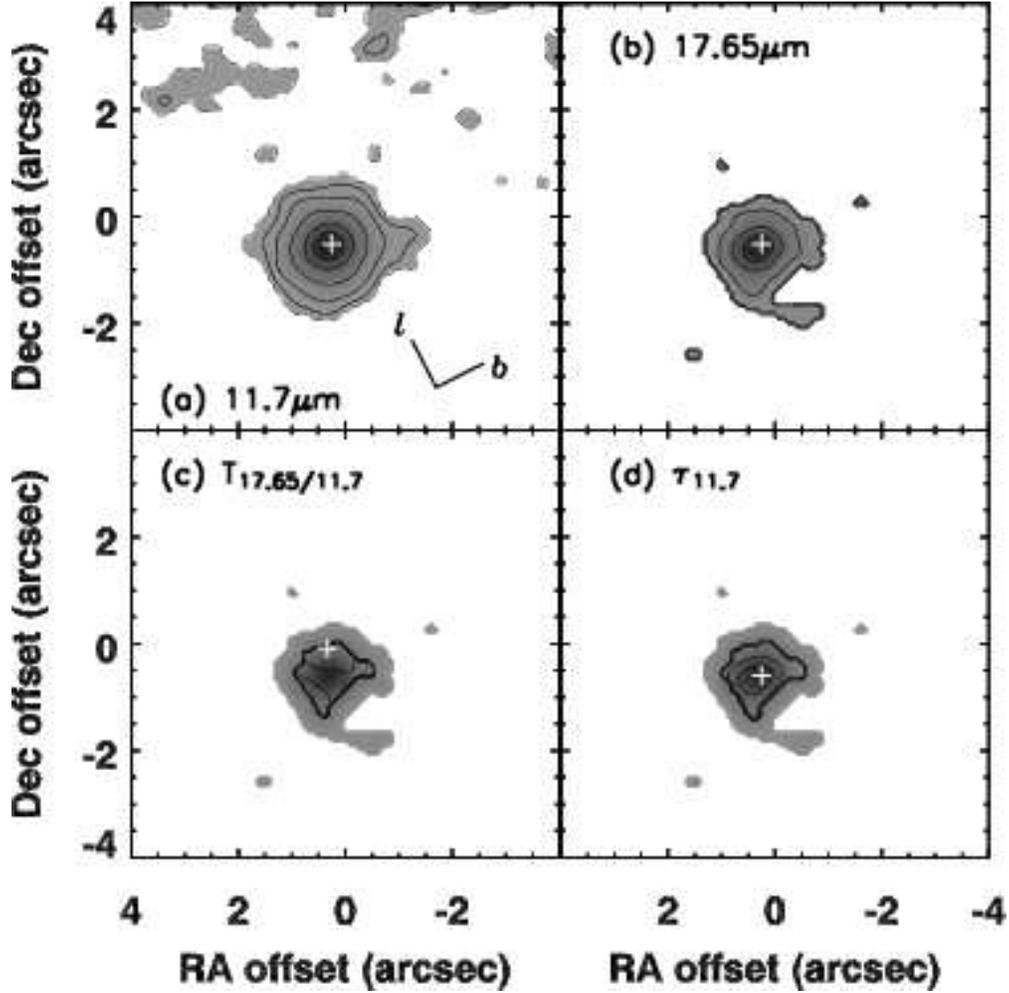}
\caption{Maps of the source 30.58991$-$0.04231: (a), (b), (c), and (d) are the same as in Figure \ref{map1_12}. The 11.7$\mu$m flux density contour levels are 0.52, 0.43, 0.29, 0.15, 0.06, 0.03, 0.006 Jansky/arcsec$^2$ with a peak of 0.58 Jansky/arcsec$^2$. The 17.65$\mu$m flux density contour levels are 0.78, 0.65, 0.44, 0.22, 0.09, 0.04, 0.009 Jansky/arcsec$^2$ with a peak of 0.87 Jansky/arcsec$^2$. Temperature contour levels are 218, 206, 183, 160, 137, 114 K, with a temperature peak of 229 K. Optical depth contour levels are 0.00021, 0.00014, $7\cdot 10^{-5}$, $2.8\cdot 10^{-5}$, $1.4\cdot 10^{-5}$, $2.8\cdot 10^{-6}$ with a peak of 0.00028.}
\label{map12}
\end{figure}

\begin{figure}
\epsscale{0.8}
\plotone{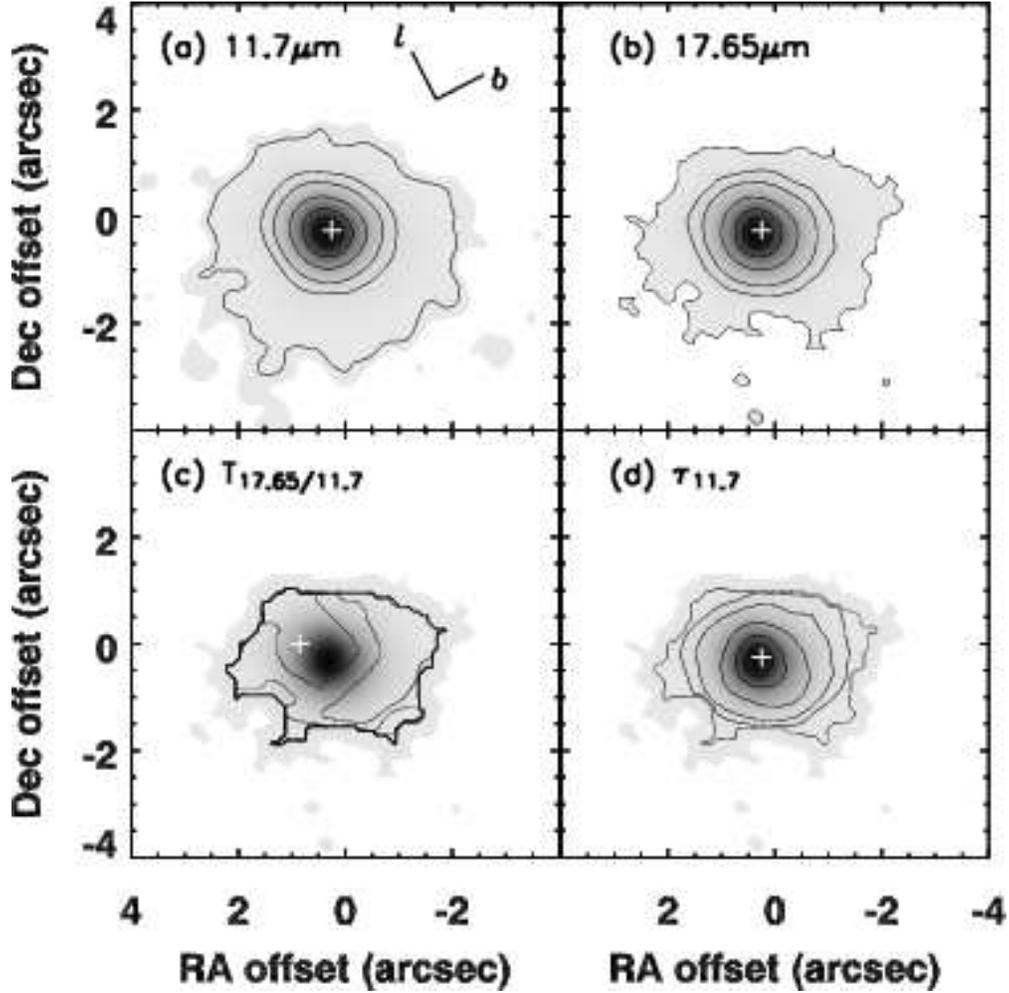}
\caption{Maps of the source 30.66808$-$0.33134: (a), (b), (c), and (d) are the same as in Figure \ref{map1_12}. The 11.7$\mu$m flux density contour levels are 1.71, 1.43, 0.96, 0.48, 0.19, 0.10, 0.02 Jansky/arcsec$^2$ with a peak of 1.91 Jansky/arcsec$^2$. The 17.65$\mu$m flux density contour levels are 5.8, 4.8, 3.2, 1.6, 0.6, 0.3, 0.06 Jansky/arcsec$^2$ with a peak of 6.4 Jansky/arcsec$^2$. Temperature contour levels are 152, 144, 128, 112, 96, 80 K, with a temperature peak of 160 K. Optical depth contour levels are 0.0073, 0.0049, 0.0024, 0.0010, 0.0005, 0.0001 with a peak of 0.0097.}
\label{map13}
\end{figure}

\begin{figure}
\epsscale{0.8}
\plotone{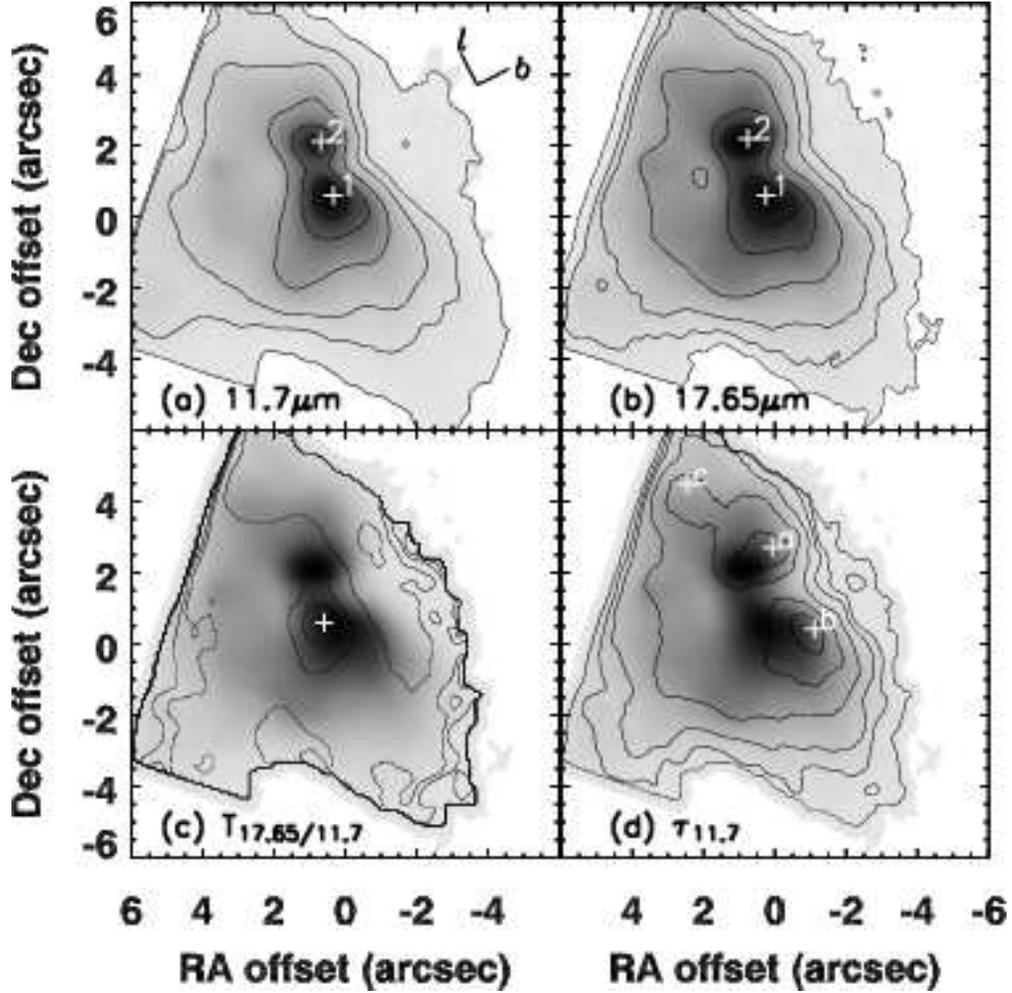}
\caption{Maps of the source 33.91585+0.11111: (a), (b), (c), and (d) are the same as in Figure \ref{map1_12}. The 11.7$\mu$m flux density contour levels are 1.2, 1.0, 0.7, 0.3, 0.1, 0.07, 0.01 Jansky/arcsec$^2$ with a peak of 1.3 Jansky/arcsec$^2$. The 17.65$\mu$m flux density contour levels are 3.7, 3.1, 2.1, 1.0, 0.4, 0.2, 0.04 Jansky/arcsec$^2$ with a peak of 4.1 Jansky/arcsec$^2$. Temperature contour levels are 152, 144, 128, 112, 96, 80 K, with a temperature peak of 160 K. Optical depth contour levels are 0.0089, 0.0060, 0.0030,   0.0012, 0.0006, 0.0001 with a peak of 0.012. Sharp edges result from array rotation to orient north up and east to the left.}
\label{map14_12}
\end{figure}

\begin{figure}
\epsscale{0.8}
\plotone{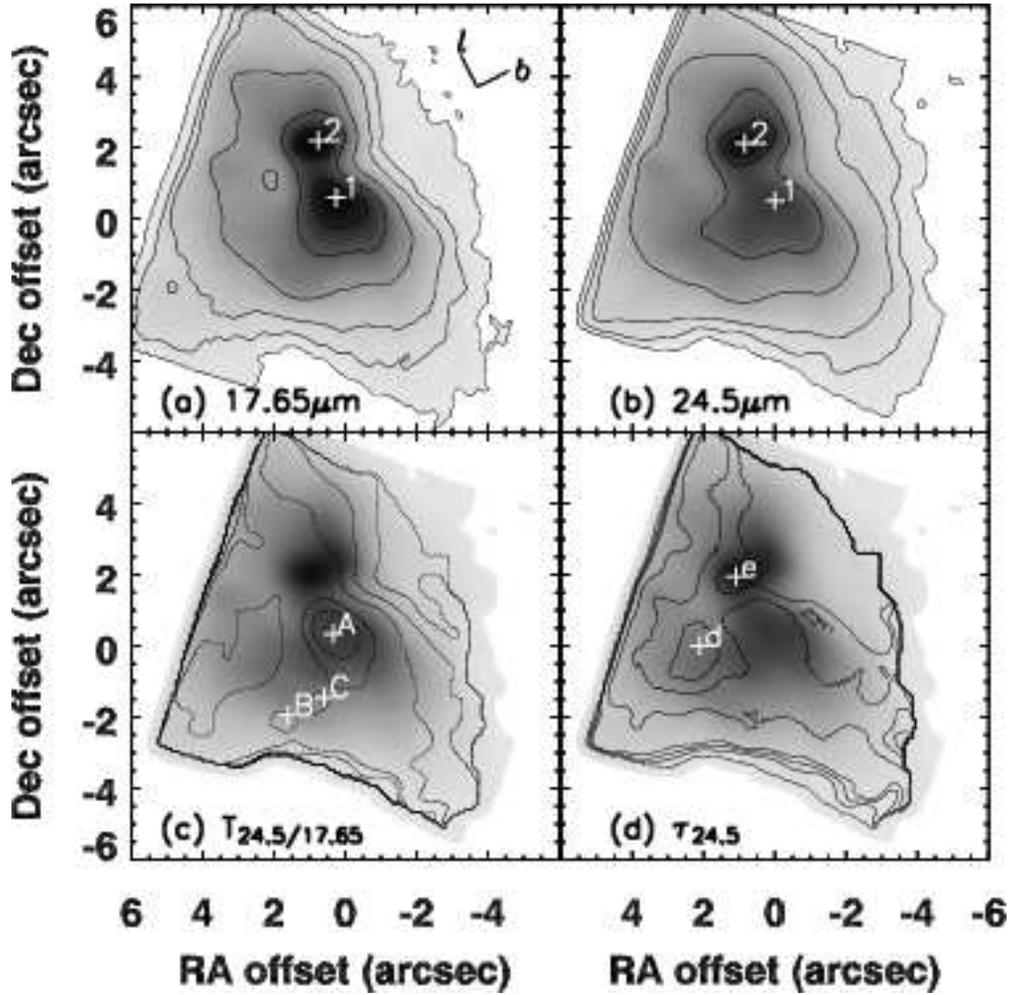}
\caption{Maps of the source 33.91585+0.11111: (a), (b), (c), and (d) are the same as in Figure \ref{map1_25}. The 17.65$\mu$m flux density contour levels are 3.7, 3.1, 2.1, 1.0, 0.4, 0.2, 0.04 Jansky/arcsec$^2$ with a peak of 4.1 Jansky/arcsec$^2$. The 24.5$\mu$m flux density contour levels are 8.1, 6.8, 4.5, 2.3, 0.9, 0.5, 0.09 Jansky/arcsec$^2$ with a peak of 9 Jansky/arcsec$^2$. Temperature contour levels are 104, 98, 87, 76, 65, 54 K, with a temperature peak of 109 K. Optical depth contour levels are 0.060, 0.040, 0.020, 0.008, 0.004, 0.0008 with a peak of 0.08. Sharp edges result from array rotation to orient north up and east to the left.}
\label{map14_25}
\end{figure}

\begin{figure}
\epsscale{0.8}
\plotone{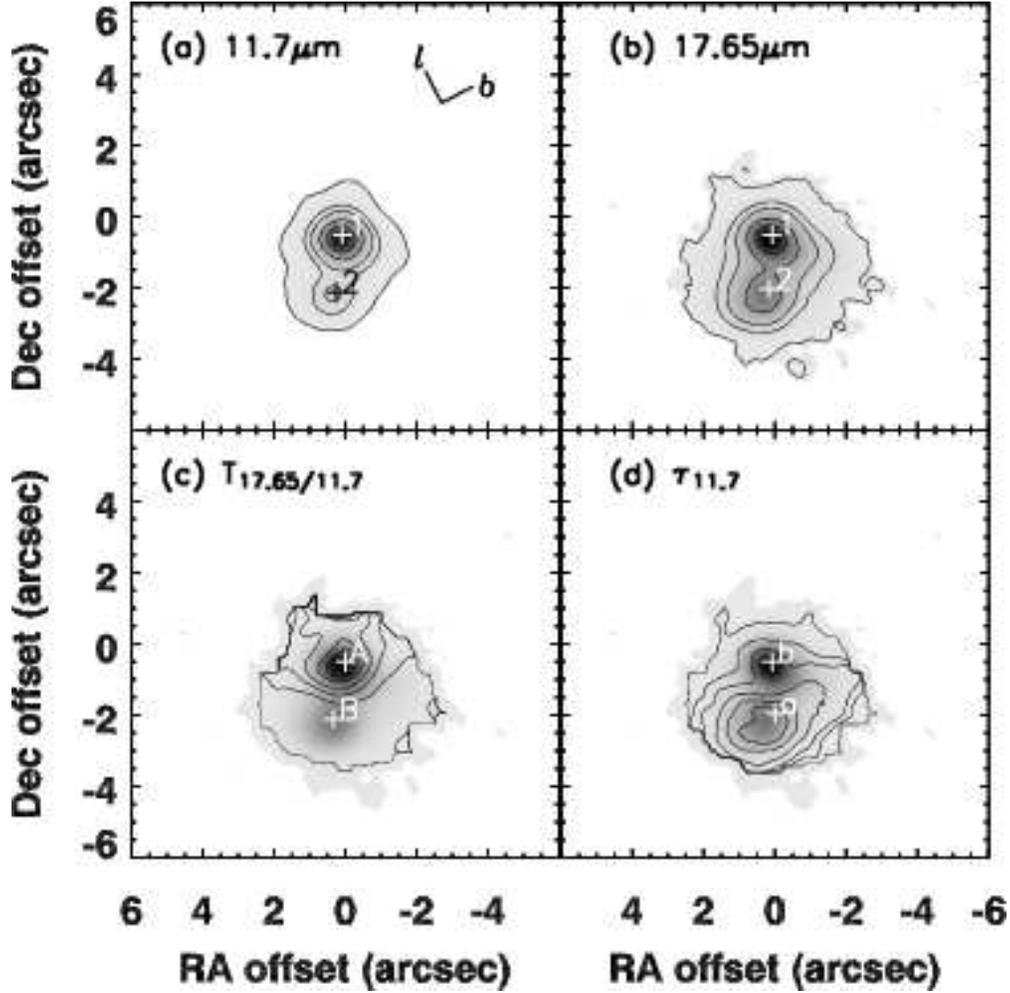}
\caption{Maps of the source 33.81104$-$0.18582: (a), (b), (c), and (d) are the same as in Figure \ref{map1_12}. The 11.7$\mu$m flux density contour levels are 7.6, 6.3, 4.2, 2.1, 0.8, 0.4, 0.08 Jansky/arcsec$^2$ with a peak of 8.4 Jansky/arcsec$^2$. The 17.65$\mu$m flux density contour levels are 9.7, 8.1, 5.4, 2.7, 1.1, 0.5, 0.1 Jansky/arcsec$^2$ with a peak of 10.8 Jansky/arcsec$^2$. Temperature contour levels are 219, 208, 185, 162, 139, 116 K, with a temperature peak of 231 K. Optical depth contour levels are 0.01065, 0.0071, 0.0035, 0.0014, 0.0007, 0.0001 with a peak of 0.014.}
\label{map15_12}
\end{figure}

\begin{figure}
\epsscale{0.8}
\plotone{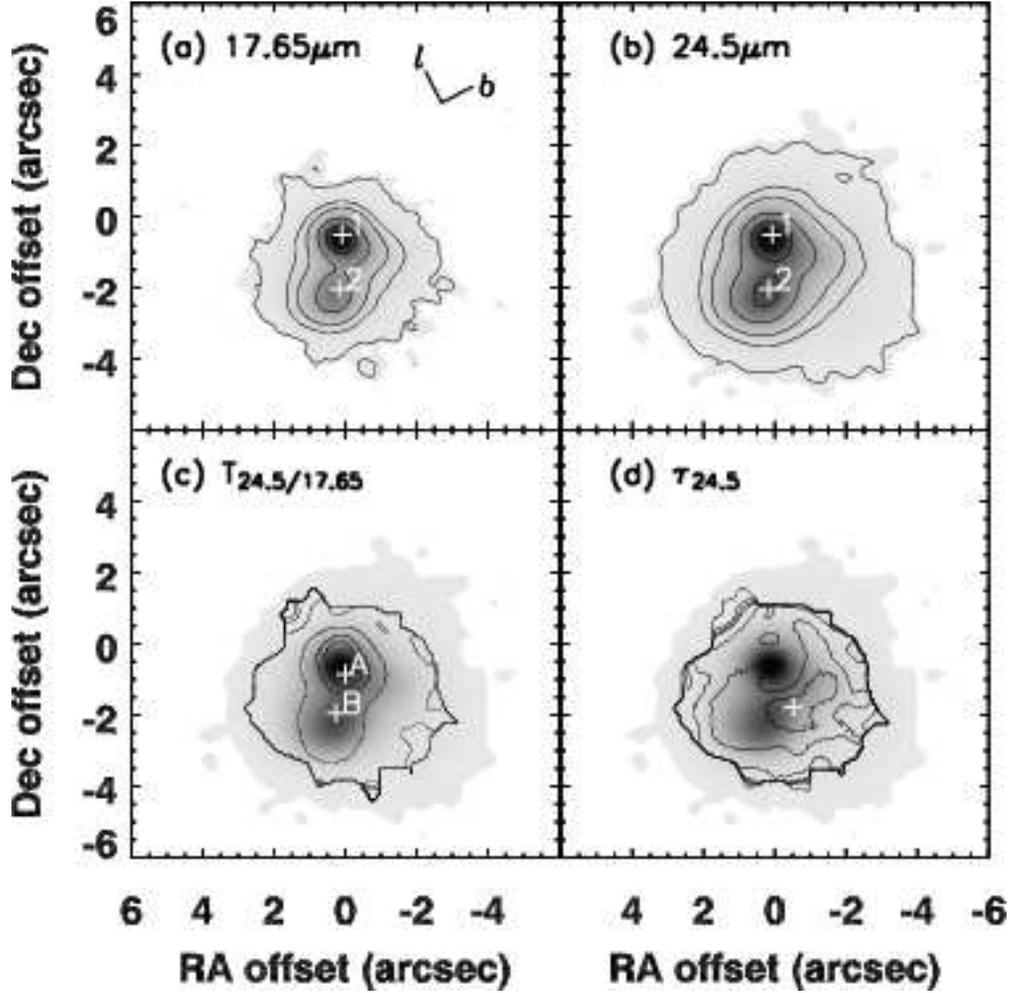}
\caption{Maps of the source 33.81104$-$0.18582: (a), (b), (c), and (d) are the same as in Figure \ref{map1_25}. The 17.65$\mu$m flux density contour levels are 9.7, 8.1, 5.4, 2.7, 1.1, 0.5, 0.1 Jansky/arcsec$^2$ with a peak of 10.8 Jansky/arcsec$^2$. The 24.5$\mu$m flux density contour levels are 24.3, 20.3, 13.5, 6.8, 2.7, 1.4, 0.3 Jansky/arcsec$^2$ with a peak of 27 Jansky/arcsec$^2$. Temperature contour levels are 85, 80, 71, 62, 53, 44 K, with a temperature peak of 89 K. Optical depth contour levels are 0.40, 0.27, 0.13, 0.05, 0.03, 0.005 with a peak of 0.53.}
\label{map15_25}
\end{figure}

\begin{figure}
\epsscale{0.8}
\plotone{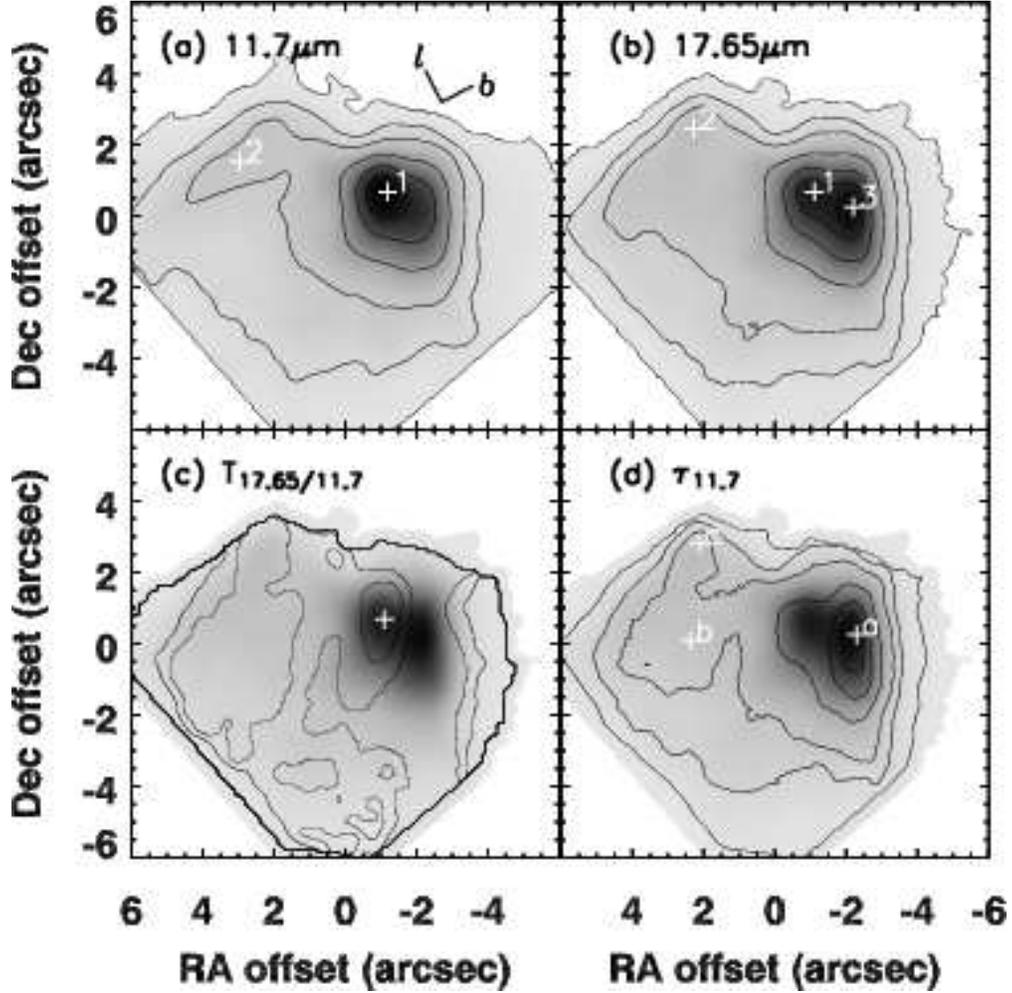}
\caption{Maps of the source 35.46832+0.13984: (a), (b), (c), and (d) are the same as in Figure \ref{map1_12}. The 11.7 $\mu$m flux density image is smoothed to a resolution of $0.5''$. The 11.7$\mu$m flux density contour levels are 1.2, 1.0, 0.7, 0.3, 0.1, 0.07, 0.01 Jansky/arcsec$^2$ with a peak of 1.38 Jansky/arcsec$^2$. The 17.65$\mu$m flux density contour levels are 5.6, 4.6, 3.1, 1.6, 0.6, 0.3, 0.06 Jansky/arcsec$^2$ with a peak of 6.2 Jansky/arcsec$^2$. Temperature contour levels are 136, 129, 114, 100, 86, 72 K, with a temperature peak of 143 K. Optical depth contour levels are 0.029, 0.019, 0.009, 0.004, 0.002, 0.0004 with a peak of 0.038. Sharp edges result from array rotation to orient north up and east to the left.}
\label{map16_12}
\end{figure}

\begin{figure}
\epsscale{0.8}
\plotone{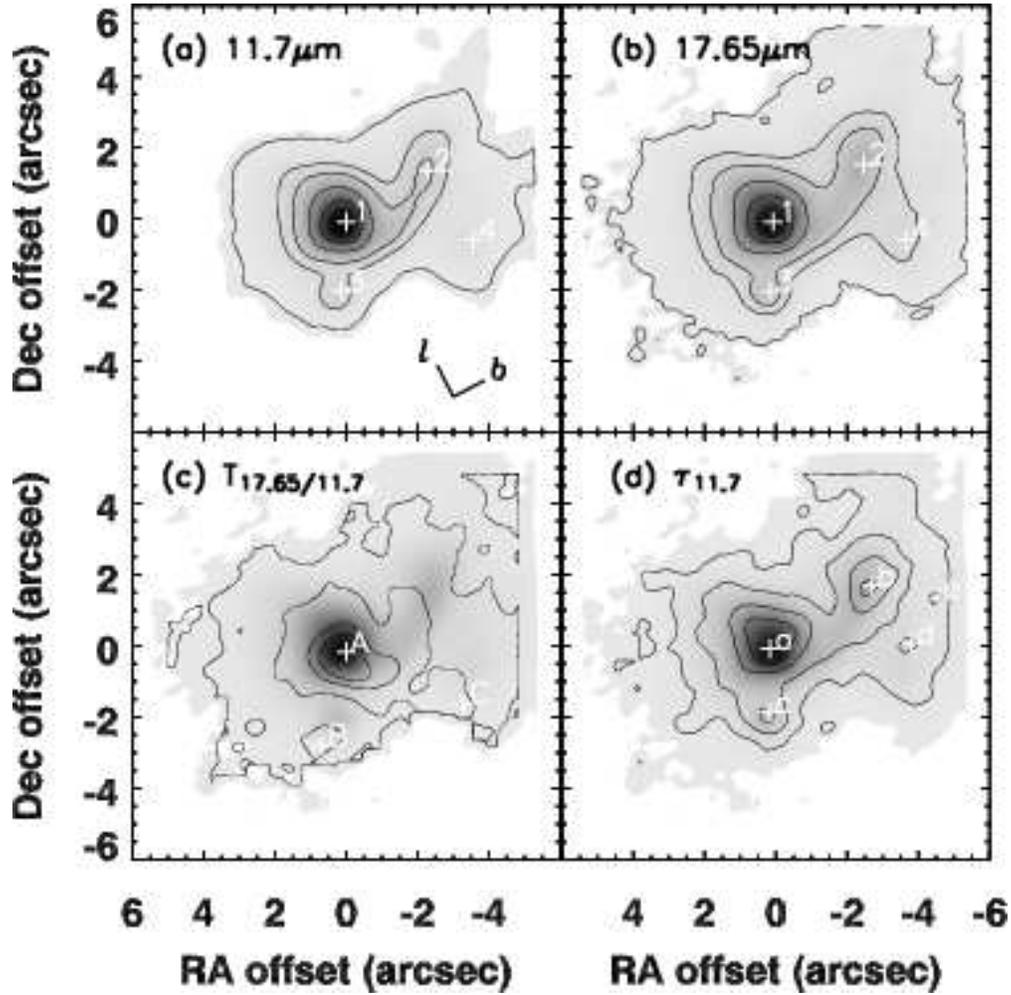}
\caption{Maps of the source 37.87411$-$0.39866: (a), (b), (c), and (d) are the same as in Figure \ref{map1_12}. The 11.7$\mu$m flux density contour levels are 7.4, 6.1, 4.1, 2.1, 0.8, 0.4, 0.08 Jansky/arcsec$^2$ with a peak of 8.2 Jansky/arcsec$^2$. The 17.65$\mu$m flux density contour levels are 13.4, 11.2, 7.4, 3.7, 1.5, 0.7, 0.1 Jansky/arcsec$^2$ with a peak of 14.9 Jansky/arcsec$^2$. Temperature contour levels are 185, 175, 156, 136 K, with a temperature peak of 195 K. Optical depth contour levels are 0.0056, 0.0038, 0.0019, 0.0008 with a peak of 0.0075. Sharp edges result from array rotation to orient north up and east to the left.}
\label{map17_12}
\end{figure}

\begin{figure}
\epsscale{0.8}
\plotone{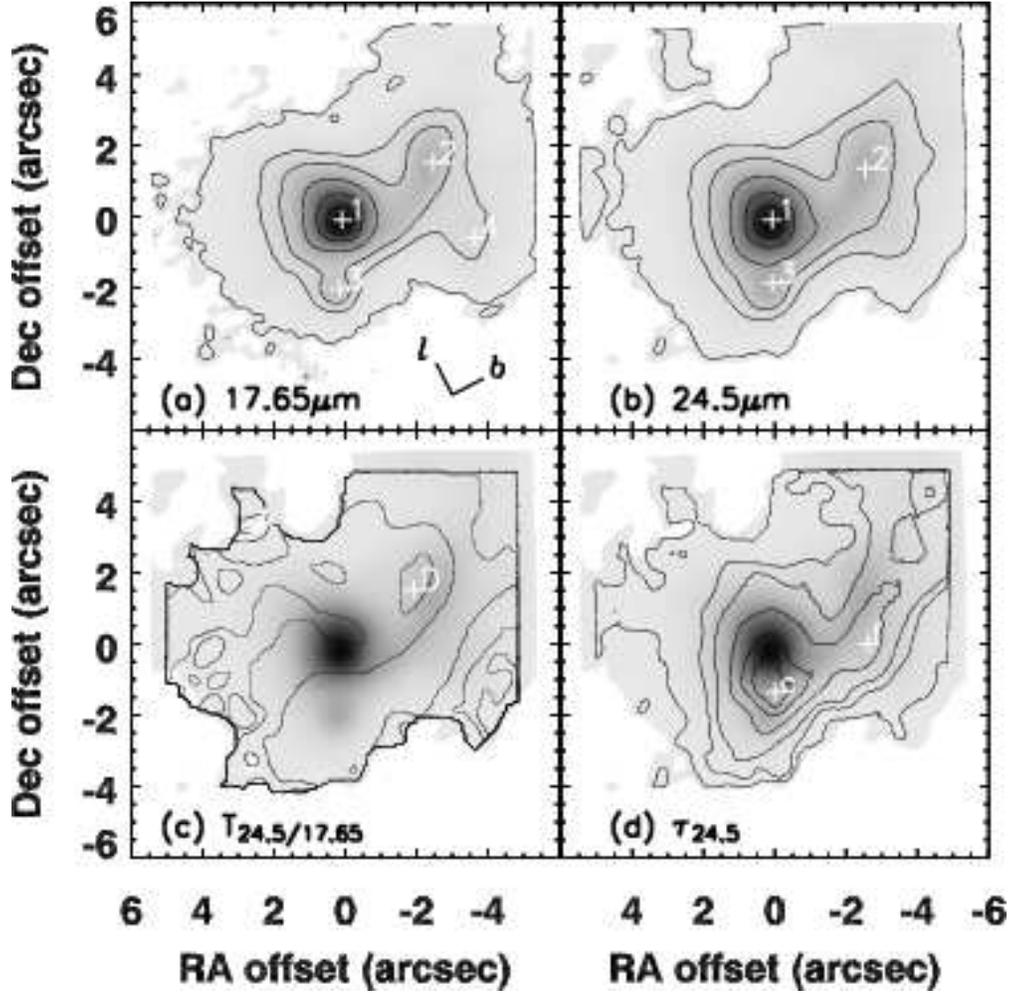}
\caption{Maps of the source 37.87411$-$0.39866: (a), (b), (c), and (d) are the same as in Figure \ref{map1_25}. The 17.65$\mu$m flux density contour levels are 13.4, 11.2, 7.4, 3.7, 1.5, 0.7, 0.1 Jansky/arcsec$^2$ with a peak of 14.9 Jansky/arcsec$^2$. The 24.5$\mu$m flux density contour levels are 44.1, 36.8, 24.5, 12.3, 4.9, 2.5 0.5 Jansky/arcsec$^2$ with a peak of 49 Jansky/arcsec$^2$. Temperature contour levels are 84, 79, 70, 62, 53, 44 K, with a temperature peak of 88 K. Optical depth contour levels are 0.90, 0.60, 0.30, 0.12, 0.06, 0.01 with a peak of 1.2. Sharp edges result from array rotation to orient north up and east to the left.}
\label{map17_25}
\end{figure}

\begin{figure}
\epsscale{0.8}
\plotone{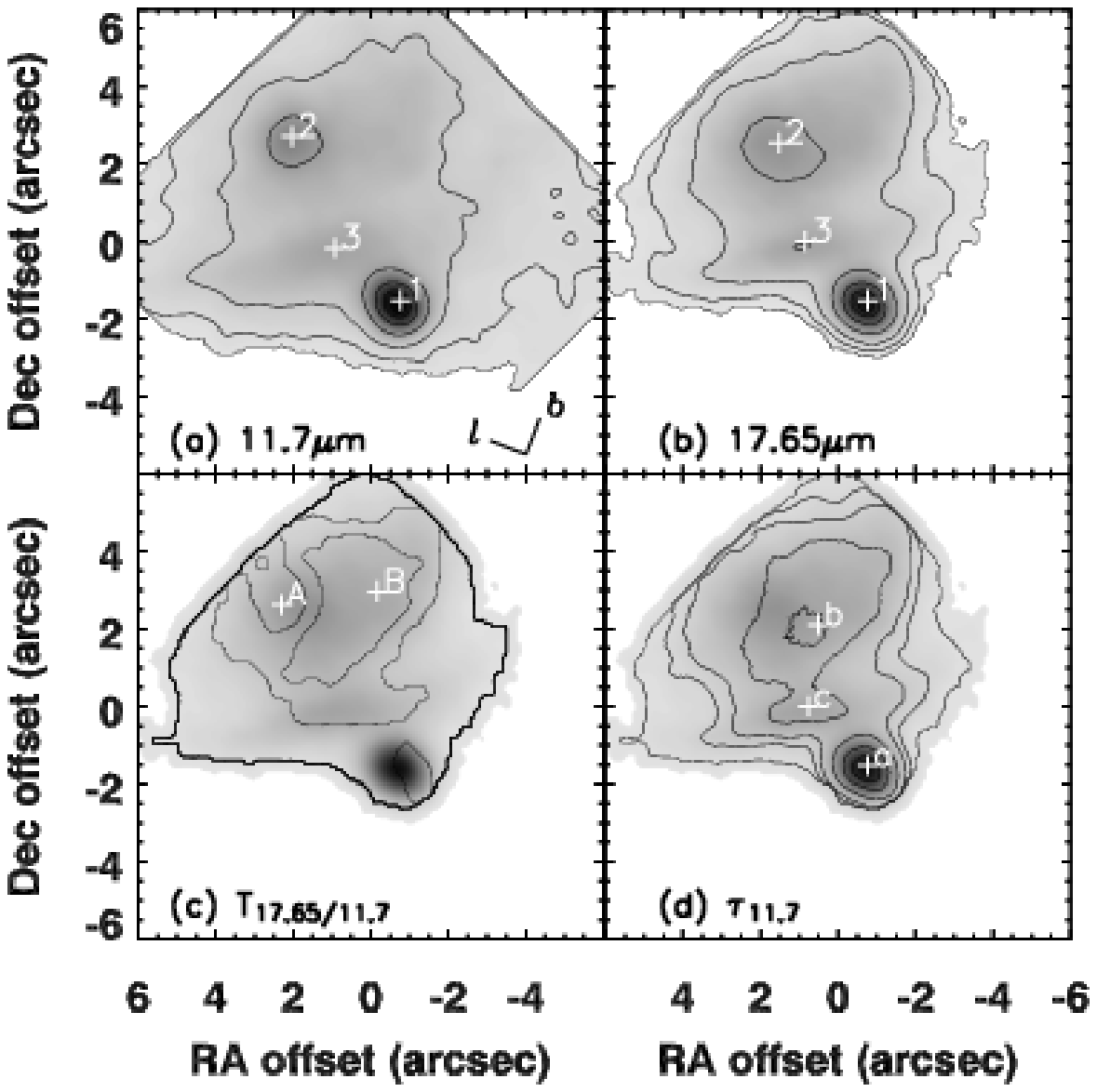}
\caption{Maps of the source 111.28293$-$0.66355: (a), (b), (c), and (d) are the same as in Figure \ref{map1_12}. The 11.7$\mu$m flux density contour levels are 0.74, 0.61, 0.41, 0.21, 0.08, 0.04, 0.008 Jansky/arcsec$^2$ with a peak of 0.82 Jansky/arcsec$^2$. The 17.65$\mu$m flux density contour levels are 4.3, 3.6, 2.4, 1.2, 0.5, 0.2, 0.05 Jansky/arcsec$^2$ with a peak of 4.8 Jansky/arcsec$^2$. Temperature contour levels are 126, 120, 106,  93,  80,  66 K, with a temperature peak of 133 K. Optical depth contour levels are 0.017, 0.012, 0.006, 0.002, 0.001, 0.0002 with a peak of 0.023. Sharp edges result from array rotation to orient north up and east to the left.}
\label{map18_12}
\end{figure}

Temperatures were computed pixel by pixel using a look-up table containing the calculated flux density ratios as a function of temperatures, assuming a modified blackbody with the dust emissivity law of Draine (2003) for $R_V=3.1$ \footnote {The extinction curve is available at http://www.astro.princeton.edu/$\sim$draine/dust/dustmix.html}, not including foreground extinction, and taking into account the filter transmission and the atmospheric transmission through the bandpass. According to the Draine (2003) emissivity law, the corresponding optical depths at the central wavelengths of our bands (incorporating the filter bandpasses) have the relations $\tau_{17.65}/\tau_{11.7}=0.71$ and $\tau_{24.5}/\tau_{17.65}=0.64$. The typical temperature uncertainty from the look-up table is 1--2 K for $T_{17.65/11.7}$, and 1--3 K for $T_{24.5/17.65}$.
Table \ref{tmap_data} lists the labels of multiple temperature peaks in column (2); the values of the temperature peaks in columns (3) and (6); the locations of the peaks (RA and Dec offsets from the coordinates of the field centers given in Table \ref{pos1}) in columns (4) and (7); and the mean temperatures in columns (5) and (8) for each source. Multiple temperature peaks are labeled by the upper-case letters `A', `B', `C', etc (column 2), and are sorted by decreasing $T_{17.65/11.7}$. The upper-case alphabetical labels at different frequencies correspond to the same peaks. We note again the higher temperature of the exceptional source 30.04343$-$0.14200 reflecting the unusual flat SED of this source. We suspect that this source is not an UC HII region (see MIR SED in Figure \ref{sed2}, \S \ref{flux}).
Figures \ref{map1_12}--\ref{map18_12} show the temperature maps as contour plots overlaid on the flux density maps for all of our sources. Peak temperatures are indicated by a cross, with sub-peaks indicated as in Table \ref{tmap_data}, where `A' is the highest peak.

\begin{deluxetable}{cccccccc}
\tabletypesize{\scriptsize}
\tablecaption{Temperature Maps Data \label{tmap_data}}
\tablewidth{0pt}
\tablehead{ &
\colhead{Peak} &
\colhead{$T_{17.65/11.7}^{\rm max}$} &
\colhead{Offset ($\Delta\alpha$,$\Delta\delta$)} &
\colhead{$<T_{17.65/11.7}>$} &
\colhead{$T_{24.5/17.65}^{\rm max}$} &
\colhead{Offset ($\Delta\alpha$,$\Delta\delta$)} &
\colhead{$<T_{24.5/17.65}>$} \\
\colhead{Source} &
\colhead{Label} &
\colhead{(K)} &
\colhead{(arcsec)} &
\colhead{(K)} &
\colhead{(K)} &
\colhead{(arcsec)} &
\colhead{(K)} \\
\colhead{(1)} & \colhead{(2)} & \colhead{(3)} & \colhead{(4)} & \colhead{(5)} & \colhead{(6)} & \colhead{(7)} & \colhead{(8)}}
\startdata
 11.11198$-$0.39795 &  & 184 & (+3.7,$-$3.4) & 143 & 81 & (+0.6,$-$0.7) & 72 \\
 11.94545$-$0.03634 & A & \nodata & \nodata & \nodata & 81 & (+0.7,+1.9) & 71 \\
 & B & 159 & ($-$1.3,+1.7) & 160 & 78 & ($-$0.9,+2.5) & \nodata \\
 & C & \nodata & \nodata & \nodata & 75 & ($-$1.4,$-$0.2) & \nodata \\
 18.71179+0.00085   &  & 154 & (+0.4,+0.6) & 149 & \nodata  & \nodata & \nodata \\
 19.75611$-$0.12775 &  & 159 & (+0.5,$-$0.1) & 142 &  91 & (+0.6,$-$0.2) & 77 \\
 21.38654$-$0.25346 &  & 159 & ($-$0.4,+0.5) & 152 & \nodata & \nodata & \nodata \\
 25.39918$-$0.14081 & A & 158 & ($-$2.1,$-$0.2) & 128 & \nodata & \nodata & \nodata \\
  & B & \nodata & \nodata & \nodata & 101 & (+0.7,0.0) & 76 \\
  & C & \nodata & \nodata & \nodata & 98 & (+0.7,$-$1.2) & \nodata \\
  & D & \nodata & \nodata & \nodata & 96 & (+0.9,$-$3.9) & \nodata \\
 25.80211$-$0.15640 &  & 143 & ($-$0.8,+0.6) & 128 & 101 & ($-$1.4,+1.2) & 85 \\
 27.18725$-$0.08095 & A & 140 & (+0.5,+0.2) & 135 & 62 & (0.0,+0.5) & 64 \\
 & B & \nodata & \nodata & \nodata & 84 & (+1.3,+3.2) & \nodata \\
 & C & \nodata & \nodata & \nodata & 63 & (+0.4,$-$0.9) & \nodata \\
 28.28875$-$0.36359 & A & 174 & (+1.0,+1.3) & 119 & 104 & (+1.1,+0.8) & 87 \\
 & B & \nodata & \nodata & \nodata & 95 & ($-$2.5,$-$3.3) & \nodata \\
 30.04343$-$0.14200 & A & 327 & ($-$0.3,+0.4) & 251 & \nodata & \nodata & \nodata \\
 & B & 243 & ($-$0.2,$-$1.8) & \nodata & \nodata & \nodata & \nodata \\
 30.86744+0.11493   & A & 215 & ($-$0.3,+0.3) & 183 & \nodata & \nodata & \nodata \\
  & B & 212 & ($-$0.6,$-$1.4) & \nodata & \nodata & \nodata & \nodata \\
 30.58991$-$0.04231 &  & 229 & (+0.3,$-$0.1) & 213 & \nodata & \nodata & \nodata \\
 30.66808$-$0.33134 &  & 160 & (+0.8,0.0) & 147 & \nodata & \nodata & \nodata \\
 33.91585+0.11111   & A & 160 & (+0.6,+0.6) & 138 & 109 & (+0.3,+0.3) & 87 \\
   & B & \nodata &  & \nodata & 100 & (+1.6,$-$1.9) & \nodata \\
   & C & \nodata &  & \nodata & 99 & (+0.6,$-$1.4) & \nodata \\
 33.81104$-$0.18582 & A & 231 & (0.0,$-$0.5) & 150 & 89 & (0.0,$-$0.8) & 70 \\
  & B & 138 & (+0.3,$-$2.1) & \nodata & 80 & (+0.4,$-$1.9) & \nodata \\
 35.46832+0.13984  &  & 143 & ($-$1.1,+0.7) & 131 & \nodata & \nodata & \nodata \\
 37.87411$-$0.39866 & A & 195 & (0.0,$-$0.2) & 149 & 85 & (+0.5,+0.7) & 78 \\
  & B & 164 & (+0.7,$-$2.8) & \nodata & \nodata & \nodata & \nodata \\
  & C & 162 & ($-$3.3,$-$1.6) & \nodata & \nodata & \nodata & \nodata \\
  & D & \nodata & \nodata & \nodata & 88 & ($-$1.9,+1.5) & \nodata \\
111.28293$-$0.66355 & A & 133 & (+2.2,+2.6) & 130 & \nodata & \nodata & \nodata \\
 & B & 117 & ($-$0.2,+3.0) & \nodata & \nodata & \nodata & \nodata \\
\enddata
\tablecomments{Temperature peaks, positions of the peaks, and temperature means for our sources. The listed values are from our best alignment of the flux density maps. The typical 1$\sigma$ uncertainties due to a possible misalignment are 30\% for the peak temperatures and 1\% for the mean temperatures. Some of the fields contain multiple temperature peaks, which are labeled `A', `B', `C', etc., and are indicated in the maps in Figures \ref{map1_12}--\ref{map18_12}.}
\end{deluxetable}

\clearpage

\subsection{Emission Optical Depths}
\label{taus}

The measured MIR intensity emitted from dust grains is related to their temperature by
\begin{equation}
I_{\nu}=\left( 1-e^{-\tau_{\nu}}\right) B_{\nu}(T_D),
\label{tau_eq}
\end{equation}
where $B_{\nu}(T_D)$ is the Planck function at dust temperature $T_D$, $\tau_{\nu}$ is the emission optical depth of the dust, and $(1-e^{-\tau_{\nu}})$ is called the emissivity function.
This emission is typically optically thin in UC \hii regions (see e.g., De Buizer et al. 2002a; and De Buizer 2005a), which are small, hot sources, so the emissivity function may be reduced to simply $\tau_{\nu}$. The optical depth may be offset by a proportion constant due to absorption from a much larger and cooler cloud. We assume that the absorption is uniform over our fields of view, since it arises on scales much larger than our sources.
The optically thinness assumption does not hold in the case of 27.18725$-$0.08095, with mean $\tau_{24.5} >1$. The peak $\tau_{24.5}$ of the source 37.87411$-$0.39866 is also larger than 1, but the average $\tau_{24.5}$ is still $< 1$.

We have generated emission optical depth maps at 11.7 $\mu$m and at 24.5 $\mu$m, pixel by pixel, using their corresponding flux density and dust temperature maps. For sources observed only in two bands, only $\tau_{11.7}$ was calculated. Flux densities were translated into intensities by normalizing them by their angular pixel size.
We have tested the robustness of our optical depth maps by shifting one of the flux density maps by $\pm 3$ pixels in RA and Dec, as was done when creating the temperature maps. The absolute values of the optical depth maxima changed by as much as $\times 2$, but the main components remained unchanged in their shapes and relative positions. The average optical depths are less sensitive to misalignments, and for the above shifts in the flux density maps, they typically vary by $50\%$.

Table \ref{taumap_data} lists the calculated 11.7 $\mu$m and 24.5 $\mu$m optical depth data: labels of multiple peaks in column (2); the values of the peaks in columns (3) and (6); the locations of the peaks (RA and Dec offsets from the coordinates of the field centers given in Table \ref{pos1}) in columns (4) and (7); and mean optical depth in columns (5) and (8). Some of the fields contain multiple optical depth peaks, which are labeled by the lower-case letters `a', `b', `c', etc, sorted by decreasing $\tau_{11.7}$. The same lower-case alphabetical labels at different frequencies correspond to the same peaks.
Figures \ref{map1_12}--\ref{map18_12} show optical depth maps as contour plots overlaid on the 17.65 $\mu$m or 24.5 $\mu$m flux density maps for all of our sources.
Peak optical depths are indicated by a cross, with sub-peaks indicated as in Table \ref{taumap_data}, where `a' is the highest peak.

In general, $\tau_{24.5}>\tau_{11.7}$ for all of our sources.
This is because first, at 24.5 $\mu$m we can see deeper into the region compared to 11.7 $\mu$m. Second, 24.5 $\mu$m probes cooler dust, which is usually distributed further out from the central source, whereas 11.7 $\mu$m probes hotter dust further in ($T_{17.65/11.7}>T_{24.5/17.65}$ as in Table \ref{tmap_data}). The overall effect is that the 24.5 $\mu$m emission comes from a larger volume.
We note that silicate absorption or emission from the strong features at 9.7 and 18 $\mu$m cannot be the reason for this effect in the optical depths. We have artificially changed the 11.7 and 17.65 $\mu$m flux densities to see the effect on the dust temperatures and optical depths. These tests show that the 18 $\mu$m feature must be a stronger absorber or a weaker emitter than the part of the 9.7 $\mu$m feature affecting our data, in order not to reduce $\tau_{11.7}$ even more, which is unlikely. However, The 18 $\mu$m feature is always shallower in absorption than the 9.7 $\mu$m feature, and 11.7 $\mu$m is still well within the influence of the 9.7 $\mu$m feature. Also, UC \hii regions generally have silicate absorption, in various depths from very shallow to very deep, but are hardly ever seen in emission.

The optical depth peaks and the temperature peaks are often at different locations -- the optical depth peaks are usually more closely-linked to the flux density peaks. The temperature peak is usually offset from the flux density peak. This means that having a large MIR flux density at a given location is usually due to a larger column of matter and/or higher density along that line-of-sight, and not necessarily due to the central ionizing source of the UC \hii region. Such offsets have been predicted by blister-type models of \hii regions, where the exciting star lies near the edge/outside the molecular cloud (Icke, Gatley, \& Israel 1980).

\clearpage

\begin{deluxetable}{cccccccc}
\tabletypesize{\scriptsize}
\tablecaption{Emission Optical Depth Maps Data \label{taumap_data}}
\tablewidth{0pt}
\tablehead{ &
\colhead{Peak} &
\colhead{$\tau_{11.7}^{\rm max}$} &
\colhead{Offset ($\Delta\alpha$,$\Delta\delta$)} &
\colhead{$<\tau_{11.7}>$} &
\colhead{$\tau_{24.5}^{\rm max}$} &
\colhead{Offset ($\Delta\alpha$,$\Delta\delta$)} &
\colhead{$<\tau_{24.5}>$} \\
\colhead{Source} &
\colhead{Label} & &
\colhead{(arcsec)} & & &
\colhead{(arcsec)} & \\
\colhead{(1)} & \colhead{(2)} & \colhead{(3)} & \colhead{(4)} & \colhead{(5)} & \colhead{(6)} & \colhead{(7)} & \colhead{(8)}}
\startdata
 11.11198$-$0.39795 & a & 0.070 & (+1.4,$-$0.6) & 0.0055 & 0.70 & (+1.6,$-$0.8) & 0.21 \\
  & b & \nodata & \nodata & \nodata & 0.60 & ($-$0.2,+0.4) & \nodata \\
 11.94545$-$0.03634 & a & 0.0065 & (+0.8,+2.6) & 0.0013 & \nodata & \nodata & \nodata \\
  & b & 0.0022 & ($-$1.2,+1.0) & \nodata & 0.33 & ($-$0.3,+1.3) & 0.24 \\
  & c & 0.0008 & (+2.4,$-$1.7) & \nodata & \nodata & \nodata & \nodata \\
  & d & \nodata & \nodata & \nodata & 0.30 & (+2.2,+1.2) & \nodata \\
 18.71179+0.00085   &  & 0.003 & (+0.2,+0.8) & 0.0012 & \nodata &  & \nodata \\
 19.75611$-$0.12775 & a & 0.029 & ($-$0.2,+0.2) & 0.0053 & 0.60 & (+0.1,$-$0.1) & 0.18 \\
 & b & \nodata & \nodata & \nodata & 0.47 & ($-$0.7,$-$1.3) & \nodata \\
 21.38654$-$0.25346 &  & 0.011 & (+0.7,+0.8) & 0.00066 & \nodata & \nodata & \nodata \\
 25.39918$-$0.14081 & a & \nodata & \nodata & \nodata & 0.80 & ($-$1.6,+2.3) & 0.18 \\
  & b & \nodata & \nodata & \nodata & 0.60 & ($-$2.4,+1.1) & \nodata \\
  & c & 0.019 & (+1.0,$-$3.8) & 0.0044 & 0.57 & (0.0,$-$3.3) & \nodata \\
  & d & 0.014 & (+0.5,0.0) & \nodata & \nodata & \nodata & \nodata \\
 & e & \nodata & \nodata & \nodata & 0.56 & (+1.9,$-$0.8) & \nodata \\
 & f & \nodata & \nodata & \nodata & 0.31 & (+3.8,$-$1.7) & \nodata \\
 & g & \nodata & \nodata & \nodata & 0.24 & (+3.3,$-$3.7) & \nodata \\
 & h & \nodata & \nodata & \nodata & 0.19 & (+4.2,$-$3.1) & \nodata \\
 25.80211$-$0.15640 & a & 0.039 & ($-$0.1,+0.2) & 0.012 & 0.33 & (0.0,+0.2) & 0.087 \\
 & b & 0.037 & ($-$0.8,+0.7) & \nodata & 0.35 & ($-$0.8,+0.5) & \nodata \\
 27.18725$-$0.08095 &  & 0.005 & (+0.3,$-$0.2) & 0.0015 & $> 1$ & (+0.6,0.0) & $> 1$ \\
 28.28875$-$0.36359 & a & 0.12 & (+0.2,$-$1.6) & 0.033 & 0.52 & ($-$0.2,$-$1.9) & 0.17 \\
 & b & 0.10 & ($-$0.3,+3.2) & \nodata & 0.57 & ($-$0.2,+2.4) & \nodata \\
 & c & 0.013 & (+1.2,$-$4.7) & \nodata & \nodata & \nodata & \nodata \\
 & d & \nodata & \nodata & \nodata & 0.49 & (+1.0,$-$1.4) & \nodata \\
 & e & \nodata & \nodata & \nodata & 0.30 & ($-$3.7,+0.8) & \nodata \\
 30.04343$-$0.14200 &  & 0.0015 & ($-$0.2,0.0) & 0.00026 & \nodata & \nodata & \nodata \\
 30.86744+0.11493   &  & 0.0034 & ($-$0.4,+0.3) & 0.00080 & \nodata & \nodata & \nodata \\
 30.58991$-$0.04231 &  & 0.00028 & (+0.3,$-$0.6) & 0.00017 & \nodata & \nodata & \nodata \\
 30.66808$-$0.33134 &  & 0.0097 & (+0.3,$-$0.3) & 0.0018 & \nodata & \nodata & \nodata \\
 33.91585+0.11111   & a & 0.012 & (+0.1,+2.7) & 0.0030 & \nodata & \nodata & \nodata \\
   & b & 0.011 & ($-$1.1,+0.4) & \nodata & \nodata & \nodata & \nodata \\
   & c & 0.008 & (+2.5,+4.5) & \nodata & \nodata & \nodata & \nodata \\
   & d & \nodata & \nodata & \nodata & 0.08 & (+2.1,0.0) & 0.054 \\
   & e & \nodata & \nodata & \nodata & 0.07 & (+1.1,+1.9) & \nodata \\
 33.81104$-$0.18582 & a & 0.014 & (0.0,$-$1.9) & 0.0030 & 0.53 & ($-$0.5,$-$1.8) & 0.24 \\
  & b & 0.0026 & (+0.1,$-$0.5) & \nodata & \nodata & \nodata & \nodata \\
 35.46832+0.13984   & a & 0.038 & ($-$2.3,+0.3) & 0.0044 & \nodata & \nodata & \nodata \\
  & b & 0.0061 & (+2.4,+0.1) & \nodata & \nodata & \nodata & \nodata \\
  & c & 0.0055 & (+2.1,+2.8) & \nodata & \nodata & \nodata & \nodata \\
 37.87411$-$0.39866 & a & 0.0075 & (+0.2,$-$0.1) & 0.0015 & \nodata & \nodata & \nodata \\
  & b & 0.0059 & ($-$2.7,+1.7) & \nodata & \nodata & \nodata & \nodata \\
  & c & 0.0042 & (+0.2,$-$1.9) & \nodata & $> 1$ & (0.0,$-$1.4) & 0.15 \\
  & d & 0.0020 & ($-$3.8,+0.1) & \nodata & \nodata & \nodata & \nodata \\
  & e & 0.0020 & ($-$4.6,+1.4) & \nodata & \nodata & \nodata & \nodata \\
  & f & \nodata & \nodata & \nodata & 0.45 & ($-$2.6,0.0) & \nodata \\
111.28293$-$0.66355 & a & 0.023 & ($-$0.8,$-$1.5) & 0.0041 & \nodata & \nodata & \nodata \\
 & b & 0.013 & (+0.5,+2.1) & \nodata & \nodata & \nodata & \nodata \\
 & c & 0.008 & (+0.8,0.0) & \nodata & \nodata & \nodata & \nodata \\
\enddata
\tablecomments{Optical depth peaks, the positions of the peaks, and optical depth means for our sources. The listed values are from our best alignment of the flux density maps. The typical 1$\sigma$ uncertainties due to a possible misalignment are 200\% for the peak optical depth and 50\% for the mean optical depth. The dominant contribution to the uncertainties is from uncertainties in aligning the flux density maps. Some of the fields contain multiple optical depth peaks, which are labeled `a', `b', `c', etc., and are indicated in the maps in Figures \ref{map1_12}--\ref{map18_12}.}
\end{deluxetable}

\clearpage

Table \ref{ext_data} lists the visible extinctions associated with the emitting dust. These were derived from the relation $A_V^{11.7}=26.88\ \tau_{11.7}$, or when possible from $A_V^{24.5}=57.82\ \tau_{24.5}$ (Draine 2003), where we use the average optical depths in columns (5) and (8) of Table \ref{taumap_data}, respectively.
Many of our sources have $A_V^{11.7}$ that are low enough to imply that many of the optical photons leak out of the region occupied by hot dust.
The $A_V^{24.5}$ values we derive for some of our sources are high enough to absorb all optical photons, thus being a better measure of the total extinction. However, the MIR colors and the MIR-derived spectral types of our sources indicate that at least for some them there is an additional extinction component that must come from cooler dust local to the source or on the line-of-sight (see \S \ref{physics} and \S \ref{colors}).
If dust and gas are well mixed, it is possible to derive the column density associated with the emitting dust by $N_H\approx 1.87\cdot 10^{21} A_V$ [cm$^{-2}$] (Draine 2003). The column densities $N_H^{11.7}$ and $N_H^{24.5}$ are given in Table \ref{ext_data} as well. See \S \ref{physics} for comparison with the column density of the ionized gas.

\clearpage

\begin{deluxetable}{ccccc}
\tablecaption{Extinction and Column Densities of Emitting Dust \label{ext_data}}
\tabletypesize{\small}
\tablewidth{0pt}
\tablehead{ &
\colhead{$A_V^{11.7}$} &
\colhead{$A_V^{24.5}$} &
\colhead{$N_H^{11.7}\times 10^{21}$} &
\colhead{$N_H^{24.5}\times 10^{21}$} \\
\colhead{Source} &
\colhead{(mag)} &
\colhead{(mag)} &
\colhead{(cm$^{-2}$)} &
\colhead{(cm$^{-2}$)} \\
\colhead{(1)} & \colhead{(2)} & \colhead{(3)} & \colhead{(4)} &  \colhead{(5)}}
\startdata
11.11198$-$0.39795 & 0.15  &   12  & 0.28  & 22    \\
11.94545$-$0.03634 & 0.03  &   14  & 0.06  & 26    \\
18.71179+0.00085   & 0.03  &\nodata& 0.06  &\nodata\\
19.75611$-$0.12775 & 0.14  &   10  & 0.26  & 19    \\
21.38654$-$0.25346 & 0.02  &\nodata& 0.04  &\nodata\\
25.39918$-$0.14081 & 0.12  &   10  & 0.22  & 19    \\
25.80211$-$0.15640 & 0.32  &    5  & 0.60  & 9     \\
27.18725$-$0.08095 & 0.04  & $>58$ & 0.07  & $>108$\\
28.28875$-$0.36359 & 0.89  &   10  & 1.66  & 19    \\
30.04343$-$0.14200 & 0.007 &\nodata& 0.01  &\nodata\\
30.86744+0.11493   & 0.02  &\nodata& 0.04  &\nodata\\
30.58991$-$0.04231 & 0.005 &\nodata& 0.009 &\nodata\\
30.66808$-$0.33134 & 0.05  &\nodata& 0.09  &\nodata\\
33.91585+0.11111   & 0.08  &    3  & 0.15  &  6    \\
33.81104$-$0.18582 & 0.08  &   14  & 0.15  & 26    \\
35.46832+0.13984   & 0.12  &\nodata& 0.22  &\nodata\\
37.87411$-$0.39866 & 0.04  &    9  & 0.07  & 17    \\
111.28293$-$0.66355& 0.11  &\nodata& 0.21  &\nodata\\
\enddata
\tablecomments{Visible extinctions and the corresponding column densities associated with the emitting dust for our sources. These figures were derived from the mean $\tau_{11.7}$ and $\tau_{24.5}$. The typical 1$\sigma$ uncertainties are 50\%.}
\end{deluxetable}

\clearpage

\subsection{Morphologies and Distances}
\label{morph}

Determining morphologies can be subjective and ambiguous, and with our sample we have only small numbers statistics, which makes it difficult to assess the statistical significance of the results. We use the morphological types that were defined by Wood \& Churchwell (1989b) and revisited by Churchwell (2002), to determine the source morphology by eye. The relatively few morphological types of UC \hii regions include: spherical or unresolved, cometary, core-halo, shell, irregular or multiple-peaked, and bipolar. Table \ref{morph_tab} lists our 18 sources with their morphology classes in column (2). We added comments explaining our choice of the morphological classes in column (3).

\begin{deluxetable}{cccccccc}
\tabletypesize{\tiny}
\tablecaption{Distances and Morphological Data \label{morph_tab}}
\tablewidth{0pt}
\tablehead{ &
\colhead{Morph.} &
\colhead{} &
\colhead{R} &
\colhead{$\Theta_{11.7\mu{\rm m}}$} &
\colhead{$D_{11.7\mu{\rm m}}$} &
\colhead{$\Theta_{5\ {\rm GHz}}$} &
\colhead{$D_{5\ {\rm GHz}}$} \\
\colhead{Source} &
\colhead{Class} &
\colhead{Comment} &
\colhead{(Kpc)} &
\colhead{(arcsec)} &
\colhead{(pc)} &
\colhead{(arcsec)} &
\colhead{(pc)} \\
\colhead{(1)} & \colhead{(2)} & \colhead{(3)} & \colhead{(4)} & \colhead{(5)} & \colhead{(6)} & \colhead{(7)} & \colhead{(8)}
}
\startdata
 11.11198$-$0.39795 & bipolar & Extensions to N, SE & $17\pm 1$\tablenotemark{1} & 1.5 & $0.12\pm 0.01$ & 2.4 & $0.20\pm 0.01$ \\
 11.94545$-$0.03634 & shell & Core at SE; Arc at NW & $4.2\pm 0.1$\tablenotemark{2,3} & 7.0\tablenotemark{a} & $0.143\pm 0.004$\tablenotemark{a} & 4.3 & $0.088\pm 0.003$ \\
 18.71179+0.00085   & spherical & \nodata & $4\pm 1$\tablenotemark{4} & $1.1$ & $0.021\pm 0.005$ & 1.1 & $0.021\pm 0.006$ \\
  & \nodata & \nodata & $12\pm 1$\tablenotemark{4} & \nodata & $0.064\pm 0.007$ & \nodata & $0.064\pm 0.008$ \\
 19.75611$-$0.12775 & spherical & \nodata & $4.2\pm 0.3$\tablenotemark{1} & 0.8 & $0.016\pm 0.002$ & 1.8 & $0.037\pm 0.003$ \\
  & \nodata & \nodata & $11.9\pm 0.3$\tablenotemark{1} & \nodata & $0.046\pm 0.005$ & \nodata & $0.104\pm 0.006$ \\
 21.38654$-$0.25346 & cometary & Off-centered core at SW & $5.6\pm 0.4$\tablenotemark{5} & 1.3 & $0.035\pm 0.003$ & 2.3 & $0.062\pm 0.005$ \\
  & \nodata & \nodata & $10.3\pm 0.3$\tablenotemark{5} & \nodata & $0.065\pm 0.005$ & \nodata & $0.115\pm 0.006$  \\
 25.39918$-$0.14081 & bipolar & Extensions to SSE and NNW & $9.8\pm 0.3$\tablenotemark{6} & 4.0\tablenotemark{a} & $0.190\pm 0.007$\tablenotemark{a} & 3.7 & $0.176\pm 0.007$ \\
 25.80211$-$0.15640 & core-halo & Extended halo & $5.5\pm 0.2$\tablenotemark{5} & 1.2 & $0.032\pm 0.003$ & 1.3 & $0.035\pm 0.003$ \\
  & \nodata & \nodata & $9.8\pm 0.2$\tablenotemark{5} & \nodata & $0.057\pm 0.004$ & \nodata & $0.062\pm 0.005$ \\
 27.18725$-$0.08095 & irregular & Weaker secondary peak at NE & $1.9\pm 0.3$\tablenotemark{1} & 0.7 & $0.006\pm 0.001$ & 2.3 & $0.021\pm 0.003$  \\
  & \nodata & \nodata & $13.3\pm 0.3$\tablenotemark{1} & \nodata & $0.045\pm 0.006$ & \nodata & $0.148\pm 0.007$  \\
 28.28875$-$0.36359 & shell & Shell with 2 peaks (17.65$\mu$m) & $3.3\pm 0.1$\tablenotemark{1,7} & 5.0\tablenotemark{a} & $0.080\pm 0.003$\tablenotemark{a} & 4.1 & $0.066\pm 0.003$ \\
 30.04343$-$0.14200 & spherical & \nodata & $6.1\pm 0.2$\tablenotemark{8} & 0.6 & $0.018\pm 0.003$ & 3.2 & $0.095\pm 0.004$ \\
  & \nodata & \nodata & $8.6\pm 0.2$\tablenotemark{8} & \nodata & $0.025\pm 0.004$ & \nodata & $0.133\pm 0.005$\\
 30.86744+0.11493   & spherical & \nodata & $12.2\pm 0.4$\tablenotemark{1,9} & 0.8 & $0.047\pm 0.005$ & 2.4 & $0.142\pm 0.008$  \\
 30.58991$-$0.04231 & spherical & \nodata & $2.5\pm 0.8$\tablenotemark{10, 11} & 0.8 & $0.010\pm 0.003$ & 2.5 & $0.03\pm 0.01$ \\
 30.66808$-$0.33134 & spherical & \nodata & $5.6\pm 0.3$\tablenotemark{4} & 0.8 & $0.022\pm 0.003$ & 1.5 & $0.041\pm 0.004$ \\
  & \nodata & \nodata & $9.1\pm 0.3$\tablenotemark{4} & \nodata & $0.035\pm 0.004$ & \nodata & $0.066\pm 0.005$ \\
 33.91585+0.11111   & bipolar & Extensions to NE, SW & $8.1\pm 0.1$\tablenotemark{12,13} & 3.0\tablenotemark{a} & $0.118\pm 0.004$\tablenotemark{a} & 3.3 & $0.129\pm 0.004$ \\
 33.81104$-$0.18582 & irregular & 2 peaks aligned N-S & $11.2\pm 0.3$\tablenotemark{1,14} & 3.0\tablenotemark{a} & $0.163\pm 0.006$\tablenotemark{a} & 1.1 & $0.060\pm 0.006$  \\
 35.46832+0.13984   & bipolar & Extension to E, W & $5.1\pm 0.3$\tablenotemark{1} & 4.0\tablenotemark{a} & $0.100\pm 0.006$\tablenotemark{a} & 5.0 & $0.124\pm 0.008$  \\
  & \nodata & \nodata & $8.8\pm 0.4$\tablenotemark{1} & \nodata & $0.171\pm 0.009$\tablenotemark{a} & \nodata & $0.21\pm 0.01$  \\
 37.87411$-$0.39866 & irregular & Sub-peaks at S,NW & $9.5\pm 0.6$\tablenotemark{1,15} & 5.0\tablenotemark{a} & $0.23\pm 0.02$\tablenotemark{a} & 2.0 & $0.092\pm 0.007$  \\
111.28293$-$0.66355 & irregular & Multi-peaks; filaments & $4.3\pm 0.5$\tablenotemark{1,6} & 2.0\tablenotemark{a} & $0.042\pm 0.005$\tablenotemark{a} & 4.7 & $0.10\pm 0.01$  \\
\enddata
\tablecomments{Morphologies (after Churchwell 2002), comments on the choice of morphological class, kinematic distances, and 11.7 $\mu$m and 5 GHz diameters (Giveon et al. 2005a). Multiple lines per object correspond to the near and far kinematic distances, in cases the distance ambiguity could not be resolved.}
\tablenotetext{a}{An upper limit on the FWHM since this object has an extended irregular or multi-peaked morphology.}
\tablecomments{References: 1 - Bronfman, Nyman, \& May (1996). 2 - Braz \& Sivagnanam (1987). 3 - Simpson \& Rubin (1990). 4 - GRS (see text). 5 - Szymczak, Hrynek, \& Kus (2000). 6 - Churchwell, Walmsley, \& Cesaroni (1990). 7 - Kurtz, Churchwell, \& Wood (1994). 8 - Lockman (1989). 9 - Palagi et al. (1993). 10 - Szymczak \& Gerard (2004). 11 - Wouterloot, Brand, \& Fiegle (1993). 12 - Wink, Wilson, \& Bieging (1983) 13 - Wood \& Churchwell (1989a). 14 - Watson et al. (2003). 15 - Afflerbach, Churchwell, \& Werner (1997).}
\end{deluxetable}

Column (4) of Table \ref{morph_tab} lists galactocentric distances ($R_{gal}$) which we derive using literature line-of-sight velocities and the Galactic rotation curve of Rohlfs \& Kreitschmann (1987).
These velocities were determined using different methods and are based on various emission features (e.g., radio recombination lines; sub-mm CS molecular lines; OH, methanol and ammonia maser emission). Each entry in Table \ref{morph_tab} has a reference to the source of the velocity. In some cases, there is also a reference to the work that resolved the distance ambiguity.
In the two cases where no direct velocity measurements are available in the literature (18.71179+0.00085 and 30.66808$-$0.33134), we used molecular emission-line data from the Galactic Ring Survey (GRS; Simon et al. 2001)\footnote{This publication makes use of molecular line data from the Boston University Galactic Ring Survey (GRS). The GRS is a joint project of Boston University and Five College Radio Astronomy Observatory, funded by the National Science Foundation under grants AST-9800334, AST-0098562, \& AST-0100793.}.
The GRS is a survey of the $^{13}$CO $J=1\rightarrow 0$ transition at 110.201 GHz,  which is known to trace embedded UC \hii regions (Koplak et al. 2003). The measurements are taken with a relatively fine sampling of $22''$ along the coordinate range $l=18^{\circ}$--$54^{\circ}$ and $|b|\le 1$, with a velocity resolution of 0.25 km s$^{-1}$. Because of the smaller line widths of $^{13}$CO compared to $^{12}$CO, it is possible to avoid velocity crowding and establish accurate kinematic distances to the clouds. For sources within the GRS coverage area, the GRS velocities and the velocities from the literature are consistent within the uncertainties implied by the line widths.

The rotation curve of Rohlfs \& Kreitschmann (1987) spans a range of 0.1 to 19.4 kpc in galactocentric radius. We have transformed this rotation curve (with $R_{\odot}=7.9$ kpc and $\theta_{\odot}=184$ km s$^{-1}$) to a rotation curve with the more updated IAU values, $R_{\odot}=8.5$ kpc and $\theta_{\odot}=220$ km s$^{-1}$. We assume that all sources lie in the Galactic plane ($b=0^{\circ}$), as all of them have very small latitudes ($|b|< 0.7^{\circ}$).
The kinematic distance ($R_{kin}$) is calculated by inverting the formula
\begin{equation}
R_{gal}^2=R_{\odot}^2-2R_{kin}R_{\odot}\cos{l}+R_{kin}^2.
\label{r_gal_eq}
\end{equation}
For sources outside the solar circle, this gives a single solution, whereas inside the solar circle there are two solutions. The choice of distance in this case is based on different methods discussed in the references.

MIR angular and physical sizes from this work are listed in columns (5)--(6). We resolve all sources. In most cases, we use the full width at half maximum (FWHM) as a measure for the source size, but in cases of extended irregular or multi-peaked sources, we estimate an upper limit on the FWHM. These cases are indicated by a footnote flag. Beam size (0.3'' at 11.7 $\mu$m) was removed in quadrature to obtain the final size values. We also list radio angular (Giveon et al. 2005a) and physical sizes in columns (7)--(8) for comparison.

\subsection{Luminosities and Spectral Types}
\label{physics}

MIR luminosities were calculated by integrating the Planck function from 1 to 1000 $\mu$m using the dust color temperatures $T_{17.65/11.7}$ and $T_{24.5/17.65}$ and the corresponding optical depths. We employ the emissivity function $(1-e^{-\tau_{\nu}})$, with $\tau_{\nu}$ taken from Draine (2003), and assume isotropic emission into $4\pi$ sr. Our MIR luminosities can be considered as lower limits to the bolometric luminosity, if we assume that most of the shorter wavelength flux from the central ionizing source is reprocessed by the dust. The results of \S \ref{taus} show that $A_V$ derived from the 24.5 $\mu$m data is sufficient to absorb all optical photons. That is usually not so for $A_V$ derived from the 11.7 $\mu$m data. An anisotropic dust distribution, very thick dust that reprocesses radiation into the FIR regime (either locally or on the line-of-sight), and the reprocessing of radiation by the ionized gas in these sources are the main reasons for underestimating the bolometric luminosity, even at 24.5 $\mu$m.

For each source, the Draine (2003) extinction curve was normalized to have the measured $\tau_{11.7}$ or $\tau_{24.5}$ of the source, when calculating  $L_{17.65/11.7}$ or $L_{24.5/17.65}$, respectively. For this calculation we derive the total emission optical depths, $\tau_{11.7}$ and $\tau_{24.5}$, for each source, based on the total flux densities at the corresponding bands.
All sources except two (19.75611$-$0.12775 and 27.18725$-$0.08095), have $\tau_{\nu} < 1$ at 11.7 and 24.5 $\mu$m. These two sources are optically thick at 24.5 $\mu$m, so we estimate their luminosities by a blackbody with the corresponding temperature.

We derive spectral types based on MIR flux densities and dynamical non-LTE model atmospheres of hot stars from Sternberg, Hoffmann, \& Pauldrach (2003), assuming these bolometric luminosities come from a single star and that the MIR luminosity is a good estimator of the bolometric luminosity. For cooler stars (later than B0.5) we use the tables of Doyon (1990), which are based on the Kurucz static LTE atmospheres (Kurucz 1992). MIR luminosities and spectral types are listed in Table \ref{lumi}, columns (2)--(3) and (9), respectively. If the bolometric luminosity heating the dust comes from a cluster of stars, the MIR-derived spectral type of the ionizing star, which is typically the most massive star in the cluster, will be later than the listed type.
\begin{deluxetable}{cccccccccc}
\tabletypesize{\tiny}
\rotate
\tablecaption{Derived Physical Properties \label{lumi}}
\tablewidth{0pt}
\tablehead{ &
\colhead{$\log{L_{17.65/11.7}}$\tablenotemark{a}} &
\colhead{$\log{L_{24.5/17.65}}$\tablenotemark{a}} &
 & \colhead{EM$\times 10^6$} &
\colhead{$n_e\times 10^4$} &
\colhead{$N_H^{ion}\times 10^{21}$} &
\colhead{$\log{N_{Lyc}}$} &
\colhead{MIR Spectral} &
\colhead{Radio Spectral} \\
\colhead{Source} &
\colhead{($L_{\odot}$)} &
\colhead{($L_{\odot}$)} &
\colhead{$\tau_{5\ {\rm GHz}}$} &
\colhead{(pc\ cm$^{-6}$)} &
\colhead{(cm$^{-3}$)} &
\colhead{(cm$^{-2}$)} &
\colhead{(s$^{-1}$)} &
\colhead{Type\tablenotemark{b}} &
\colhead{Type} \\
\colhead{(1)} & \colhead{(2)} & \colhead{(3)} & \colhead{(4)} & \colhead{(5)} & \colhead{(6)} & \colhead{(7)} & \colhead{(8)} & \colhead{(9)} & \colhead{(10)}}
\startdata
 11.11198$-$0.39795 & $4.3\pm 0.2$ & $5.2\pm 0.3$ & $0.15\pm 0.05$ & $14\pm 5$ & $0.8\pm 0.1$ & $2.5\pm 0.3$ & $48.5\pm 0.1$ & B1--B0.5 / B0--O6.5 & O8.5--O9 \\
 11.94545$-$0.03634 & $3.4\pm 0.1$ & $4.3\pm 0.2$ & $0.19\pm 0.07$ & $17\pm 6$ & $1.4\pm 0.2$ & $1.9\pm 0.3$ & $47.9\pm 0.1$ & B3--B2.5 / B1 & B0.5--B0 \\
 18.71179+0.00085   & $2.6\pm 0.4$ & \nodata & $0.9\pm 0.5$ & $80\pm 50$ & $6\pm 2$ & $1.9\pm 0.9$ & $47.3\pm 0.3$ & B8--B5 & B0.5 \\
  & $3.5\pm 0.4$ & \nodata & \nodata & \nodata & $4\pm 1$ & $4\pm 1$ & $48.3\pm 0.2$ & B4--B1.5 & O9.5--O9 \\
 19.75611$-$0.12775 & $3.3\pm 0.3$ & $4.0\pm 0.3$\tablenotemark{c} & $0.07\pm 0.02$ & $6\pm 2$ & $1.3\pm 0.2$ & $0.7\pm 0.1$ & $46.7\pm 0.1$ & B4--B2.5 / B2--B1 & B0.5 \\
  & $4.2\pm 0.3$ & $4.9\pm 0.3$\tablenotemark{c} & \nodata & \nodata & $0.8\pm 0.1$ & $1.3\pm 0.2$ & $47.6\pm 0.1$ & B1.5--B0.5 / B0.5--O8.5 & B0.5 \\
 21.38654$-$0.25346 & $3.2\pm 0.2$ & \nodata & $0.15\pm 0.05$ & $14\pm 5$ & $1.5\pm 0.3$ & $1.4\pm 0.3$ & $47.5\pm 0.1$ & B5--B2.5 & B0.5--B0 \\
  & $3.7\pm 0.2$ & \nodata & \nodata & \nodata & $1.1\pm 0.2$ & $2.0\pm 0.4$ & $48.0\pm 0.1$ & B2 & B0 \\
 25.39918$-$0.14081 & $4.3\pm 0.1$ & $5.2\pm 0.2$ & $0.5\pm 0.2$ & $50\pm 20$ & $1.7\pm 0.3$ & $4.6\pm 0.8$ & $48.9\pm 0.1$ & B1--B0.5 / O9--O7 & O8--O7 \\
 25.80211$-$0.15640 & $3.5\pm 0.2$ & $4.2\pm 0.3$ & $0.16\pm 0.06$ & $14\pm 5$ & $2.0\pm 0.4$ & $1.1\pm 0.2$ & $47.0\pm 0.1$ & B3--B2 / B1.5--B0.5 & B0.5 \\
  & $4.0\pm 0.2$ & $4.7\pm 0.3$ & \nodata & \nodata & $1.5\pm 0.3$ & $1.4\pm 0.3$ & $47.5\pm 0.1$ & B2--B1 / B0.5--O9.5 & B0.5 \\
 27.18725$-$0.08095 & $1.9\pm 0.4$ & $2.1\pm 0.2$\tablenotemark{c} & $0.019\pm 0.007$ & $1.7\pm 0.6$ & $0.9\pm 0.2$ & $0.29\pm 0.08$ & $45.6\pm 0.2$ & A1--B8 / B9.5--B8 & B1.5--B1 \\
  & $3.6\pm 0.4$ & $3.8\pm 0.2$\tablenotemark{c} & \nodata & \nodata & $0.34\pm 0.06$ & $0.8\pm 0.1$ & $47.3\pm 0.1$ & B3--B1.5 / B2.5--B1.5 & B0.5 \\
 28.28875$-$0.36359 & $4.2\pm 0.1$ & $4.7\pm 0.2$ & $0.25\pm 0.09$ & $23\pm 8$ & $1.8\pm 0.3$ & $1.8\pm 0.3$ & $47.7\pm 0.1$ & B1 / B0.5--B0 & B0.5 \\
 30.04343$-$0.14200 & $3.7\pm 0.7$ & \nodata & $0.0014\pm 0.0005$ & $0.13\pm 0.05$ & $0.12\pm 0.02$ & $0.18\pm 0.03$ & $45.8\pm 0.1$ & B5--B0.5 & B1.5--B1 \\
  & $4.0\pm 0.7$ & \nodata & \nodata & \nodata & $0.10\pm 0.02$ & $0.21\pm 0.04$ & $46.1\pm 0.1$ & B3--B0.5 & B1 \\
 30.86744+0.11493   & $4.1\pm 0.4$ & \nodata & $0.4\pm 0.1$ & $36\pm 9$ & $1.6\pm 0.2$ & $3.5\pm 0.5$ & $48.6\pm 0.1$ & B1.5--B0.5 & O9--O8 \\
 30.58991$-$0.04231 & $1.7\pm 0.5$ & \nodata & $0.06\pm 0.02$ & $5\pm 2$ & $1.3\pm 0.3$ & $0.6\pm 0.2$ & $46.4\pm 0.2$ & A4--B8 & B1--B0.5 \\
 30.66808$-$0.33134 & $3.2\pm 0.3$ & \nodata & $0.7\pm 0.3$ & $60\pm 30$ & $4\pm 1$ & $2.5\pm 0.7$ & $47.8\pm 0.2$ & B5--B2.5 & B0.5--B0 \\
  & $3.6\pm 0.3$ & \nodata & \nodata & \nodata & $3.0\pm 0.8$ & $3.1\pm 0.8$ & $48.2\pm 0.2$ & B3--B1.5 & B0--O9 \\
 33.91585+0.11111   & $3.9\pm 0.1$ & $4.7\pm 0.3$ & $0.3\pm 0.1$ & $27\pm 9$ & $1.4\pm 0.2$ & $2.8\pm 0.4$ & $48.4\pm 0.1$ & B2--B1.5 / B0.5--O9.5 & O9.5--O9 \\
 33.81104$-$0.18582 & $4.2\pm 0.1$ & $4.8\pm 0.3$ & $0.6\pm 0.3$ & $50\pm 30$ & $2.9\pm 0.9$ & $2.7\pm 0.9$ & $48.0\pm 0.2$ & B1 / B0.5--O9 & B0.5--O9.5 \\
 35.46832+0.13984   & $4.0\pm 0.1$ & \nodata & $0.10\pm 0.03$ & $9\pm 3$ & $0.8\pm 0.1$ & $1.5\pm 0.2$ & $47.9\pm 0.1$ & B1.5 & B0.5--B0 \\
  & $4.5\pm 0.1$ & \nodata & \nodata & \nodata & $0.6\pm 0.1$ & $1.9\pm 0.3$ & $48.4\pm 0.1$ & B0.5 & O9.5--O9 \\
 37.87411$-$0.39866 & $4.5\pm 0.1$ & $5.1\pm 0.3$ & $1.3\pm 0.7$\tablenotemark{d} & $200\pm 100$ & $5\pm 1$ & $7\pm 2$ & $49.0\pm 0.2$ & B0.5 / B0.5--O7 & O8--O6.5 \\
111.28293$-$0.66355 & $3.6\pm 0.2$ & \nodata & $0.6\pm 0.3$\tablenotemark{e} & $4\pm 2$ & $0.6\pm 0.2$ & $0.9\pm 0.3$ & $47.3\pm 0.2$ & B2.5--B2 & B0.5 \\
\enddata
\tablecomments{MIR luminosities, free-free optical depths at 5 GHz (or at 1.4 GHz for sources with no 5 GHz detection), emission measures, electron densities, column densities, Lyman continuum photon rates, and MIR and radio-derived spectral types for our sample. The MIR luminosities and spectral types should be considered to be lower limits (see text). Multiple lines per object correspond to the near and far kinematic distances, as in Table \ref{morph_tab}.}
\tablenotetext{a}{When a source was observed with all three bands, luminosities are derived for the two color temperatures calculated.}
\tablenotetext{b}{Spectral types are derived for the two luminosities calculated - $L_{17.65/11.7}$ / $L_{24.5/17.65}$.}
\tablenotetext{c}{Optically thick cases where luminosities were estimated using a blackbody spectrum.}
\tablenotetext{d}{$T_e=15\ 000$ K was assumed instead of 10 000.}
\tablenotetext{e}{This is actually $\tau_{1.4\ {\rm GHz}}$.}
\end{deluxetable}

An independent way to estimate stellar types is possible using our radio data, assuming they are dominated by free-free emission. Radio emission does not suffer from extinction at all, and thus allows us to test our assumptions regarding the MIR emission. We calculate the radio properties similarly to Wood \& Churchwell (1989a) and list them in Table \ref{lumi}.
For ionized gas, the brightness temperature at the Rayleigh-Jeans approximation at a frequency $\nu$ is
\begin{equation}
T_{b\nu}={{c^2 S_{\nu}\times 10^{-26}}\over{2k\ \nu^2\ \Omega}}\ [K],
\label{tbright}
\end{equation}
where $S_{\nu}$ is the integrated flux density in milliJanskys and $\Omega$ is the radio source solid angle in sr.

The free-free optical depth, listed in column (4) of Table \ref{lumi}, is then calculated from
\begin{equation}
T_{b\nu}=\left(1-e^{-\tau_{\nu}}\right)\ T_e,
\label{tau_rad}
\end{equation}
assuming an electron temperature $T_e=10^4$ K with a 50\% uncertainty. One exception for that is the source 37.87411$-$0.39866, which has a higher brightness temperature. In that case we assumed $T_e=1.5\cdot 10^4$ K.
The emission measure, listed in column (5), is given by
\begin{equation}
EM={{\tau_{\nu}}\over{8.235\times 10^{-2}\ a_{\nu}\ T_e^{-1.35}\ \nu^{-2.1}_{\rm GHz}}}\ [{\rm pc\ cm}^{-6}],
\label{em_rad}
\end{equation}
where $a_{\nu}$ is a unity order constant that equals 0.9938 for $T_e=10^4$ K and $\nu=5$ GHz (Mezger \& Henderson 1967).
We can estimate the electron density by the rms value (Wood \& Churchwell 1989a),
\begin{equation}
n_e=990\ \sqrt{\left({EM\over{10^6\ {\rm pc\ cm}^{-6}}}\right)\left({{\rm pc}\over D}\right)}\ [{\rm cm}^{-3}],
\label{dens}
\end{equation}
where $D$ is the source diameter in pc. The densities are listed in column (6), and the column densities, $N_H^{ion}$ derived from the densities are listed in column (7).
Assuming a homogeneous, ionization-bounded \hii region, in which none of the hydrogen ionizing photons ($h\nu \ge 13.6$ eV) is absorbed by dust, we use the relation
\begin{equation}
N_{\rm Lyc}=\alpha_B \int n_e^2\ dV=\alpha_B\ EM \int d\Omega\ R_{kin}^2
\label{lyc1}
\end{equation}
to calculate the Lyman continuum photon rate $N_{\rm Lyc}$,
\begin{equation}
N_{\rm Lyc}=1.7\times10^{44}\left({EM\over{10^6\ {\rm pc\ cm}^{-6}}}\right)\left({{\Omega}\over{{\rm arcsec}^2}}\right)\left({{R_{kin}}\over{\rm kpc}}\right)^2\ [s^{-1}],
\label{lyc2}
\end{equation}
which is listed in column (8).
No assumption was made here regarding the optical depth of the cloud.
Assuming that these ionizing photons come from a single star, it is possible to determine the spectral type, again, using the stellar model atmospheres of Sternberg, Hoffmann, \& Pauldrach (2003) for hot stars and of Doyon (1990) for cooler stars. The radio-derived spectral types are listed in column (10) of Table \ref{lumi}. It is unlikely that the central sources are star clusters, since our single-star estimates are already giving relatively late type stars. A central cluster would imply an even later type for the most massive star in the cluster.

Tables \ref{morph_tab} and \ref{lumi} confirm that our sources, selected according to the combination of their radio and MIR continuum emissions are indeed UC \hii regions, as they generally have very small physical sizes ($D\ltorder 0.1$ pc), high densities ($n_e\gtorder 10^4$ cm$^{-3}$), and have bright ($EM\gtorder 10^6$ pc cm$^{-6}$, but in most cases $\gtorder 10^7$ pc cm$^{-6}$) photoionized gas (Wood \& Churchwell 1989a; Kurtz, Churchwell, \& Wood 1994).

Comparing dust-emitting column densities, $N_H^{11.7}$ and $N_H^{24.5}$, derived from the LWS observations in this work (Table \ref{ext_data}), and the ionized gas column density, $N_H^{ion}$, derived from our previous radio observations (White, Becker, \& Helfand 2005; Giveon et al. 2005a) shows that $N_H^{11.7}<N_H^{ion}<N_H^{24.5}$. This supports the common assumption that the hot emitting dust is a relatively thin layer surrounding the Str\"omgren sphere of ionized gas, while the cooler dust that absorbs all of the optical photons consists of a much thicker column, further away from the central source.

An interesting result that emerges from these calculations is that even according to the radio-derived ionizing luminosities, 50\% of our objects are excited by B-stars and not by O-stars, and all our objects have derived types that are later than an O6 star. A possible explanation for missing the earliest O-stars might be a bias imposed by selecting the most compact radio sources (5 GHz diameters $\le 5''$; see also \S \ref{correlate}). In 50\% of the sources, the MIR-derived spectral types agree with the radio-derived spectral types. The 24.5 $\mu$m-derived spectral types always show a better agreement with the radio ones compared to the 11.7 $\mu$m-derived types. Two other sources (11.94545$-$0.03634 and 27.18725$-$0.08095) show clear deficiency in MIR photons (even at 24.5 $\mu$m) to account for the radio-derived ionizing source. Both cases may be explained simply by the 24.5 $\mu$m luminosities underestimating the real luminosities due to the reasons given above. Alternatively, this may be explained by an $A_V=30$ foreground dust exctinction, or by a comparable extinction by dust cooler than $\sim 60$ K that is still local to the source. Seven sources that show a MIR-radio inconsistency in the derived spectral types were not observed at 24.5 $\mu$m. Their MIR luminosities, derived only using the 11.7 $\mu$m data, significantly underestimate the real luminosities. This inconsistency might be resolved by obtaining 24.5 $\mu$m observations of these sources.

On average, assuming the far distances for all sources that still have ambiguous distances, the spectral type derived from $L_{17.65/11.7}$ are later than the ones derived from $L_{24.5/17.65}$ -- B2 ($\pm 2$) compared to B0 ($\pm 1$, i.e., in the range B1--O9).
Assuming the near kinematic distance for the ambiguous sources pushes the radio-derived spectral types to even later types, which are already quite late compared to what one expects from ionized compact HII regions. Non of the results changed significantly due to the choice of distance.
The longer wavelength MIR-derived spectral types agree with the average radio-derived spectral type, which is also B0 ($\pm 1$). The later types derived from the 11.7 $\mu$m luminosities (as late as A stars for 25.80211$-$0.15640 and 30.58991$-$0.04231), suggest that the derivation using the longer wavelengths may better match the real bolometric luminosities.
The consistency of the spectral types derived from $L_{24.5/17.65}$ and from the radio suggest that the contribution from plausible lower-luminosity non-ionizing companion stars in a cluster is not significant.
This better consistency seems to be due to the 24.5 $\mu$m emission coming from a thicker layer of dust compared to the 11.7 $\mu$m -- better representing the entire surrounding cloud -- as is evident from our calculated emission optical depths (\S \ref{taus}).
Another reason for underestimating the luminosities with $L_{17.65/11.7}$ may be absorption by the broad 9.7 $\mu$m silicate feature going into the 11.7 $\mu$m band, and by the 18 $\mu$m silicate feature.

\section{Discussion}
\label{discuss}

\subsection{Correlations}
\label{correlate}

We have tested for the existence of correlations among all measured quantities described above, both averages and peaks. We chose Spearman's rank-correlation coefficient over Pearson's correlation coefficient because it tests for a general monotonic relation rather than only a linear one. However, in cases of bimodality Spearman's coefficient may fail to detect correlations.

\begin{figure}
\plotone{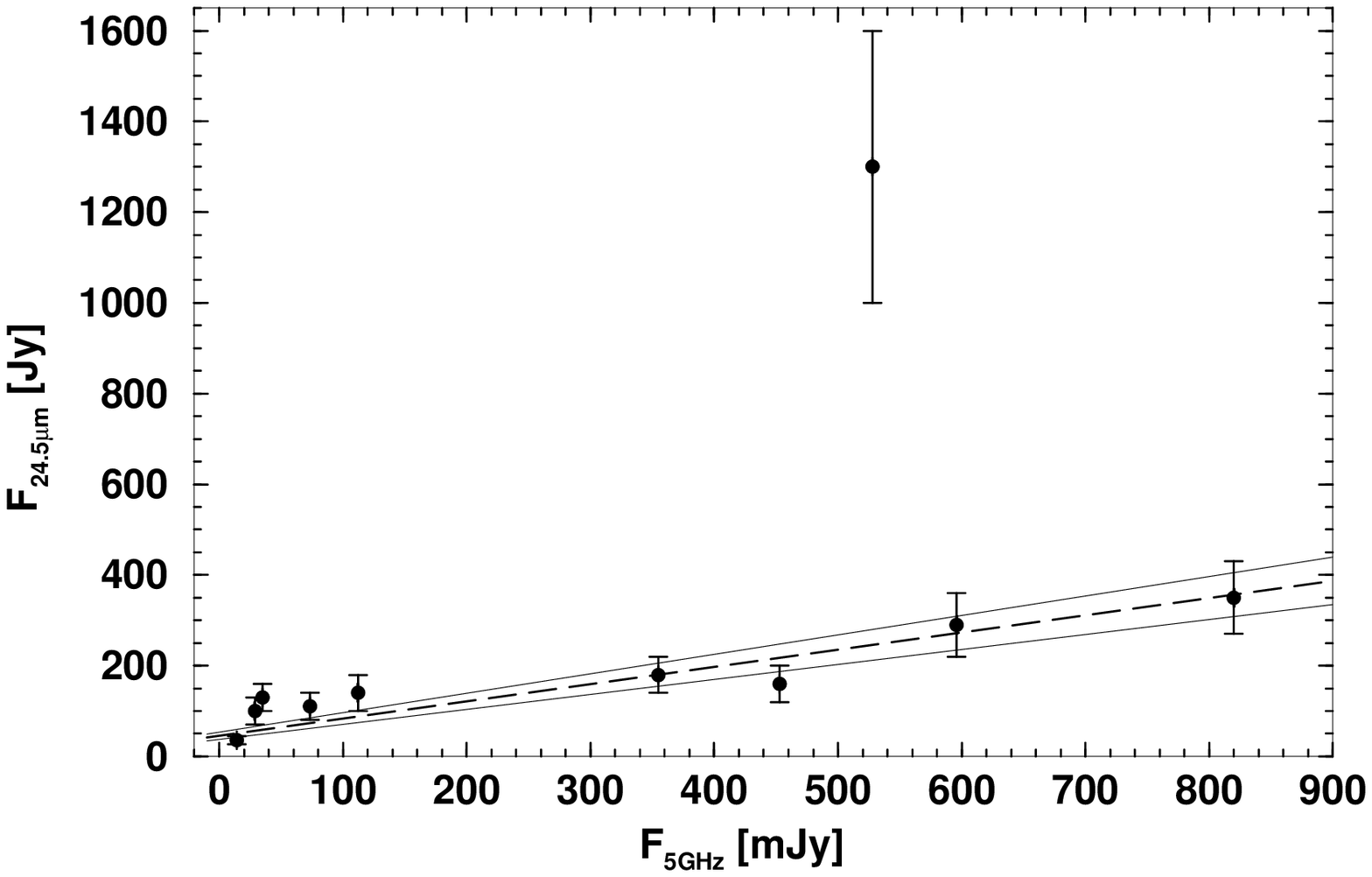}
\caption{Flux density at 24.5 $\mu$m as a function of the 5 GHz flux density with 1$\sigma$ errors. The horizontal error bars are smaller than the plotted circles. The linear fit given in eq. \ref{eq_f6_f25} is overplotted as a dashed line. The solid lines mark the 1$\sigma$ uncertainty around this fit.}
\label{f6_f25}
\end{figure}

The strongest correlation that emerges from our data is between the MIR and the radio integrated flux densities. Figure \ref{f6_f25} shows the 24.5 $\mu$m flux densities as a function of the 5 GHz flux densities with our best fit.

\clearpage

The correlation coefficient is $r_S=0.94$ with probability $P_r=5\cdot 10^{-5}$ of being random. This strong correlation reflects the tight connection between the ionizing luminosity of the central star or stars, and the star's total luminosity reprocessed by the surrounding dust into the IR. However, the correlations of the 11.7 and 17.65 $\mu$m bands are much weaker -- $r_S=0.54$ ($P_r=0.02$) and $r_S=0.55$ ($P_r=0.02$), respectively. These correlations remain weaker than that of the 24.5 $\mu$m even when only the same 10 sources are considered -- $r_S=0.87$ ($P_r=0.001$) and $r_S=0.74$ ($P_r=0.014$) for 11.7 and 17.65 $\mu$m, respectively. The fact that of the three bands, the 24.5 $\mu$m shows the strongest correlation with the radio further supports our result from \S \ref{physics} that the 24.5 $\mu$m flux density is a better estimator of the total flux density reprocessed by the dust. The source 28.28875$-$0.36359 is exceptionally bright in the MIR -- perhaps part of the MIR emission in this extended object is not related to the ionizing source. Fitting a linear relation with 1$\sigma$ error weighting, we obtain,
\begin{equation}
F_{\rm 24.5\mu m}[{\rm Jy}]=(0.38\pm 0.05)\cdot F_{\rm 5GHz}[{\rm mJy}] + (45\pm 8).
\label{eq_f6_f25}
\end{equation}
The source 28.28875$-$0.36359 does not affect the fit significantly due to the large uncertainty in its MIR flux density. Increasing the errors to 3$\sigma$ increases the errors of the fit -- $F_{\rm 24.5\mu m}[{\rm Jy}]=(0.4\pm 0.2)\cdot F_{\rm 5GHz}[{\rm mJy}] + (50\pm 20)$ -- but the slope remains significantly larger than zero.

\begin{figure}
\plotone{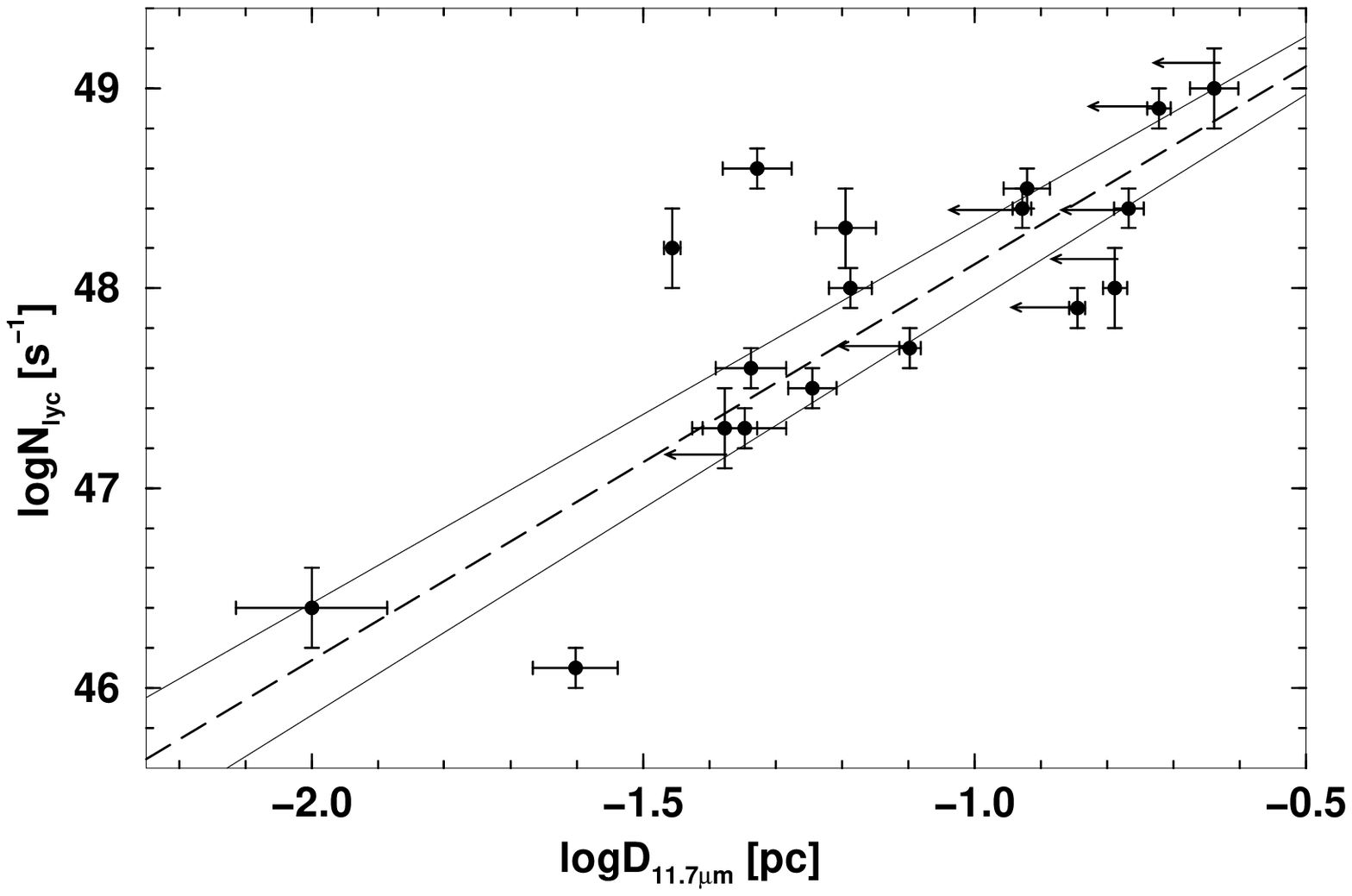}
\caption{The ionizing photon rate as a function of the 11.7 $\mu$m physical size with 1$\sigma$ errors. Source MIR size increases with the ionizing rate of the central star. The fit to this relationship, expressed by eq. \ref{eq_dnlyc} is overplotted as a dashed line. The solid lines mark the 1$\sigma$ uncertainty around this fit.}
\label{size_nlyc}
\end{figure}

Another correlation we find is between the 11.7 $\mu$m size $D_{11.7}$ as defined in \S \ref{morph} and the ionizing photon rate, $N_{lyc}$: $r_S=0.73$  ($P_r=6\cdot 10^{-4}$).
$D_{11.7}$ and $N_{lyc}$ have the fitted power-law relation
\begin{equation}
\log N_{lyc}\ [s^{-1}]=(1.98\pm 0.09)\log D_{11.7}\ [pc]+(50.1\pm 0.1).
\label{eq_dnlyc}
\end{equation}
The physical MIR size show weaker correlations with the MIR luminosities: $r_S=0.63$  ($P_r=0.005$) with $L_{17.65/11.7}$ and $r_S=0.61$  ($P_r=0.008$) with  $L_{24.5/17.65}$.
Even though the 24.5 $\mu$m flux densities show a stronger correlation with the radio flux densities, and the spectral types derived from the 24.5 $\mu$m flux densities are more consistent with the radio-derived types compared to the types derived from the 11.7 $\mu$m flux densities, there is a smaller number of sources at 24.5 $\mu$m and the range of their sizes is more limited. This makes the probability of their correlation with $N_{lyc}$ smaller than the 11.7um-$N_{lyc}$ correlation ($r_S=0.87$, $P_r=0.001$), and the uncertainty in the fit larger: $\log N_{lyc}\ [s^{-1}]=(2.4\pm 0.2)\log D_{11.7}\ [pc]+(50.3\pm 0.2)$. Also, most of the sources observed at 24.5 $\mu$m are still upper limits (6 out of 10) compared to a smaller fraction at 11.7 $\mu$m (8 out of 18).
We plot $N_{lyc}$ vs. $D_{11.7}$ with their 1$\sigma$ uncertainties in Figure \ref{size_nlyc}, together with the best fit and its 1$\sigma$ uncertainty.

This relation in eq. \ref{eq_dnlyc} implies that dust at the same temperature will be further away from the central source as $N_{lyc}$ increases. This is because as the ionizing and heating power of the central star(s) increases, the dust destruction radius increases and the surviving dust is heated at larger distances. The true power-law index is probably larger than 2, since a few of the sizes are upper limits (see \S \ref{morph}). When the upper limits are not included, it is $2.8\pm 0.2$. Garay \& Lizano (1999) found a correlation between the radio luminosity and the radio size for a collection of samples comprising a few hundreds UC and compact \hii regions. Paladini, Davies, \& DeZotti (2004) find a similar trend in a sample of 250 \hii regions. The ionizing photon rate $N_{lyc}$ is proportional to the radio luminosity (see eqs. \ref{tbright}--\ref{lyc1}), but other variables, such as the temperature and the angular size, introduce an additional scatter in $N_{lyc}$.

Another important implication of this result is that the more compact sources are actually due to later spectral types rather than young age. This was suggested using dynamic considerations by e.g., De Pree, Rodr\'{i}guez \& Goss (1995), Garc\'{i}a-Segura \& Franco (1996), Xie et al. (1996), and Garay \& Lizano (1999), and was known as the age problem of UC \hii regions (Kim \& Koo 2001): their number is about an order of magnitude greater than expected from other indicators of massive star-formation rate based on their dynamical age (Garay \& Lizano 1999). If most of the UC \hii regions are in fact B-stars and not O-stars, as suggested by the present work, the age problem might be resolved.
 This is why we might be missing the earliest O-stars: selecting the most compact radio sources (5 GHz diameters $\le 5''$) may have biased our sample towards later types.

\subsection{MIR Color Criteria}
\label{colors}

Wood \& Churchwell (1989b) suggested IR color criteria for selecting UC \hii regions, based on IRAS data. The $\sim 150$ factor in angular resolution between IRAS and LWS have an impact on these criteria, since in some cases, a significant fraction of the flux density comes from diffuse emission or sources unrelated to the compact object, that could not be separated with the IRAS beam.

\begin{figure}
\plotone{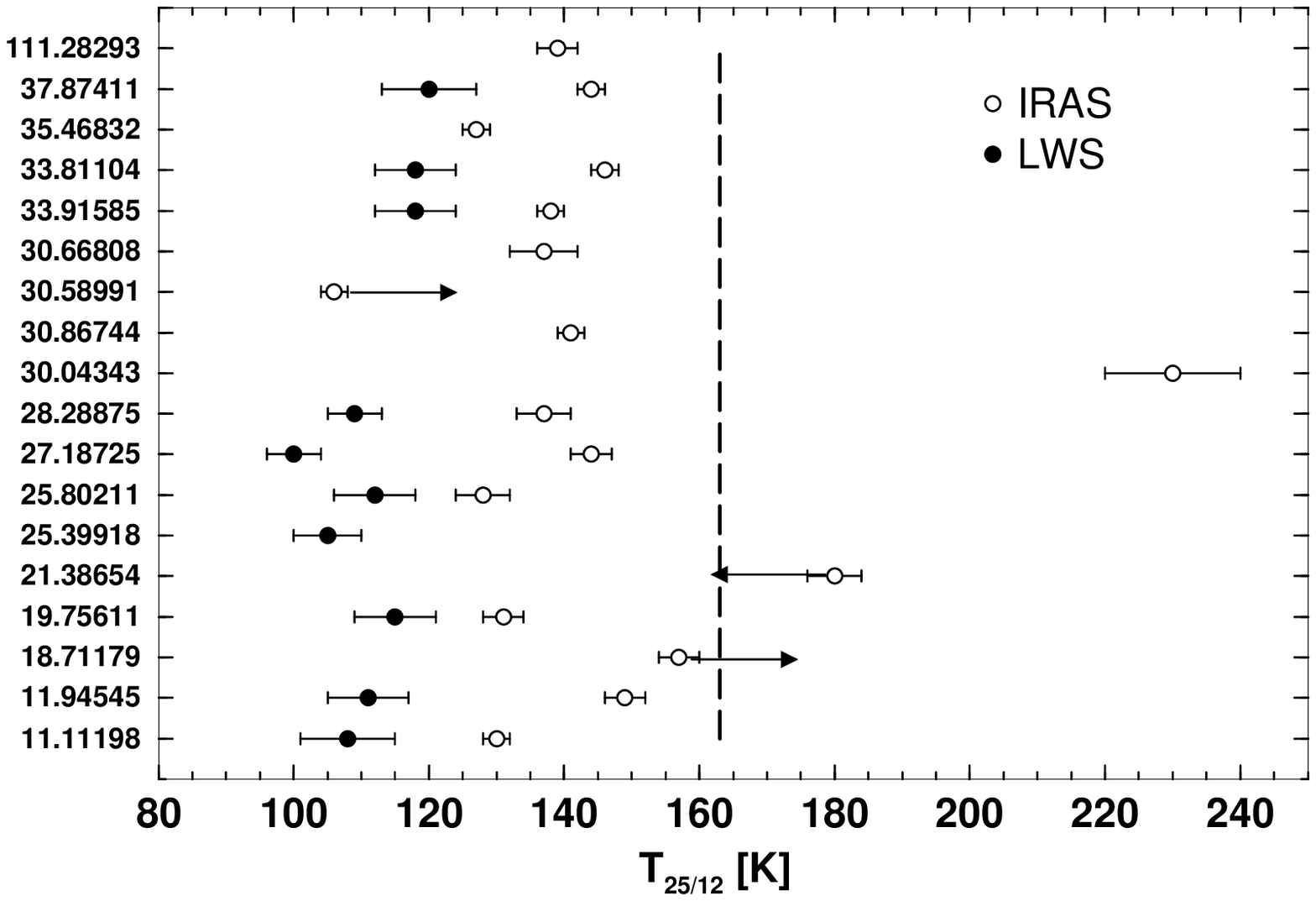}
\caption{A comparison between $T_{25/12}$ derived from the $F_{\nu}(25\mu{\rm m})/F_{\nu}(12\mu{\rm m})$ ratios of sources in our sample (when available) as measured by IRAS (empty circles) and by LWS (filled circles), with 1$\sigma$ error bars. Arrows indicate limits on the ratios. The vertical dashed line is the lower limit of 163 K for UC \hii regions according to Wood \& Churchwell (1989b).}
\label{cdiag}
\end{figure}

We compare the $T_{25/12}$ temperature derived from the $F_{\nu}(25\mu{\rm m})/F_{\nu}(12\mu{\rm m})$ ratio of IRAS (Wood \& Churchwell 1989b) to that we obtain from LWS ($T_{24.5/11.7}$), assuming a modified blackbody with the Draine (2003) emissivity law as described in \S \ref{temp}. These temperatures are shown in Figure \ref{cdiag}. We preferred to compare the temperatures and not the raw flux density ratios since they are easier to interpret.
The lower limit on the temperature for UC \hii regions (163 K) is overplotted as a dashed line. This limit was derived from the limit on the $F_{\nu}(25\mu{\rm m})/F_{\nu}(12\mu{\rm m})$ ratio (=3.7) from Wood \& Churchwell (1989b), taking into account the bandpass response of IRAS and the assumed emissivity law. There is a significant decrease in the limiting temperature between IRAS and LWS to $\sim$125 K. This decrease may be explained in part by the steep slopes of modified blackbody spectra at 100-200 K, and the different band widths of IRAS and LWS, making the flux density in the IRAS 12 $\mu$m band (8.0--14.5 $\mu$m band width) be dominated by emission at 14.5 $\mu$m -- outside the LWS 11.7 $\mu$m band. However, simulating modified blackbody spectra at the temperature range 100-200 K show that for 2 out of the 10 sources with LWS $T_{24.5/11.7}$ temperatures (11.94545$-$0.03634 and 27.18725$-$0.08095), the bandpass difference alone cannot explain the lower LWS temperatures.

We conclude that at least in these two cases, this decrease is a result of the improved angular resolution:
the larger beam of IRAS averaged out many lines-of-sights with both low and high extinctions, but the higher resolution of LWS means looking at the peak of the column density. Extinction local to the UC \hii region may be caused, for example, by a broad silicate absorption feature centered at 9.7 $\mu$m, reaching into the 11.7 $\mu$m bandpass. A larger extinction at 11.7 $\mu$m compared to 24.5 $\mu$m will make the \hii region appear cooler in the higher resolution observations.

The original location of the limiting line in Figure \ref{cdiag} could have let in contamination from other source populations, which can explain the broader latitude distribution found using the lower resolution IRAS catalog (Wood \& Churchwell 1989b) compared to works incorporating the better resolution of the MSX catalog (Giveon et al. 2005a). Such a change should be established using a larger sample of UC \hii regions. A change in the Wood \& Churchwell color criteria for selecting UC \hii regions, will change their estimation of the formation rate of O stars in the Galaxy, and may lead to alleviation of the age problem of UC \hii regions (\S \ref{correlate}).

\section{Conclusions}
\label{conc}

We have presented high-resolution observations of a sub-sample of UC \hii regions candidates drawn from the sample of Giveon et al. (2005a) in the first Galactic quadrant. We have presented sub-arcsecond resolution flux density maps, dust temperature and emission optical depth maps of these sources, and have studied the relations between their central stellar objects and the dust properties. The improved angular resolution of our observations ($\times 50$ compared to the MSX; $\times 150$ compared to IRAS) lead to better pinpointing of the UC \hii regions within the MIR sources, and thus, to better flux density and morphological determination. Some of the sources are still compact, but the peaks of all sources are resolved.

Our main findings are:
\begin{itemize}
\item[1.] Half of our objects are excited by B-stars and not by O-stars, and all our objects have derived types that are later than an O6 star, even for radio-derived luminosities and assuming far kinematic distances when ambiguous.
Only two sources (10\%) show a significant inconsistecy between the MIR and radio-derived spectral type, which may be explained by the MIR luminosities ($L_{25/18}$) strongly underestimating the bolometric luminosities, or by foreground extinction of the MIR wavelengths (equivalent to $A_V=30$). In general, approximating the bolometric luminosities using the longer wavelengths (17.65 and 24.5 $\mu$m) is better compared to the shorter wavelengths (11.7 and 17.65 $\mu$m).
The average spectral type derived from $L_{25/18}$ is B0 ($\pm 1$, i.e., in the range B1--O9) -- the same as the radio-derived average spectral type.

\item[2.] The majority of our sources are optically thin at MIR wavelengths, with a significantly optically thinner ($\tau_{11.7}^{max}<0.07$ for all sources) hot dust (150--250 K), compared to an optically thicker ($\tau_{24.5}^{max}>0.08$ for all sources and $>1$ in two cases) component of warm dust (60--110 K). Most of the optical photons are absorbed by the warm dust.

\item[3.] The 24.5 $\mu$m flux densities of our sources are correlated with the 6cm radio flux densities. This relation is fitted by $F_{\rm 24.5\mu m}[{\rm Jy}]=(0.38\pm 0.05)\cdot F_{\rm 5GHz}[{\rm mJy}] + (45\pm 8)$. The correlations of the 11.7 and 17.65 $\mu$m bands are weaker, further supporting our result that the 24.5 $\mu$m flux density is a better estimator of the total flux density reprocessed by the dust.

\item[4.] The MIR sizes are correlated with the source ionizing photon rates, confirming the picture of dust cocoons enveloping Str\"omgren spheres of ionized gas. We find the relation $\log N_{lyc}\ [s^{-1}]=(1.98\pm 0.09)\log D_{11.7}\ [pc]-(50.1\pm 0.1)$. This correlation suggests that the more compact sources are actually due to later spectral types rather than young age, which may lead to alleviation of the age problem of UC \hii regions. A possible explanation for missing the earliest O-stars might be a bias imposed by selecting the most compact radio sources (5 GHz diameters $\le 5''$).

\item[5.] The new flux density estimations lead on average to redder MIR colors or cooler temperature for our sources compared to IRAS observations. For at least two of our sources, the reddening effect is too big to be exclusively explained by the different bandpass response of IRAS and LWS. We conclude that in these cases the significantly smaller beam of LWS focuses on larger dust columns towards the more accurate positions of the UC \hii regions, while the larger beam of IRAS averaged out many lines-of-sight. Our sources are thus shifted to redder colors in the Wood \& Churchwell (1989b) color diagram that is used for selecting UC \hii regions based on their MIR colors. This may account for some contaminations in selecting UC \hii regions using the IRAS criteria, and may lead to a change in the estimated star-formation when larger samples are observed at sub-arcsecond resolution.

\item[6.] The MIR maps show a good overall match in shape when compared to existing radio images with comparable resolution.
 \end{itemize}

The significant improvement in angular resolution and the ability to pinpoint the UC \hii regions lead us to conclude that such an analysis applied to a larger sample would improve our understanding of star-formation in the Milky Way: what are the effects of including a possible large population of B-stars, and the reddening effect we discovered on color-selection criteria of UC \hii regions in both the Galactic and extragalactic objects?
Our observations had a 25\% rate of non-detections with LWS of sources with high MIR-radio matching probabilities based on the MSX and the VLA data. We suspect the reason is the large difference in angular resolution between LWS and MSX, combined with the small field of view of LWS. This either caused a false match between the IR source and a radio source, or even if the match is real, the IR source might be a low surface brightness source, below the detection limit of LWS. The GLIMPSE survey is most suitable for confirming the MIR-radio associations of sources in our original sample (Giveon et al. 2005a), since it has higher resolution than the MSX. However, it is limited in wavelength. Additional observations which will compensate for that deficiency are thus needed.

\section*{Acknowledgments}

Special thanks to James De Buizer for his valuable comments and support.
MJR acknowledges the support by NASA under award NNG04GG92G and by NSF under award AST-0307497.
R.H.B. acknowledges the support of the National Science Foundation under grants AST-02-655. R.H.B.'s work was supported in part under the auspices of the US Department of Energy by Lawrence Livermore National Laboratory under contract W-7405-ENG-48.
R.L.W. acknowledges the support of the Space Telescope Science Institute, which is operated by the Association of Universities for Research in Astronomy, Inc., under NASA contract NAS5-26555.

\end{document}